\documentclass[aps,prd,reprint,superscriptaddress,showpacs,nofootinbib]{revtex4-1}
\usepackage{graphicx}
\usepackage{amsfonts}
\usepackage{amsmath}
\usepackage{mathrsfs}
\usepackage{epsfig}
\usepackage{slashed}
\usepackage{dcolumn}%

\usepackage{ulem}
\usepackage{color}
\usepackage{caption}
\usepackage{subcaption}

\def\aap{Astron.\ Astrophys.\ }
\def\apj{Astrophys.\ J.\ }
\def\apjl{Astrophys.\ J.\ Lett.\ }

\def\mnras{Mon.\ Not.\ Roy.\ Astron.\ Soc.\ }

\def\prd{Phys.\ Rev.\ D\ }

\def\jcap{J.\ Cosmol.\ Astropart.\ Phys.\ }
\def\ssr{Space\ Sci.\ Rev.\ }
\def\aaps{Astron.\ Astrophys.\ Supp.\ }
\def\pasp{Publications\ of\ the\ Astronomical\ Society\ of\ the\ Pacific}

\newcommand\diff{\,\mathrm{d}}
\def\kpc{\,\mathrm{kpc}}
\def\km{\,\mathrm{km}}

\def\GeV{\,\mathrm{GeV}}
\def\MeV{\,\mathrm{MeV}}
\def\GV{\,\mathrm{GV}}
\def\TV{\,\mathrm{TV}}

\def\cm{\,\mathrm{cm}}
\def\s{\,\mathrm{s}}
\def\p{\,\mathrm{p}}
\def\He{\,\mathrm{He}}
\def\A{\,\mathrm{A}}
\def\sr{\,\mathrm{sr}}
\newcolumntype{p}{D{,}{\pm}{-1}}
\def\pbar{\,\bar{\text{p}}}
\def\pbarp{\,\bar{\text{p}}/\text{p}}

\begin{document}


\title{Galactic cosmic-ray model in the light of AMS-02 nuclei data}

\author{Jia-Shu Niu}
\email{jsniu@itp.ac.cn}
\affiliation{CAS Key Laboratory of Theoretical Physics, Institute of Theoretical Physics, Chinese Academy of Sciences, Beijing, 100190, China}
\affiliation{School of Physical Sciences, University of Chinese Academy of Sciences, No.~19A Yuquan Road, Beijing 100049, China}

\author{Tianjun Li}%
\email{tli@itp.ac.cn}

\affiliation{CAS Key Laboratory of Theoretical Physics, Institute of Theoretical Physics, Chinese Academy of Sciences, Beijing, 100190, China}
\affiliation{School of Physical Sciences, University of Chinese Academy of Sciences, No.~19A Yuquan Road, Beijing 100049, China}

\date{\today}

\begin{abstract}

Cosmic ray (CR) physics has entered a precision-driven era. With the latest AMS-02 nuclei data 
(boron-to-carbon ratio, proton flux, helium flux and antiproton-to-proton ratio),
we perform a global fitting and constrain the primary source and propagation parameters of cosmic rays 
in the Milky Way by considering 3 schemes with different data sets (with and without $\pbarp$ data) 
and different propagation models (diffusion-reacceleration and diffusion-reacceleration-convection models). 
We find that the data set with $\pbarp$ data can remove the degeneracy between the propagation parameters effectively 
and it favors the model with a very small value of convection (or disfavors the model with convection). 
The separated injection spectrum parameters are used for proton and other nucleus species, which reveal 
the different breaks and slopes among them. Moreover, the helium abundance, antiproton production cross sections 
and solar modulation are parametrized in our global fitting.
Benefited from the self-consistence of the new data set, the fitting results show a little bias, 
and thus the disadvantages and limitations of the existed propagation models appear.
Comparing to the best fit results for the local interstellar spectra ($\phi = 0$) with the VOYAGER-1 data, 
we find that the primary sources or propagation mechanisms should be 
different between proton and helium (or other heavier nucleus species). Thus, how to explain these results properly
is an interesting and challenging question.

\end{abstract}


                              
\maketitle


\section{Introduction}

Galactic cosmic rays (CRs) carry abundant information about their sources and the propagation environments, which provide us a useful tool to probe the properties of the structure of the galaxy, the interstellar medium (ISM) and even dark matter (DM) in the galaxy. During the propagation, the spatial information of CRs' source lost because of the charged CRs diffusive propagation the turbulence of stochastic magnetic field in the galaxy, and they experience possibly the reacceleration, convection, spallation, and energy loss processes \citep{Strong2007}. As a result, the propagation of CRs in the Milky Way becomes a fundamental theme to understand the origin and interactions of galactic CRs. 

The propagation process can be described by the diffusive transport equation \citep{Strong2007}. Based on different simplifications,  the transport equation can be solved analytically \citep{Webber1992,Bloemen1993,Maurin2002a,Shibata2004}. Alternatively, some numerical packages developed to include most of the relevant processes and the observation-based astrophysical inputs to solve the propagation equation in a self-consistent way, e.g., {\sc galprop} \citep{Strong1998}, {\sc dragon} \citep{Evoli2008} and {\sc picard} \citep{Kissmann2014}. Based on these numerical codes, we could set the relevant parameters of the propagation model and get the results according to calculation. These results can be compared with the observational data, and improve the propagation parameters inversely.

The propagation of CRs couples closely with the source, leading to the entanglement between source parameters and propagation parameters.  Fortunately, the secondary-to-primary ratios of nuclei are almost independent of the source injection spectrum. They are always employed to constrain the propagation parameters in 
 the propagation equation~\citep{Strong2007}. Generally used are the Boron-to-Carbon ratio (B/C) and unstable-to-stable Beryllium ratio ($^{10}$Be/$^9$Be) (see, e.g., \citep{Trotta2011,Johannesson2016,Lin2015,Yuan2017}). But the $^{10}$Be/$^9$Be data are always with large uncertainties and from different experiment, which always bring large systematics into the subsequent fitting. Recently, \citet{Jin2015} claimed that the combination of B/C ratio and the proton flux can lift the degeneracy in $z_h$ (the half-height of the propagation region) and $D_0$ (the normalization of the diffusion coefficient), and both parameters can be determined by the AMS-02 data alone, which seriously depends on the precision of the data.

The space station experiment Alpha Magnetic Spectrometer (AMS-02), which was launched in May 2011, improve the measurement precision of the CR fluxes by an order of the systematics \citep{AMS2013}. With the results of AMS-02, we could study the CR physics more quantitatively than qualitatively \citep{Jin2013,Feng2014,Mauro2014,Yuan2015,Lin2015,Jin2015}.
The AMS-02 collaboration has already released its nucleus data for proton \citep{AMS02_proton}, helium \citep{AMS02_helium}, B/C \citep{AMS02_b_c}, $\pbarp$ and $\pbar$ \citep{AMS02_pbar_proton}, which provide us the opportunity to study the primary source and propagation models effectively and precisely.

Considering the situations of  high-dimensional parameter space of propagation model and precise data sets, we employ a Markov Chain Monte Carlo (MCMC \citep{Lewis2002}) method (embed by {\sc galprop}) to do global fitting and sample the parameter space of CR propagation and nuclei injections \citep{Liu2010,Lin2015,Yuan2017}.
In this work, we use the AMS-02 nuclei data only, to study 
 3 schemes with different data sets and different propagation models. 
Specifically, the propagation models include the diffusion-reacceleration (DR) model \citep{Trotta2011,Johannesson2016} and the diffusion-reacceleration-convection (DRC) model \citep{Yuan2017}.
Thus the systematics between different experiments are avoided. 
Additionally, because of the significant difference in the slopes of proton and helium, of about $\sim 0.1$ \citep{Adriani2011,AMS02_proton,AMS02_helium}, has been observed, we use separate primary source spectra settings for proton 
and other nuclei (all $Z >1$ nuclei have the same injection parameters).

The paper is organized as follows. We first introduce the theoretical aspects on the propagation of CRs in the Galaxy in Sec. II. The fitting procedure is give in Sec. III. After analysis the fitting results in Sec. IV, 
 we present some discussions in Sec. V and conclusions in Sec. VI.

\section{Theory}
\label{sec:theory}
Galactic CR particles diffuse in the Galaxy after being accelerated, 
 experiencing the fragmentation and energy loss in the ISM 
and/or the interstellar radiation field (ISRF) and magnetic field, as well as
decay and possible reacceleration or convection. Denoting the density of CRs per unit momentum interval as $\psi$ (which is related to the phase space density $f(\mathbf{r},\mathbf{p}, t)$ as $\psi(\mathbf{r},p,t)=4\pi p^{2}f(\mathbf{r},\mathbf{p},t) $), the propagation can be described by the
propagation equation \citep{Strong2007}

\begin{align}
\label{eq:propagation_equation}
\frac{\partial \psi}{\partial t} = \ & Q(\mathbf{r}, p) + \nabla \cdot \left( D_{xx}\nabla\psi - \mathbf{V}_{c}\psi \right) + \frac{\partial}{\partial p}p^2D_{pp}\frac{\partial}{\partial p}\frac{1}{p^2}\psi \notag \\
& \ - \frac{\partial}{\partial p}\left[ \dot{p}\psi - \frac{p}{3}\left( \nabla\cdot\mathbf{V}_c\psi \right) \right]- \frac{\psi}{\tau_f} - \frac{\psi}{\tau_r},
\end{align}
where $Q(\mathbf{r}, p)$ is the source distribution, $D_{xx}$ is the spatial diffusion coefficient, $\mathbf{V}_c$ is the convection velocity, $D_{pp}$ is diffusion coefficient in the momentum-space, $\tau_f$ and $\tau_r$ are the characteristic time scales used to describe the fragmentation and radioactive decay.

The convection velocity $\mathbf{V}_c$ is generally assumed to linearly depend on the distance away from the Galaxy disk, $\mathbf{V}_c = \mathbf{z} \cdot \diff V_c / \diff z$, where $\mathbf{z}$ is the position vector in the vertical direction to the galactic disk.  Such a configuration can avoid the discontinuity at the galactic plane.

The diffusion coefficient can be parametrized as 
\begin{equation}
\label{eq:diffusion_coefficient}
D_{xx} = D_0\beta \left( R/R_0 \right)^{\delta}~,~\,
\end{equation}
where $\beta$ is the velocity of the particle in unit of light speed $c$,  $R_{0}$ is the reference rigidity,
and $R\equiv pc/Ze$ is the rigidity.

 The reacceleration effect is always used to describe with the diffusion in momentum space. Considering the scenario in which the CR particles are reaccelerated by colliding with the interstellar random weak hydrodynamic waves, the relation between the spatial diffusion coefficient $D_{xx}$ and the momentum diffusion coefficient $D_{pp}$ can be expressed as
\citep{Seo1994}
\begin{equation}
  \label{eq:reacceleration}
  D_{pp}D_{xx}=\frac{4p^2v^2_{A}}{3\delta(4-\delta^2)(4-\delta)\omega},
\end{equation}
where $v_{A}$ is the Alfven velocity and the parameter $\omega$ is used to characterize the level of the interstellar turbulence. Because only $v^2_A/\omega$ plays a role, we adopt $\omega = 1$ and use $v_A$ to characterize the reacceleration. Free escape is assumed at boundaries, $r_h$ and $z_h$, for the cylindrical coordinate system.

The injection spectra of all kinds of nuclei are assumed to be a broken power law form
\begin{equation}
  q_\mathrm{i}(p)=  N_\mathrm{i}\times\left\{ \begin{array}{ll}
    \left( \dfrac{R}{R\mathrm{_{A}}} \right)^{-\nu_{\A1}} & R \le R_{\A}\\
    \left( \dfrac{R}{R\mathrm{_{A}}} \right)^{-\nu_{\A2}} & R > R_{\A}
  \end{array}
  \right.,
  \label{eq:injection_spectra}
\end{equation}
where $\mathrm{i}$ denotes the species of nuclei, $N_\mathrm{i}$ is the normalization constant proportional to the relative abundance of the corresponding nuclei, and $\nu_{\A}=\nu_{\A1}(\nu_{\A2})$ for the nucleus rigidity $R$ below (above) a reference rigidity $R_{\A}$. In this work, we use independent proton injection spectrum, and 
the corresponding parameters are $R_{\p}$, $\nu_{\p1}$, and $\nu_{\p2}$. 
 All the $Z > 1$ nuclei are assumed to have the same value of injection parameters.

The radial distribution of the source term can be determined by independent observables. Based on the distribution of SNR, the spatial distribution of the primary sources is assumed to have the following form \citep{Case1996}
\begin{equation}
\label{eq:source-distribution}
f(r,z) = q_{0} \left(  \frac{r}{r_{\odot}} \right)^{a} \exp 
\left[
-b \cdot \frac{r-r_{\odot}}{r_{\odot}}
-\frac{|z|}{|z_{s}|}
 \right] \ ,
\end{equation}
where $a=1.25$ and $b=3.56$ are adapted to reproduce the Fermi-LAT gamma-ray data of the 2nd Galactic quadrant \citep{Strong1998,Trotta2011,Tibaldo2009}, $z_{s} \approx 0.2$ kpc is the characteristic height of Galactic disk, and $q_{0}$ is a normalization parameter. In the 2D diffusion model, one can use the realistic nonuniform interstellar gas distribution of $\text{H}_{\text{I,II}}$ and $\text{H}_{2}$ determined from 21cm and CO surveys. Thus, the injection source function for a specific CR species can be written as follows
\begin{equation}
\label{eq:primary_source}
Q(\mathbf{r}, p) = f(r,z) \cdot q_{i}(p)  .
\end{equation}

The secondary  cosmic-ray particles are produced in collisions of primary cosmic-ray particles with ISM. 
And the secondary antiprotons are generated dominantly from inelastic pp-collisions and pHe-collisions. 
The corresponding source term  is 
\begin{equation}
  \label{eq:secondary_source}
q(\pbar)=
\beta c n_{i} \sum_{i=\text{H,He}}
\int \diff p'   \frac{\diff \sigma_{i}(p,p')}{\diff p'} n_{\p}(p')
\end{equation}
where $n_{i}$ is the number density of interstellar hydrogen (helium), $n_{\p}$ is the number density of primary cosmic-ray proton per total momentum, and$\diff \sigma_{i}(p,p')/\diff p'$ is the differential cross section for  $\p+\text{H(He)}\to \pbar + X$. Because there are uncertainties from the antiproton production cross section \citep{Tan1983,Duperray2003,Kappl2014,diMauro2014}, we employ an energy-independent  factor $c_{\pbar}$, which has been suggested to approximate the ratio of antineutron-to-antiproton production cross sections \citep{diMauro2014}, to rescale the antiproton flux. The energy dependence of $c_{\pbar}$ is unclear at present \citep{diMauro2014,Kappl2014}. We expect that a constant factor is a simple  assumption.

The  interstellar flux of the cosmic-ray particle is related to its density function as 
\begin{equation}
\label{eq:phi}
\Phi= \frac{v}{4\pi} \psi(\boldsymbol{r},p)~.
\end{equation}

For high energy nuclei $v\approx c$. We adopt the force-field approximation \citep{Gleeson1968} to describe the effects of solar wind and helioshperic magnetic field in the solar system, which contains only one parameter the so-called solar-modulation $\phi$.  In this approach, the cosmic-ray nuclei flux at the top of the atmosphere of the Earth which is observed by the experiments $\Phi_{\text{obs}}$  is related to the interstellar flux as follows
\begin{equation}
\label{eq:solar_modulation}
\Phi_{\text{obs}}(E_{\text{obs}})=\left(\frac{2m  E_{\text{obs}}+E_{\text{obs}}^{2}}{2m  E_{\text{kin}}+E_{\text{kin}}^{2}}\right)\Phi(E_{\text{kin}}) ,
\end{equation}
where $E_{\text{obs}}=E_{\text{kin}}-|Z| e \phi$ is the kinetic energy of the cosmic-ray nuclei measured by the experiments, where $Z$ is the charge number of the cosmic ray particles.

The public code  {\sc galprop} v54 \footnote{http://{\sc galprop}.stanford.edu} r2766 \footnote{https://sourceforge.net/projects/{\sc galprop}/} \citep{Strong1998,Moskalenko2002,Strong2001,Moskalenko2003,Ptuskin2006} was used to solve the diffusion equation of 
Eq.~(\ref{eq:propagation_equation}) numerically. {\sc galprop} utilizes  the realistic astronomical information on the distribution of interstellar gas and other data as input, and considers various kinds of data including primary and secondary nuclei, electrons and positrons, $\gamma$-rays, synchrotron  radiation, etc, in a self-consistent way. Other approaches based on simplified assumptions on the Galactic gas distribution which  allow  for fast  analytic solutions can be found in Refs.~\citep{Donato2001,Maurin2002b,Donato2004,Putze2010,Cirelli2011}. Some custom modifications are performed in the original code, such as the possibility to use specie-dependent injection spectra, which is not allowed by default in {\sc galprop}.

The {\sc galprop} primary source (injection) isotopic abundances are taken first as the solar system abundances, which are iterated to achieve an agreement with the propagated abundances as provided by ACE at $\sim$ 200 MeV/nucleon \citep{Wiedenbeck2001,Wiedenbeck2008} assuming a propagation model. The source abundances derived for two propagation models, diffusive reacceleration and plain diffusion, were used in many {\sc galprop} runs. In view of some discrepancies when fitting with the new data which use the default abundance in {\sc galprop} \citep{Johannesson2016}, we use a factor $c_{\He}$ to rescale the helium-4 abundance (which has a default value of $7.199 \times 10^4$) which help us to get a global best fitting.

\section{Fitting Procedure}
\label{sec:fitting_pro}

\subsection{Bayesian inference}
\label{sec:Bayes}
 In this work, we use Bayesian inference to get the posterior probability distribution function (PDF), which is based on the following formula

 \begin{equation}
 \label{eq:Bayes}
p(\boldsymbol{\theta}|D)=\frac{\mathcal{L}(D|\boldsymbol\theta)\pi(\boldsymbol\theta)}{p(D)} ,
\end{equation}
where $\boldsymbol{\theta}=\{\theta_{1},\dots,\theta_{m}\}$ is the free parameter set, $D$ is the experimental data set, $\mathcal{L}(D|\boldsymbol\theta)$ is the likelihood function, and $\pi(\boldsymbol\theta)$ is the prior PDF which represents our state of knowledge on the values of the parameters before taking into account of the new data. 
(The quantity $p(D)$ is the Bayesian evidence which is not that important in this work but it is important for Bayesian model comparison.)

 We take the prior PDF as a uniform distribution
 \begin{equation}
 \label{eq:priors}
\pi(\theta_{i}) \propto
\left\{
\begin{tabular}{ll}
1, &  \text{for } $\theta_{i,\text{min}}<\theta_{i}<\theta_{i,\text{max}}$
\\
0, & \text{otherwise}
\end{tabular}
\right. 
,
\end{equation}
 and the likelihood function as a Gaussian form
 \begin{equation}
 \label{eq:likelihood}
\mathcal{L}(D|\boldsymbol\theta)=
\prod_{i}
\frac{1}{\sqrt{2\pi \sigma_{i}^{2}}}
\exp\left[
  -\frac{(f_{\text{th},i}(\boldsymbol\theta)-f_{\text{exp},i})^{2}}{2\sigma_{i}^{2}}  
\right]  ,  
\end{equation}
where $f_{\text{th},i}(\boldsymbol\theta)$ is the predicted $i$-th observable from the model which depends on the parameter set $\boldsymbol\theta$, and $f_{\text{exp},i}$ is the one measured by the experiment with uncertainty $\sigma_{i}$.

Here we use the algorithms such as the one by \citet{Goodman2010} instead of classical Metropolis-Hastings for its excellent performance on clusters. The algorithm by \citet{Goodman2010} was slightly altered and implemented as the {\tt Python} module {\tt emcee}\footnote{http://dan.iel.fm/emcee/} by \citet{Mackey2013}, which makes it easy to use by the advantages of {\tt Python}. Moreover, {\tt emcee} could distribute the sampling on the multiple nodes of modern cluster or cloud computing environments, and then increase the sampling efficiency observably.

\subsection{Data sets and parameters for different schemes}

 In our work, we propose 3 schemes which utilizes the AMS-02 data (proton \citep{AMS02_proton}, helium \citep{AMS02_helium}, B/C \citep{AMS02_b_c}, and $\pbarp$ \citep{AMS02_pbar_proton}) only to determine the primary source and propagation parameters. The benefits are as follows: (i): the statistics of the AMS-02 data on charged cosmic-ray particles are now much higher than 
the other experiments and will continue to increase; (ii) these data can constitute a complete data set to determine the related parameters; (iii) this scheme can avoid the complicities involving the combination of the systematics of different type of experiments.

 These 3 schemes are given in Table \ref{tab:schemes}.

\begin{table*}[htb]
\begin{center}
\begin{tabular}{c | c | c | c}
  \hline\hline
  Schemes & Propagation Models  & Data Sets \footnote{Considering the degeneracy between $D^{\text{AMS}}_{\pbarp}$ and $D^{\text{AMS}}_{\pbar}$, we just use $D^{\text{AMS}}_{\pbarp}$ to do MCMC fitting and use them together to show the fitting result.}  & Parameters             \\
\hline
I & DR & $\{D^{\text{AMS}}_{\p}, D^{\text{AMS}}_{\He},  D^{\text{AMS}}_{B/C} \}$ & $\{ D_{0}, \delta, z_{h}, v_{A}, | N_{\p},  R_{\p}, \nu_{\p1}, \nu_{\p2},  R_{\A}, \nu_{\A1}, \nu_{\A2}, |  c_{\He},  \phi \}$ \\

  II & DR & $\{D^{\text{AMS}}_{\p}, D^{\text{AMS}}_{\He},  D^{\text{AMS}}_{B/C}, D^{\text{AMS}}_{\pbarp} \}$ & $\{ D_{0}, \delta, z_{h}, v_{A}, | N_{\p},  R_{\p}, \nu_{\p1}, \nu_{\p2},  R_{\A}, \nu_{\A1}, \nu_{\A2}, | c_{\He}, c_{\pbar}, \phi \}$  \\

  III & DRC & $\{D^{\text{AMS}}_{\p}, D^{\text{AMS}}_{\He},  D^{\text{AMS}}_{B/C}, D^{\text{AMS}}_{\pbarp} \}$ & $\{ D_{0}, \delta, z_{h}, v_{A}, \diff V_c/ \diff z, | N_{\p},  R_{\p}, \nu_{\p1}, \nu_{\p2},  R_{\A}, \nu_{\A1}, \nu_{\A2}, | c_{\He}, c_{\pbar}, \phi \}$ \\

\hline\hline

\end{tabular}
\end{center}
\caption{ The propagation models, data sets and parameters of the 3 schemes.}
\label{tab:schemes}
\end{table*}

For DR model, the convection velocity $\mathbf{V}_{c}=0$. We consider the case $R=20 \kpc$ and spatial independent diffusion coefficient. Thus, the major parameters to describe the propagation are $(D_{0}, \delta, v_{A}, z_{h})$. 
For DRC model, the convection velocity is described as  $\mathbf{V}_c = \mathbf{z} \cdot \diff V_c / \diff z$. Thus, the propagation parameters for DRC model are  $(D_{0}, \delta, v_{A}, z_{h}, \diff V_c / \diff z)$.

The primary source term can be determined by Eq.~(\ref{eq:primary_source}), from which we get free parameters: the power-law indices $\nu_{\p1}$ and $\nu_{\p2}$ (for proton), as well as $\nu_{\A1}$ and $\nu_{\A2}$ (for other nuclei); the break 
in rigidity $R_{\p}$ and $R_{\A}$; the normalization factor $N_{\p}$ at a reference kinetic energy $E_{kin}=100 \GeV$; the solar modulation is described by $\phi$. Additionally, as described in Sec. \ref{sec:theory}, we employ a factor $c_{\He}$ to rescale the isotopic abundance of helium \citep{Johannesson2016,Korsmeier2016} and a factor $c_{\pbar}$ to rescale the calculated secondary flux to fit the data (which in fact account for the antineutron-to-antiproton production ratio \citep{diMauro2014,Cui2017}).

The radial and $z$ grid steps are chosen as $\Delta r = 1 \kpc$, and $\Delta z = 0.2 \kpc$. The grid in kinetic energy per nucleon is logarithmic between $10^2$ and $10^7 \MeV$ with a step factor of $1.2$. The free escape boundary conditions are used by imposing $\psi$ equal to zero outside the region sampled by the grid.

These parameters can be separated into three groups: the propagation parameters, the source parameters and nuisance parameters. And their priors are chosen to be uniform distributions according to Eq.~(\ref{eq:priors}) with the prior intervals given in Tables \ref{tab:propagation_params_I}, \ref{tab:propagation_params_II},
 and \ref{tab:propagation_params_III}.

\section{Fitting Results}

\begin{figure*}[!htbp]
  \centering
  \includegraphics[width=0.3\textwidth]{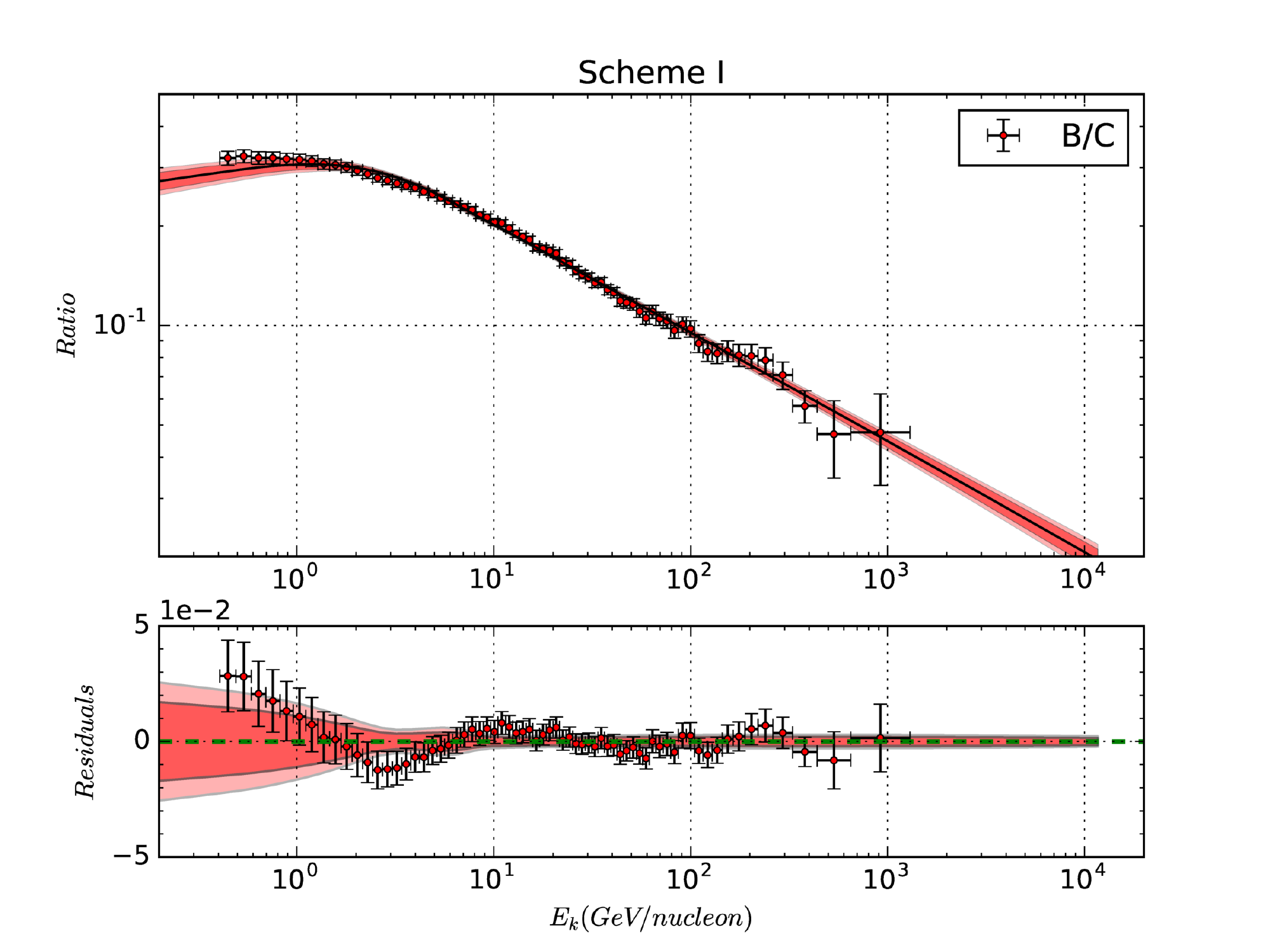}
  \includegraphics[width=0.3\textwidth]{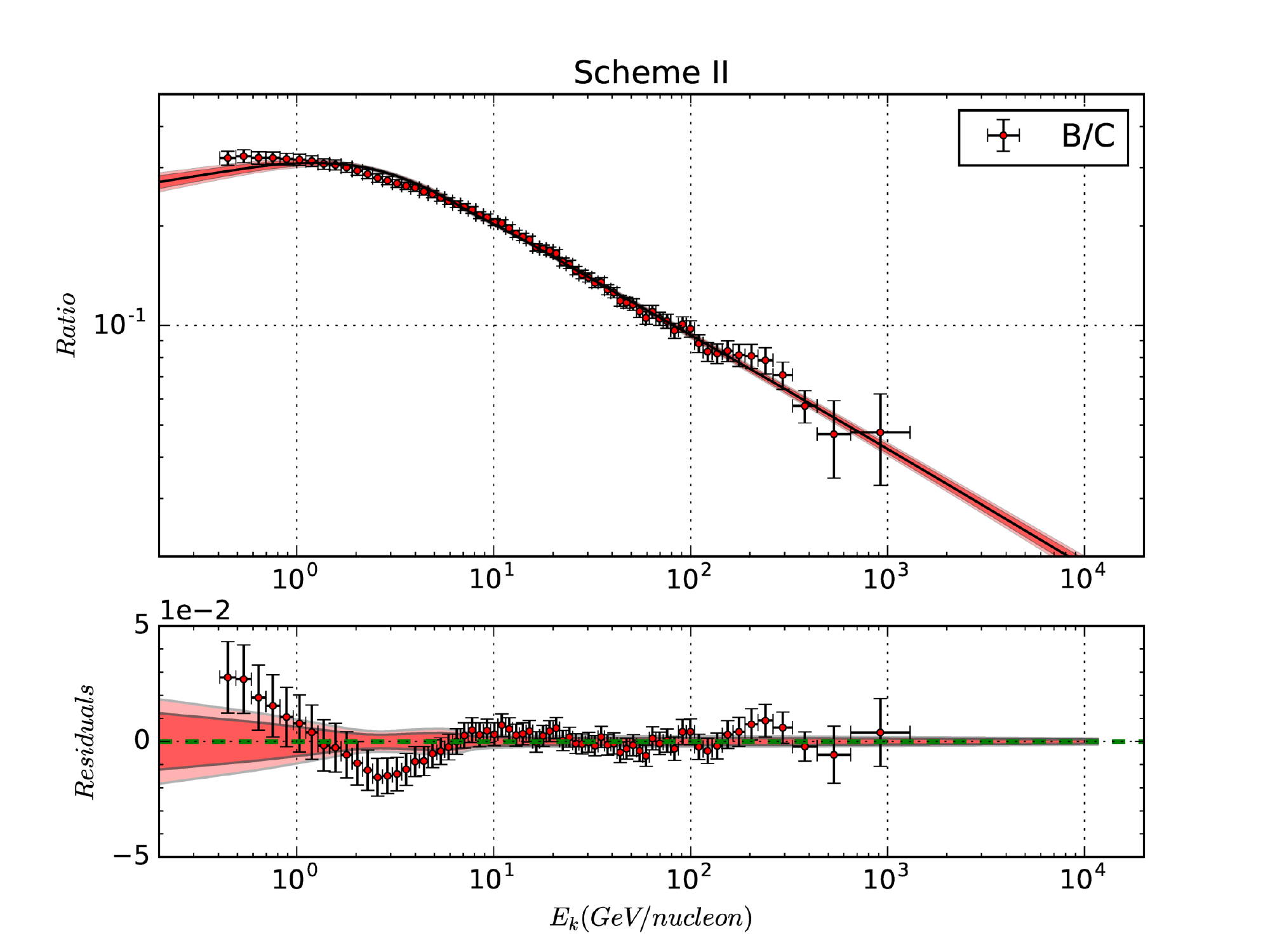}
  \includegraphics[width=0.3\textwidth]{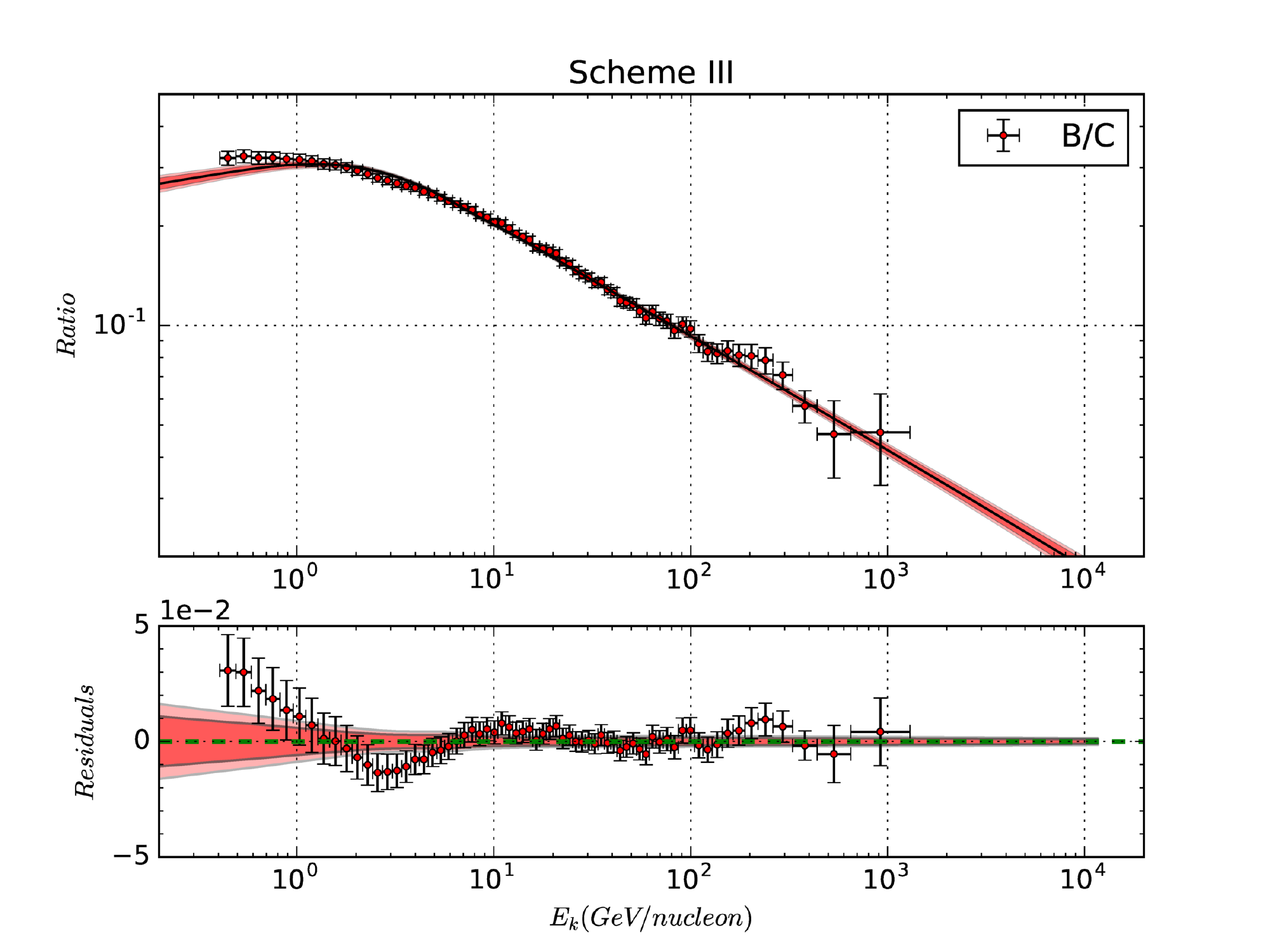}
  \includegraphics[width=0.3\textwidth]{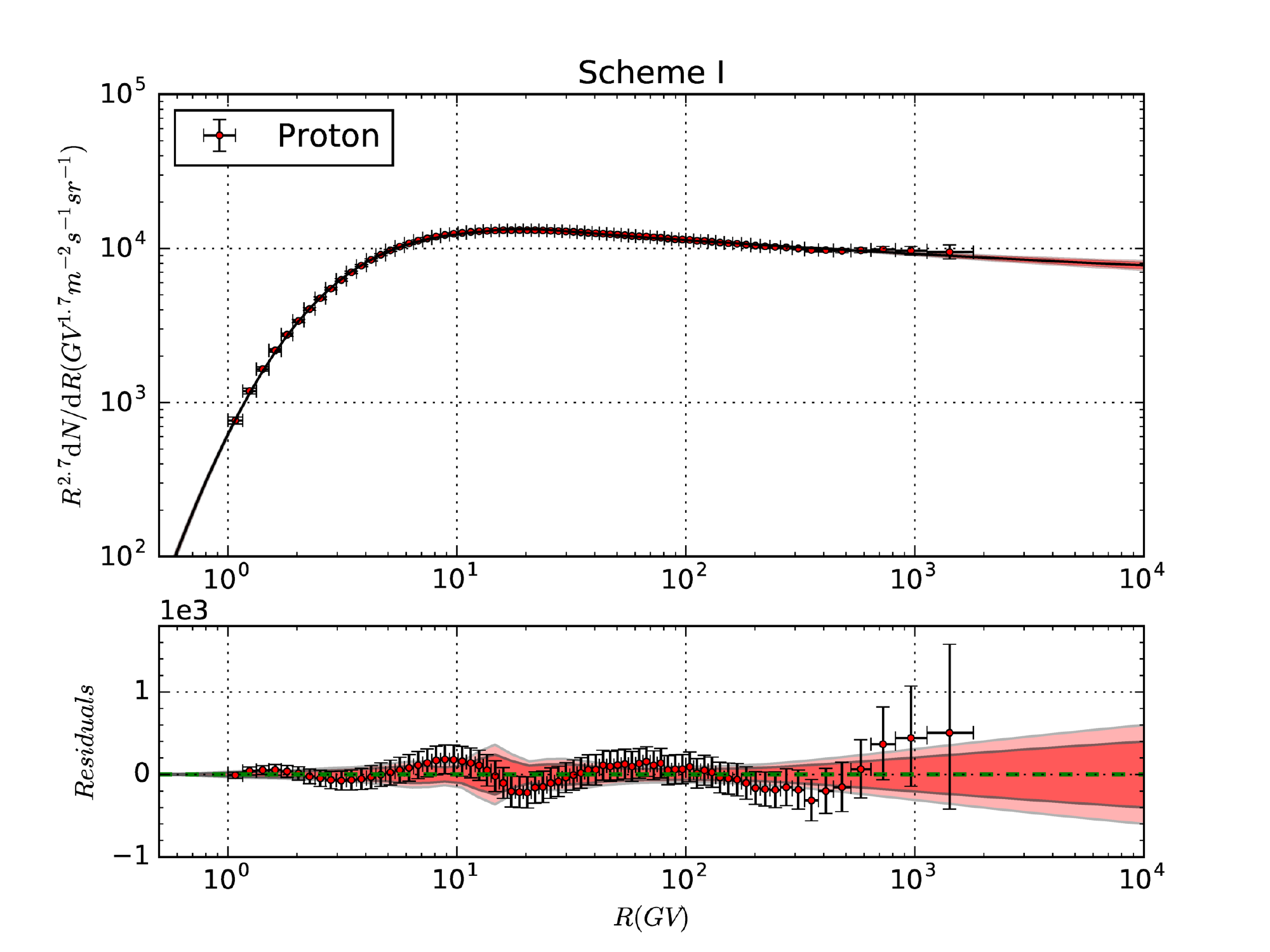}
  \includegraphics[width=0.3\textwidth]{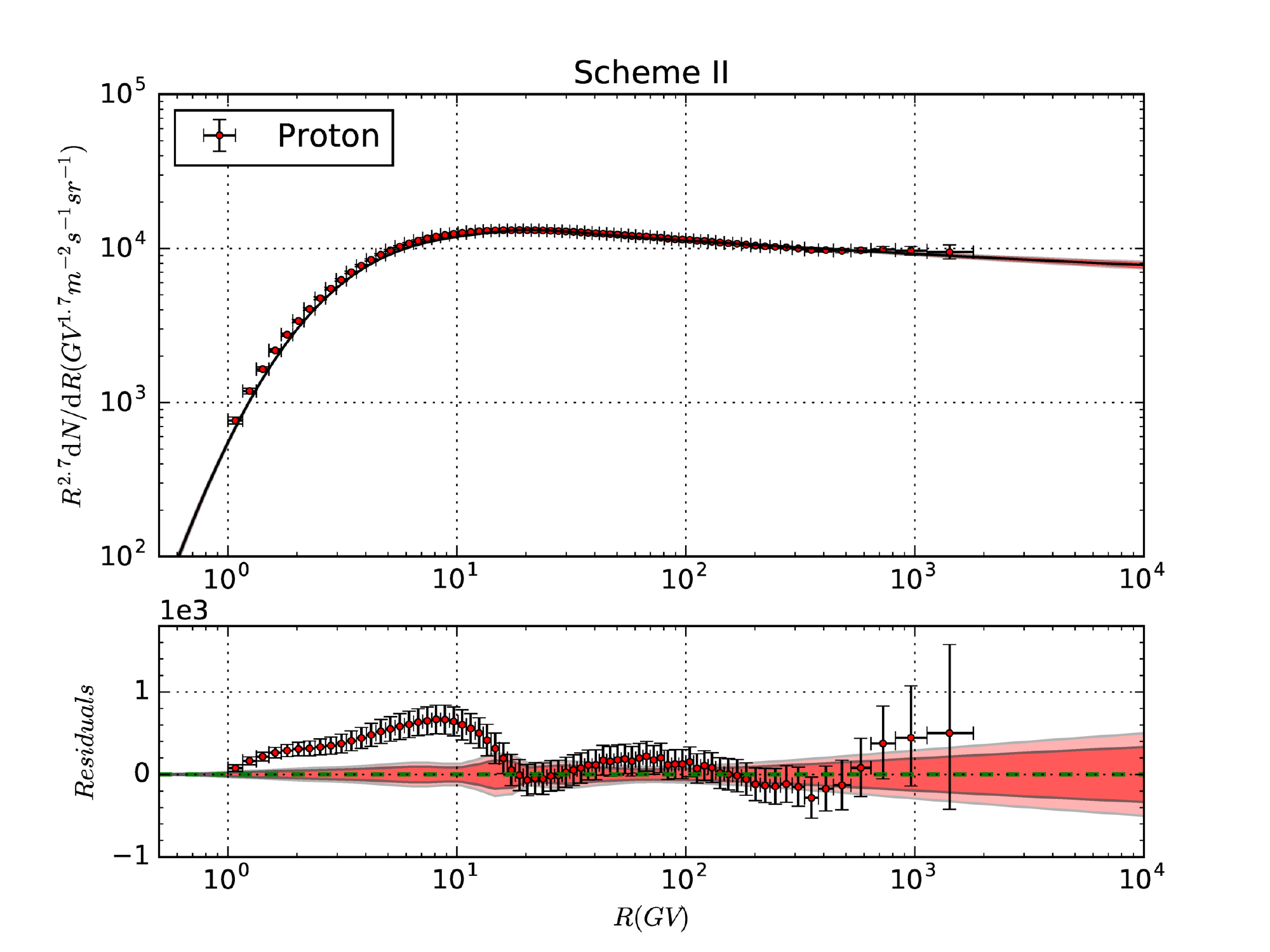}
  \includegraphics[width=0.3\textwidth]{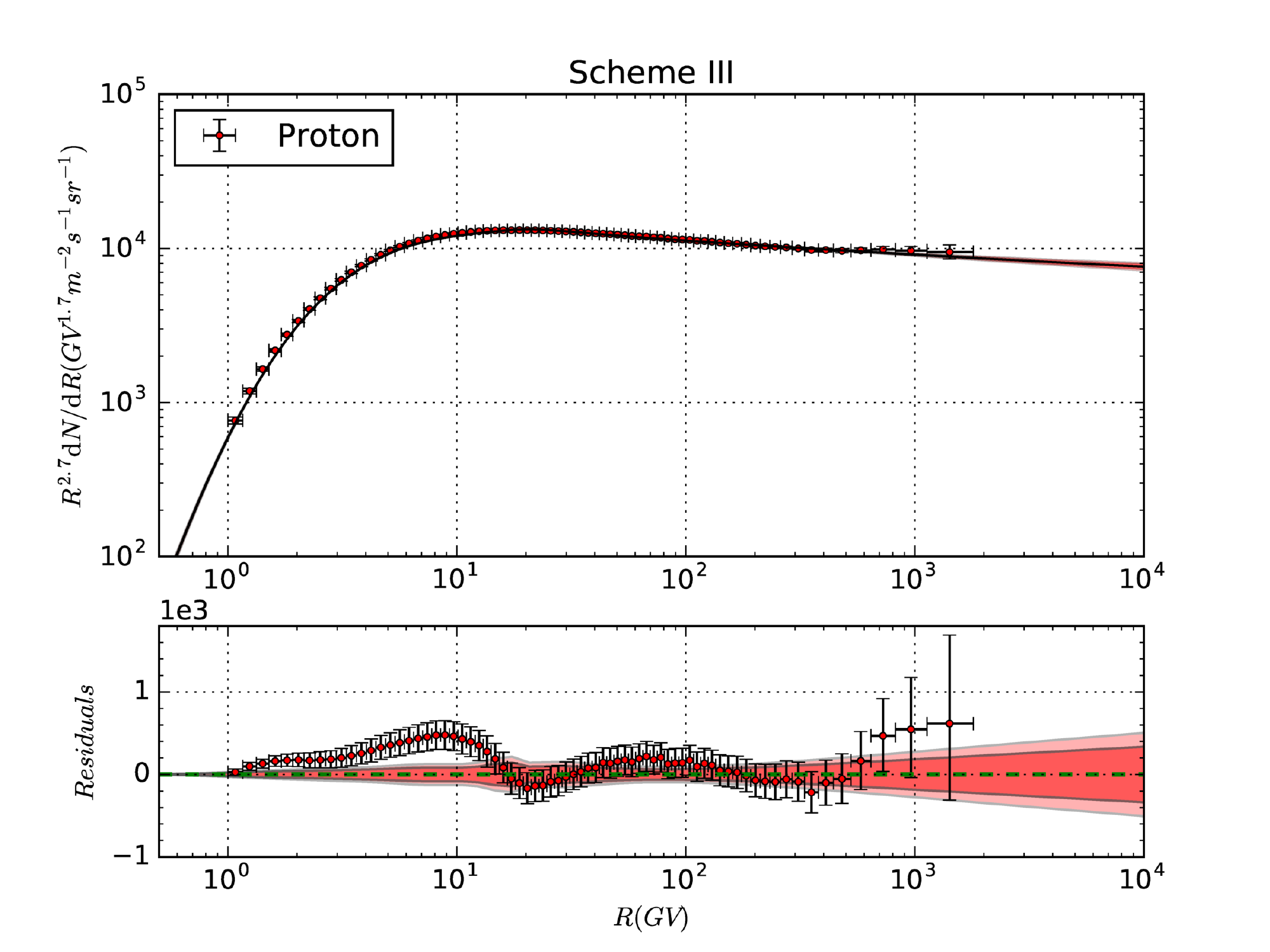}
  \includegraphics[width=0.3\textwidth]{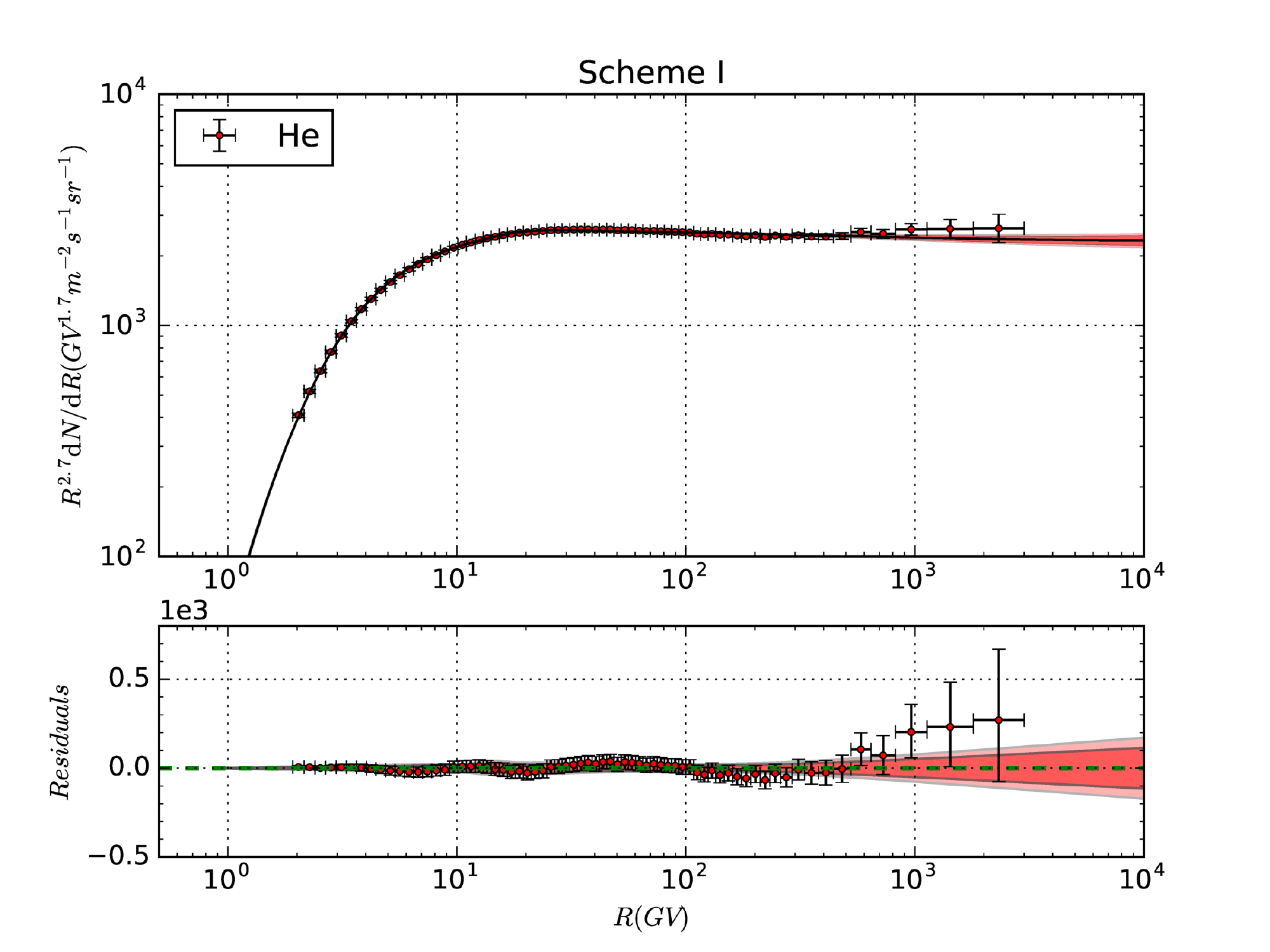}
  \includegraphics[width=0.3\textwidth]{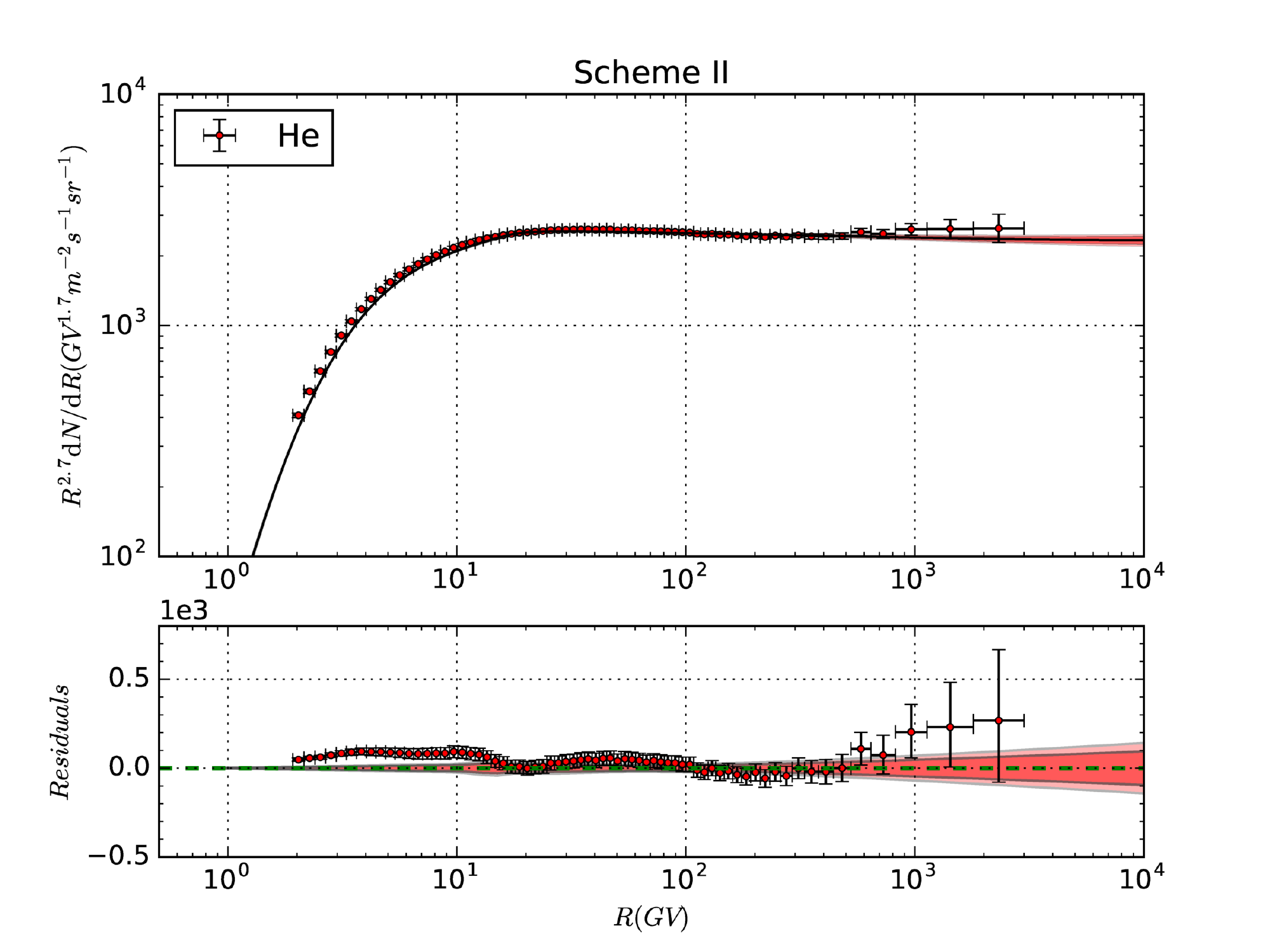}
  \includegraphics[width=0.3\textwidth]{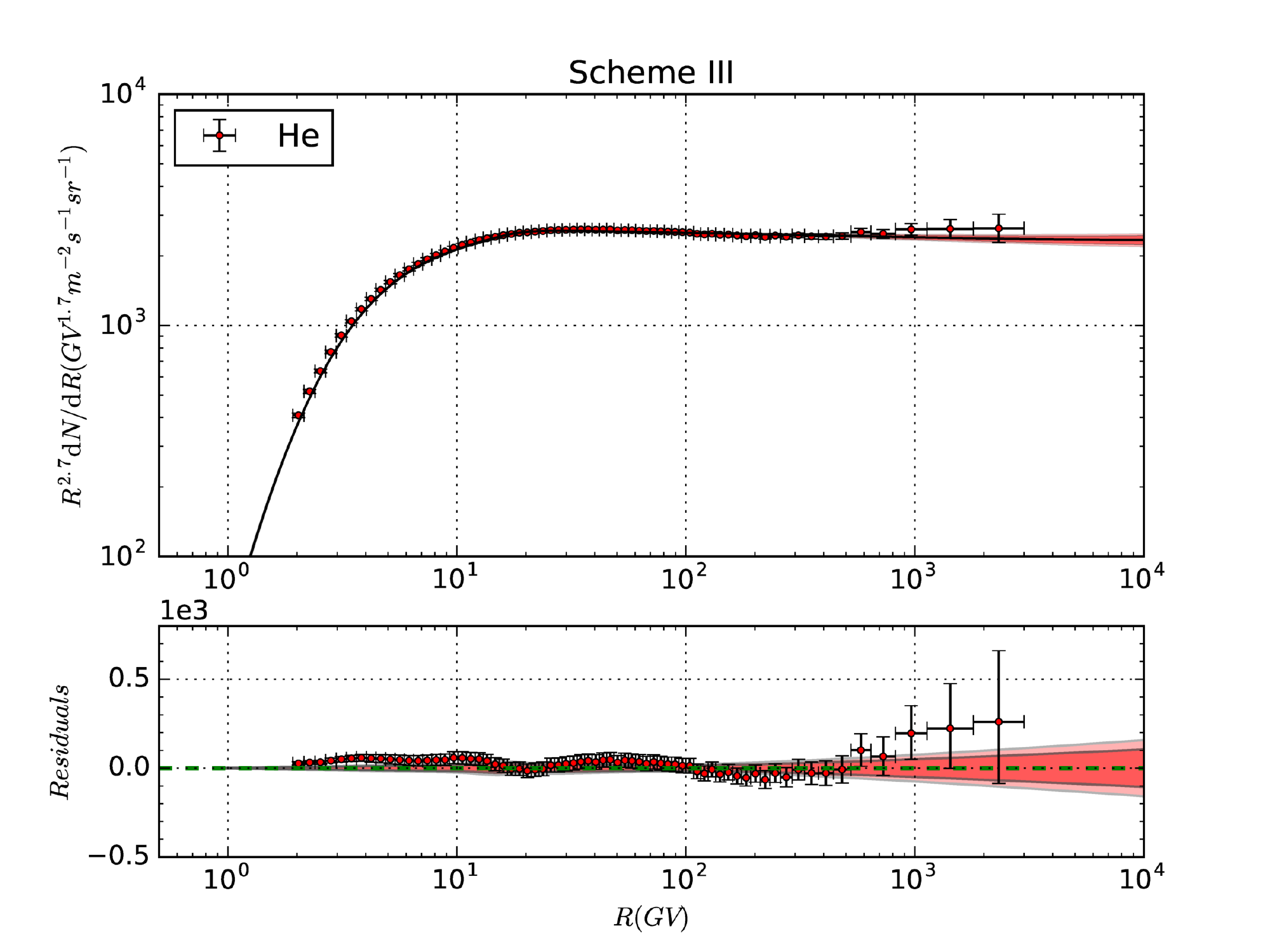}
  \includegraphics[width=0.3\textwidth]{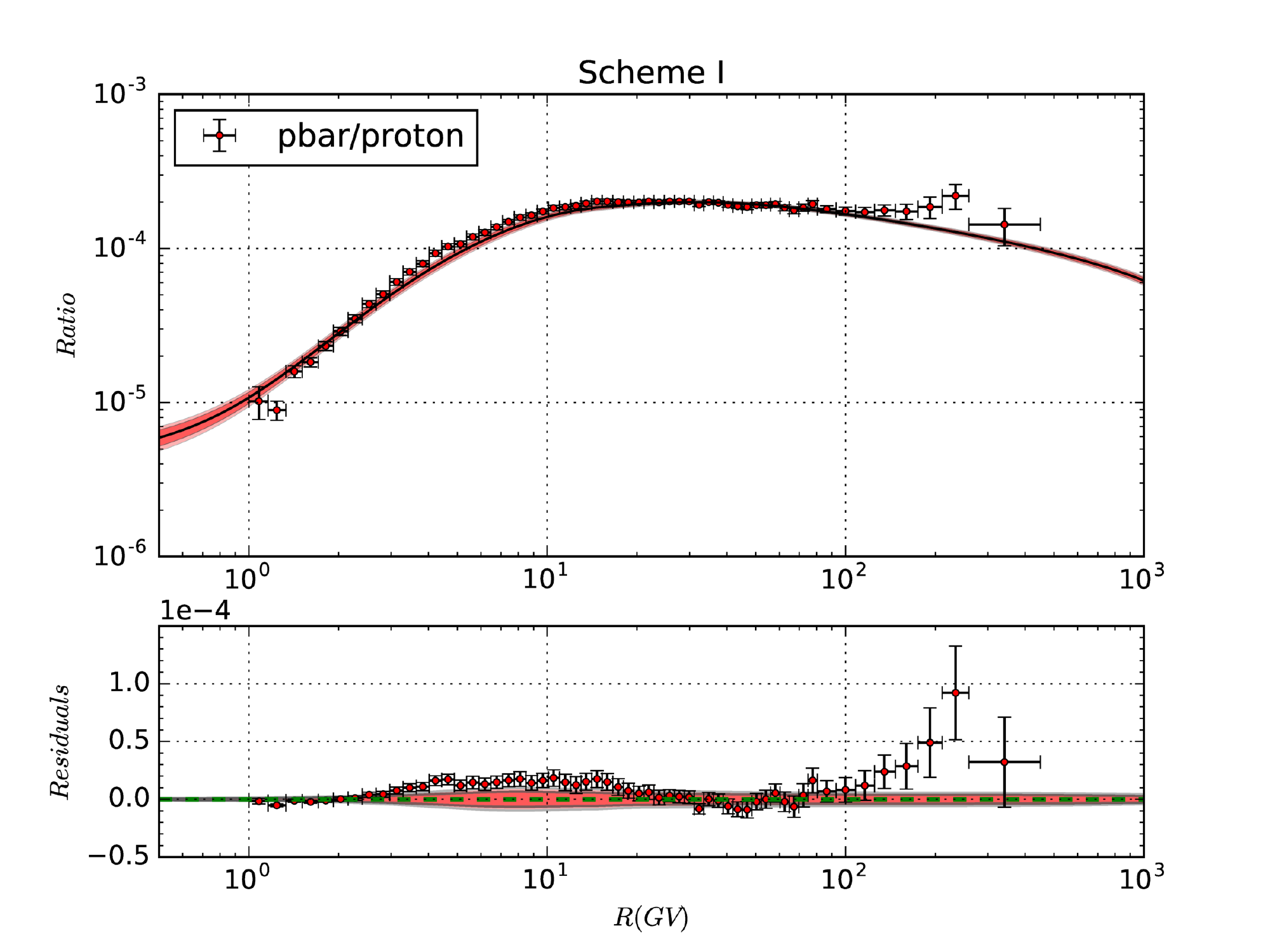}
  \includegraphics[width=0.3\textwidth]{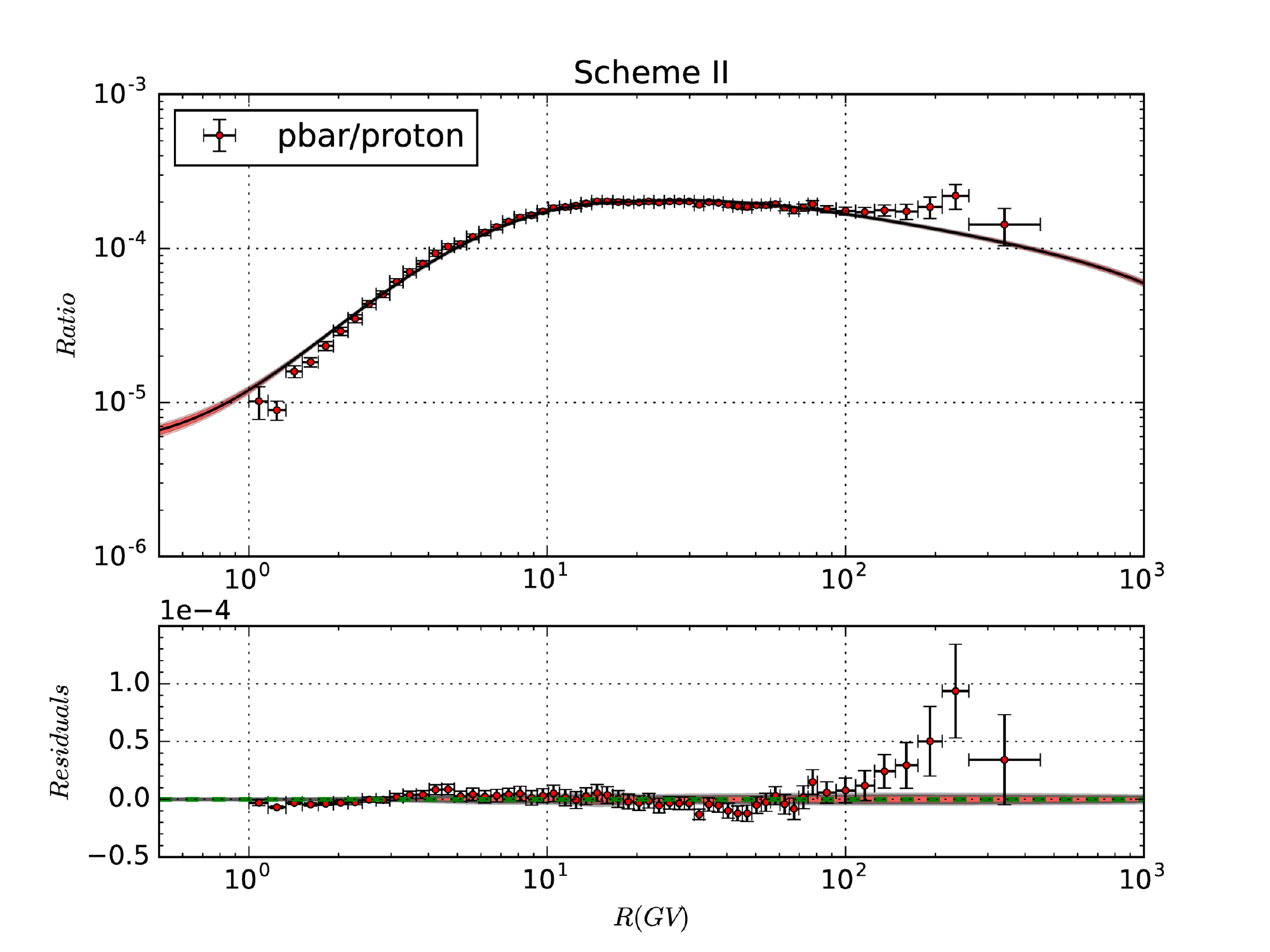}
  \includegraphics[width=0.3\textwidth]{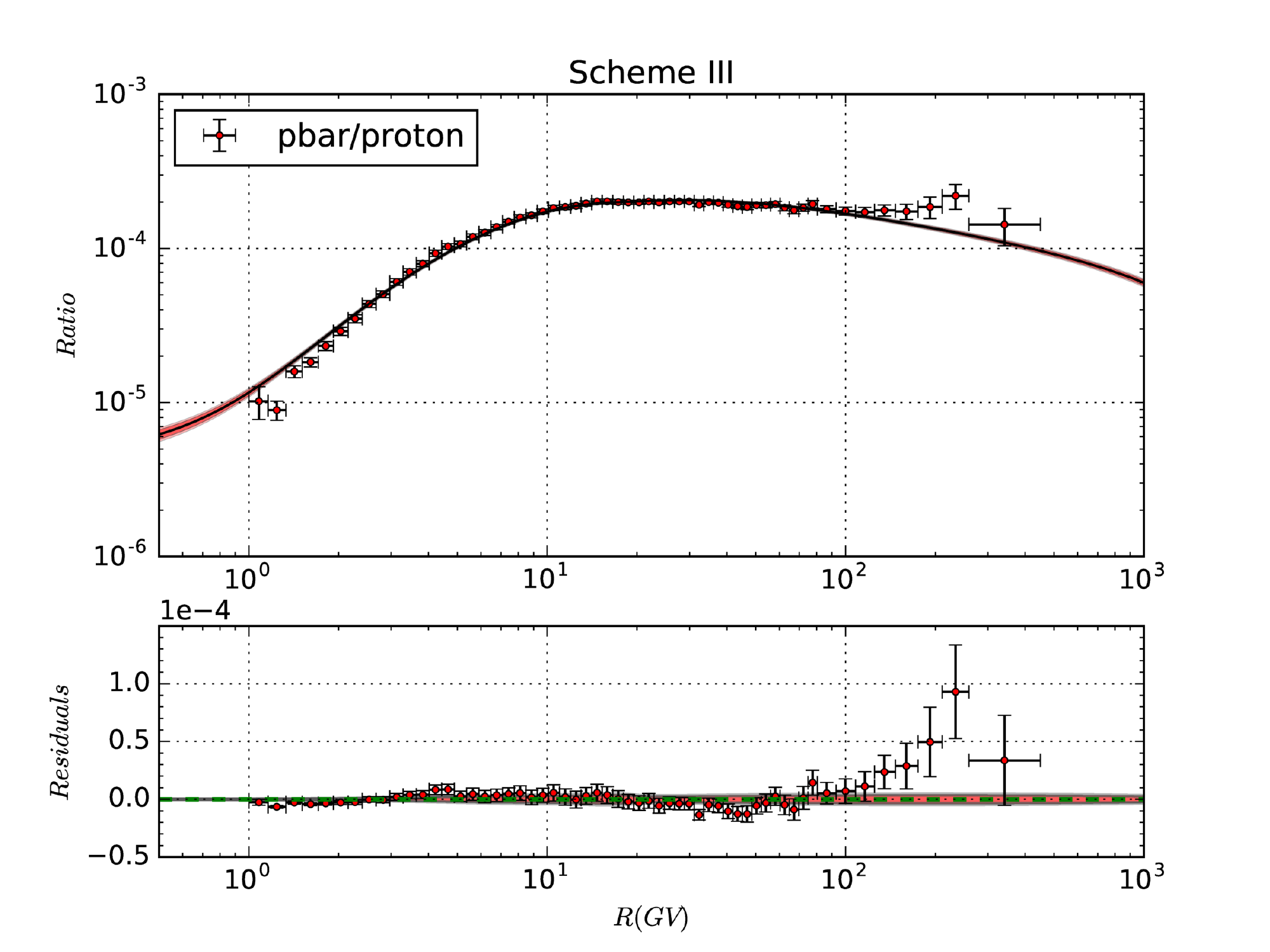}
  \includegraphics[width=0.3\textwidth]{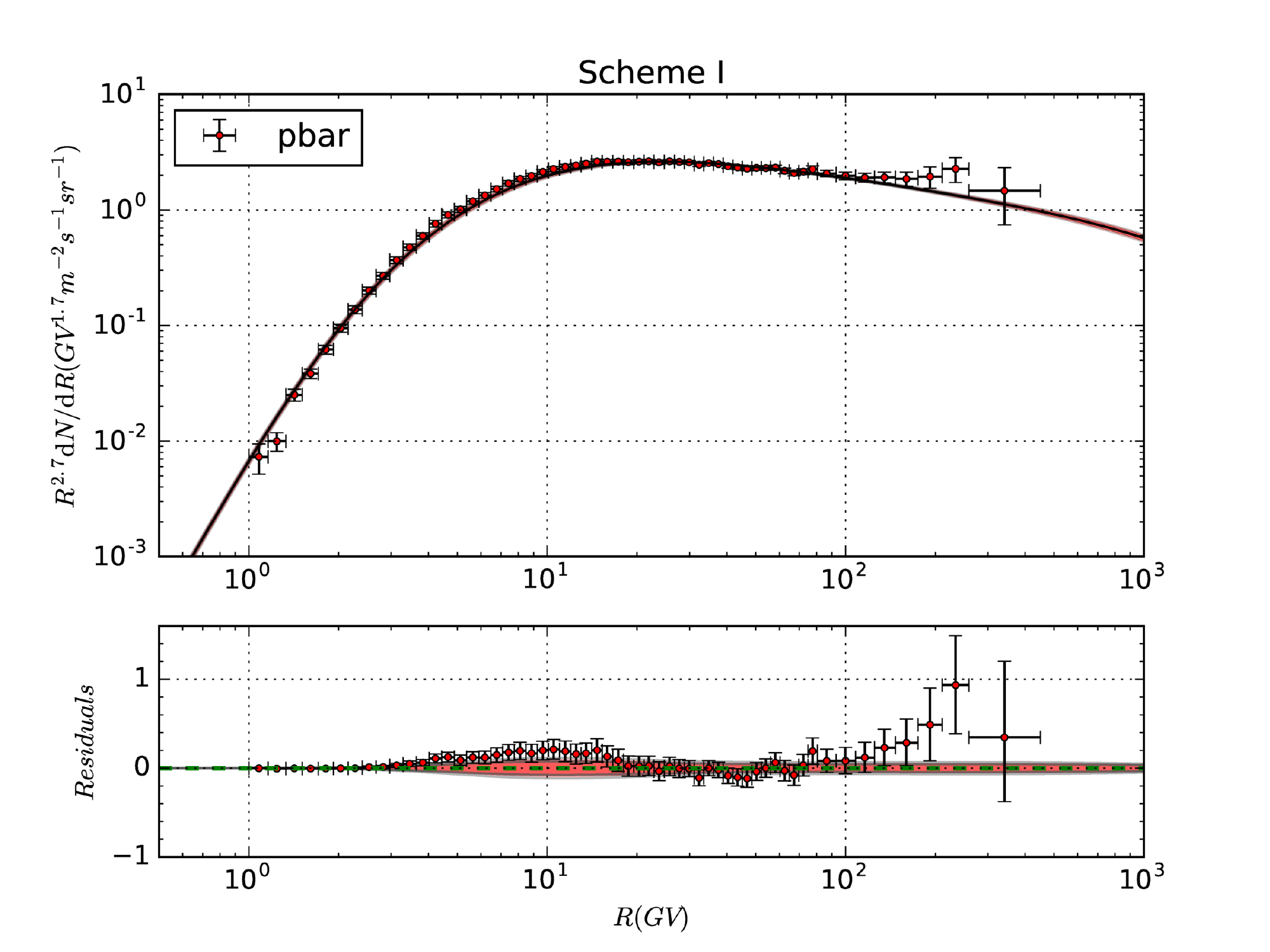}
  \includegraphics[width=0.3\textwidth]{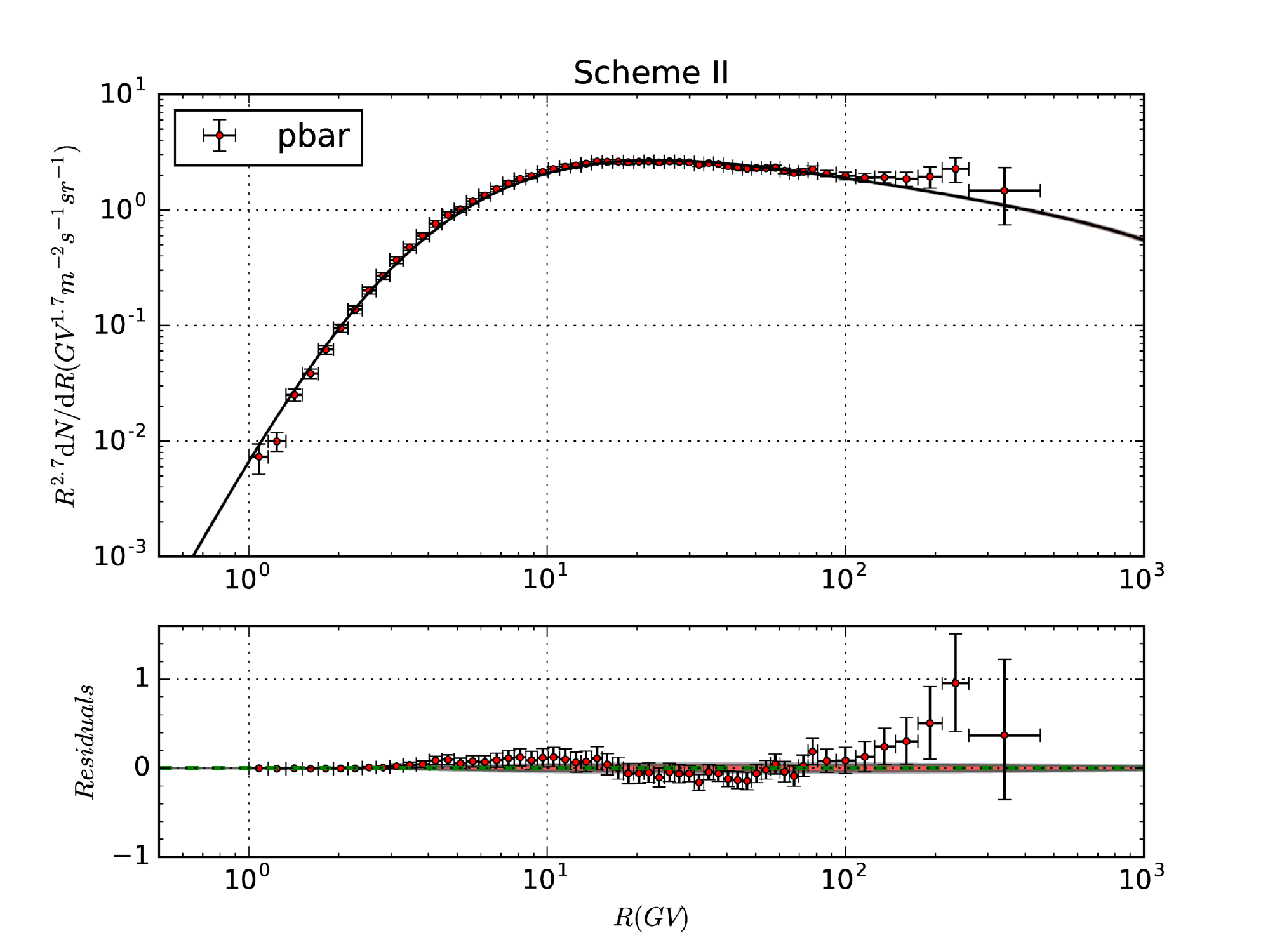}
  \includegraphics[width=0.3\textwidth]{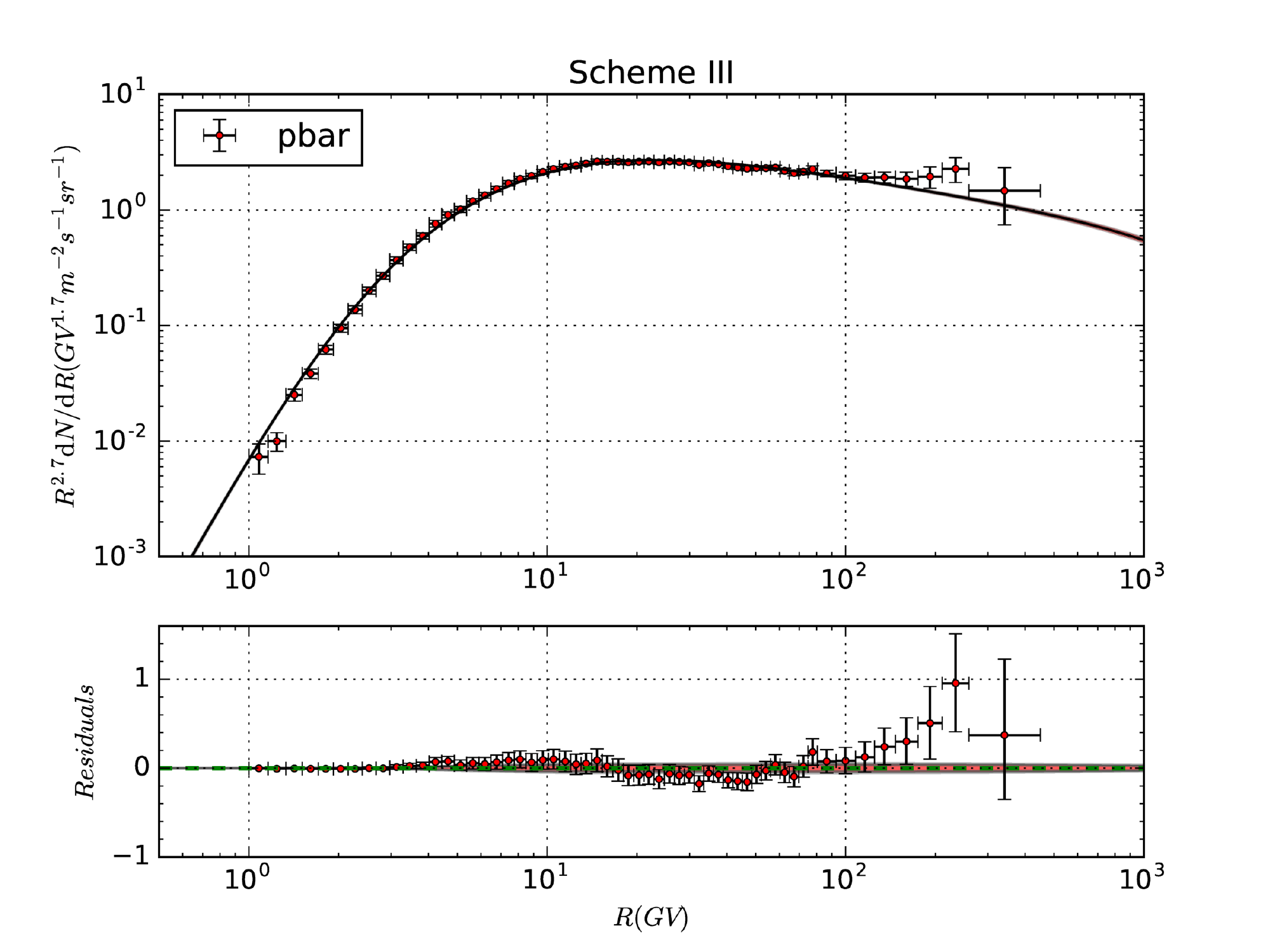}

  \caption{The global fitting results and the corresponding residuals to the AMS-02  B/C ratio, proton flux, helium flux, $\pbarp$ ratio  and $\pbar$ flux data for Scheme I, II and III. The $2\sigma$ (deep red) and $3\sigma$ (light red) bound are also showed in the figures. Note that the $\pbar$ flux data is not used in the global fitting and we show it for a cross validation.  For Scheme I, we did not use the $\pbarp$ ratio data either, and fix the value $c_{\pbar} = 1.34$ to show the results of $\pbarp$ and $\pbar$. }
\label{fig:prop_results}
\end{figure*}

We use the MCMC algorithm to determine the parameters of the three schemes as described in Sec. \ref{sec:fitting_pro} through fitting to the data set. When the Markov Chains have reached their equilibrium state， we take the samples of the parameters as their posterior PDFs. The best-fitting results and the corresponding residuals of the spectra and ratios are showed in Fig. \ref{fig:prop_results}. The best-fit values, statistical mean values, standard deviations and allowed intervals at $95 \%$ CL for these parameters are shown in Tables \ref{tab:propagation_params_I}, 
\ref{tab:propagation_params_II}, and \ref{tab:propagation_params_III} for
Schemes I, II, and III, respectively.

\begin{table*}[htb]
\begin{center}
\begin{tabular}{lllll | l}
  \hline\hline
ID  &Prior & Best-fit &Posterior mean and   &Posterior 95\% &Ref. \citet{Yuan2017}    \\
    &range&value  &Standard deviation & range   &$\chi^{2}/d.o.f. = 438.8/462$             \\
\hline
$D_{0}\ (10^{28}\cm^{2}\s^{-1})$
    &[1, 16]  &11.94  &10.55$\pm$0.77     &[8.99, 12.26] &7.24$\pm$0.97   \\

$\delta$
  &[0.1, 1.0] &0.359  &0.366$\pm$0.008     &[0.350, 0.376] &0.380$\pm$0.007    \\

$z_h\ (\kpc)$
  &[0.5, 20.0]  &12.40  &9.89$\pm$1.27     &[7.74, 12.97] &5.93$\pm$1.13    \\

$v_{A}\ (\km/\s)$
  &[0, 50]  &37.0  &37.6$\pm$1.6     &[34.0, 38.9] &38.5$\pm$1.3    \\

\hline

$N_{\p}\ \footnote{Post-propagated normalization flux of protons at 100 GeV in unit $10^{-9}\cm^{-2}\s^{-1}\sr^{-1}\MeV^{-1}$}$
  &[1, 8] &4.46   &4.46$\pm$0.02    &[4.44, 4.50] &4.50$_{-0.02}^{+0.02}$ \\

$R_{\p}\ (\GV)$
  &[1, 30]  &18.3  &17.5$\pm$1.6      &[15.1, 20.4] &12.9$_{-0.6}^{+0.6}$    \\
  
$\nu_{\p1}$
  &[1.0, 4.0] &2.074  &2.051$\pm$0.026     &[2.031, 2.101] &1.69$\pm$0.02    \\

$\nu_{\p2}$
  &[1.0, 4.0] &2.425  &2.421$\pm$0.009     &[2.413, 2.436] &2.37$\pm$0.01    \\
  
$R_{\A}\ (\GV)$
  &[1, 30]  &18.6  &17.5$\pm$1.1      &[15.8, 19.8] &12.9$_{-0.6}^{+0.6}$   \\

$\nu_{\A1}$
  &[1.0, 4.0] &2.081  &2.055$\pm$0.025     &[2.039, 2.099] &1.69$\pm$0.02    \\

$\nu_{\A2}$
  &[1.0, 4.0] &2.367  &2.364$\pm$0.009     &[2.356, 2.379] &2.37$\pm$0.01    \\

\hline

$c_{\He}$
  &[0.1, 5.0] &0.57  &0.60$\pm$0.07     &[0.48, 0.70] &---    \\
$\phi\ (\GV)$
    &[0, 1.5] &0.71   &0.70$\pm$0.04    &[0.66, 0.77] &0.86$\pm$0.02  \\
  \hline\hline
\end{tabular}
\end{center}
\caption{
 Constraints of the parameters on Scheme I. The prior interval, best-fit value, statistic mean, standard deviation and the allowed range at $95\%$ CL are listed for each propagation parameter. For a comparison,  we also list the posterior mean and $68\%$ credible uncertainties of  these parameters from  a previous analysis in \citet{Yuan2017}. For best-fit values, $\chi^{2}/d.o.f. = 133.41/192$. }
\label{tab:propagation_params_I}
\end{table*}

\begin{table*}[htb]
\begin{center}
\begin{tabular}{lllll | l}
  \hline\hline
ID  &Prior & Best-fit &Posterior mean and   &Posterior 95\% &Ref. \citet{Yuan2017}    \\
    &range&value  &Standard deviation & range   &$\chi^{2}/d.o.f. = 438.8/462$             \\
\hline
$D_{0}\ (10^{28}\cm^{2}\s^{-1})$
    &[1, 16]  &9.97  &8.81$\pm$0.72     &[8.03, 10.48] &7.24$\pm$0.97   \\

$\delta$
  &[0.1, 1.0] &0.376  &0.376$\pm$0.009     &[0.366, 0.380] &0.380$\pm$0.007    \\

$z_h\ (\kpc)$
  &[0.5, 20.0]  &9.12  &7.37$\pm$0.62     &[6.22, 9.37] &5.93$\pm$1.13    \\

$v_{A}\ (\km/\s)$
  &[0, 50]  &38.6  &38.5$\pm$3.2     &[37.4, 41.7] &38.5$\pm$1.3    \\

\hline

$N_{\p}\ \footnote{Post-propagated normalization flux of protons at 100 GeV in unit $10^{-9}\cm^{-2}\s^{-1}\sr^{-1}\MeV^{-1}$}$
  &[1, 8] &4.44   &4.44$\pm$0.02    &[4.41, 4.46] &4.50$_{-0.02}^{+0.02}$ \\

$R_{\p}\ (\GV)$
  &[1, 30]  &18.8  &17.4$\pm$1.9      &[16.5, 20.2] &12.9$_{-0.6}^{+0.6}$    \\
  
$\nu_{\p1}$
  &[1.0, 4.0] &2.004  &1.990$\pm$0.022     &[1.981, 2.028] &1.69$\pm$0.02    \\

$\nu_{\p2}$
  &[1.0, 4.0] &2.404  &2.409$\pm$0.008     &[2.400, 2.423] &2.37$\pm$0.01    \\
  
$R_{\A}\ (\GV)$
  &[1, 30]  &17.9  &16.1$\pm$1.7      &[15.8, 18.8] &12.9$_{-0.6}^{+0.6}$   \\

$\nu_{\A1}$
  &[1.0, 4.0] &2.002  &1.985$\pm$0.027     &[1.979, 2.025] &1.69$\pm$0.02    \\

$\nu_{\A2}$
  &[1.0, 4.0] &2.348  &2.350$\pm$0.006     &[2.343, 2.361] &2.37$\pm$0.01    \\

\hline

$c_{\He}$
  &[0.1, 5.0] &0.69  &0.73$\pm$0.07     &[0.63, 0.84] &---    \\
$c_{\pbar}$
  &[0.1, 5.0] &1.34  &1.34$\pm$0.05     &[1.33, 1.39] &---    \\
$\phi\ (\GV)$
    &[0, 1.5] &0.62   &0.62$\pm$0.03    &[0.58, 0.67] &0.86$\pm$0.02  \\
  \hline\hline
\end{tabular}
\end{center}
\caption{ Same as Table \ref{tab:propagation_params_I}, bur for Scheme II. For best-fit values, $\chi^{2}/d.o.f. = 251.48/248$. }
\label{tab:propagation_params_II}
\end{table*}

\begin{table*}[htb]
\begin{center}
\begin{tabular}{lllll | l}
  \hline\hline
ID  &Prior & Best-fit &Posterior mean and   &Posterior 95\% &Ref. \citet{Yuan2017}    \\
    &range&value  &Standard deviation & range   &$\chi^{2}/d.o.f. = 380.5/461$             \\
\hline
$D_{0}\ (10^{28}\cm^{2}\s^{-1})$
    &[1, 16]  &10.82  &9.70$\pm$0.67     &[8.62, 11.20] &6.14$\pm$0.45   \\

$\delta$
  &[0.1, 1.0] &0.378  &0.376$\pm$0.006     &[0.368, 0.389] &0.478$\pm$0.013    \\

$z_h\ (\kpc)$
  &[0.5, 20.0]  &11.15  &9.05$\pm$1.05     &[7.57, 11.62] &12.70$\pm$1.40    \\

$v_{A}\ (\km/\s)$
  &[0, 50]  &38.1  &40.2$\pm$1.3     &[37.3, 41.7] &43.2$\pm$1.2    \\

$\diff V_c/ \diff z \ (\km \s^{-1} \kpc^{-1})$
  &[0, 30]  &0.56  &2.01$\pm$1.31     &[0.09, 3.48] &11.99$\pm$1.26    \\

\hline

$N_{\p}\ \footnote{Post-propagated normalization flux of protons at 100 GeV in unit $10^{-9}\cm^{-2}\s^{-1}\sr^{-1}\MeV^{-1}$}$
  &[1, 8] &4.42   &4.44$\pm$0.02    &[4.41, 4.46] &4.52$_{-0.02}^{+0.02}$ \\

$R_{\p}\ (\GV)$
  &[1, 30]  &19.1  &18.8$\pm$0.9      &[18.0, 20.6] &16.6$_{-1.1}^{+1.2}$    \\
  
$\nu_{\p1}$
  &[1.0, 4.0] &2.015  &2.022$\pm$0.015     &[1.997, 2.047] &1.82$\pm$0.02    \\

$\nu_{\p2}$
  &[1.0, 4.0] &2.409  &2.416$\pm$0.012     &[2.403, 2.424] &2.37$\pm$0.01    \\
  
$R_{\A}\ (\GV)$
  &[1, 30]  &18.4  &17.6$\pm$0.8      &[16.7, 19.4] &16.6$_{-1.1}^{+1.2}$    \\

$\nu_{\A1}$
  &[1.0, 4.0] &2.018  &2.020$\pm$0.015     &[1.998, 2.051] &1.82$\pm$0.02    \\

$\nu_{\A2}$
  &[1.0, 4.0] &2.348  &2.355$\pm$0.009     &[2.344, 2.363] &2.37$\pm$0.01    \\

\hline

$c_{\He}$
  &[0.1, 5.0] &0.66  &0.70$\pm$0.07     &[0.59, 0.85] &---    \\
$c_{\pbar}$
  &[0.1, 5.0] &1.37  &1.38$\pm$0.04     &[1.34, 1.40] &---    \\
$\phi\ (\GV)$
    &[0, 1.5] &0.62   &0.62$\pm$0.02    &[0.57, 0.67] &0.89$\pm$0.03  \\
  \hline\hline
\end{tabular}
\end{center}
\caption{ Same as Tab. \ref{tab:propagation_params_I}, bur for Scheme II. Note that add a propagation parameter $\diff V_c/ \diff z$. For best-fit values, $\chi^{2}/d.o.f. = 246.69/247$.}
\label{tab:propagation_params_III}
\end{table*}

Because the data are precise enough and from the same experiment, we obtain statistically the 
good constraints on the model parameters. Some of the model parameters, such as the injection spectral indices, are constrained to a level of $\lesssim 1 \%$. The propagation parameters are constrained to be about $\lesssim 10 \%$ (in Scheme II), which are relatively large due to the degeneracy among some of them but obtained an obvious improvement compared with Scheme I and previous studies~\citep{Trotta2011,Jin2015,Lin2015,Johannesson2016,Korsmeier2016,Yuan2017}. For the rigidity-dependent slope of the diffusion coefficient, $\delta$, the statistical error is only a few percent ($\lesssim 2 \%$). 
The uncertainties of three nuisance parameters  are $\lesssim 10 \%$, which give us 
an opportunity to read the relevant information behind these parameters.

For a comparison, we also present the posterior mean and $68 \%$ credible uncertainties determined from a previous analysis in \citet{Yuan2017} and  which is based on data of B/C (from AMS-02 \citep{AMS02_b_c} and ACE-CRIS \footnote{http://www.srl.caltech.edu/ACE/ASC/level2/lvl2DATA\_CRIS.html}), $^{10}$Be/$^{9}$Be (from Ulysses \citep{Ulysses}, ACE \citep{ACE}, Voyager \citep{Voyager}, IMP \citep{IMP_ISEE3} , ISEE-3 \citep{IMP_ISEE3}, and ISOMAX \citep{ISOMAX}) and proton flux (from AMS-02 \citep{AMS02_proton} and PAMELA \citep{PAMELA2013}) for each Schemes.

 From Fig. \ref{fig:prop_results}, the major discrepancy comes from the fitting results of B/C ratio, 
proton and helium flux below $\sim 10 \GeV$, and $\pbarp$ ratio and $\pbar$ flux larger 
than $\sim 100 \GeV$. Comparing with the results of Schemes I and II, we can see that 
the $\pbarp$ data effectively relieve the degeneracy of the classical correlation between $D_0$ and $z_h$. 
In these Schemes,  $\pbar$s have been entirely produced as the secondary products of proton and helium.
The $\pbarp$ data play a crucial role in reducing the uncertainty of $z_h$.  Moreover, the comparison 
between Schemes II and III shows that the data set disfavors a large value of $\diff V_c / \diff z$, 
or the DRC model, although the fitting result of Scheme III seems a little better than that of Scheme II.

In consideration of the relatively independent among three groups of the models' parameters (the propagation parameters, the source parameters and nuisance parameters), we would analyze the results of these three groups separately. At the same time, 
we compare the different aspects of these two models.

\subsection{Propagation parameters}

\begin{figure}
\centering
\includegraphics[width=0.5\textwidth]{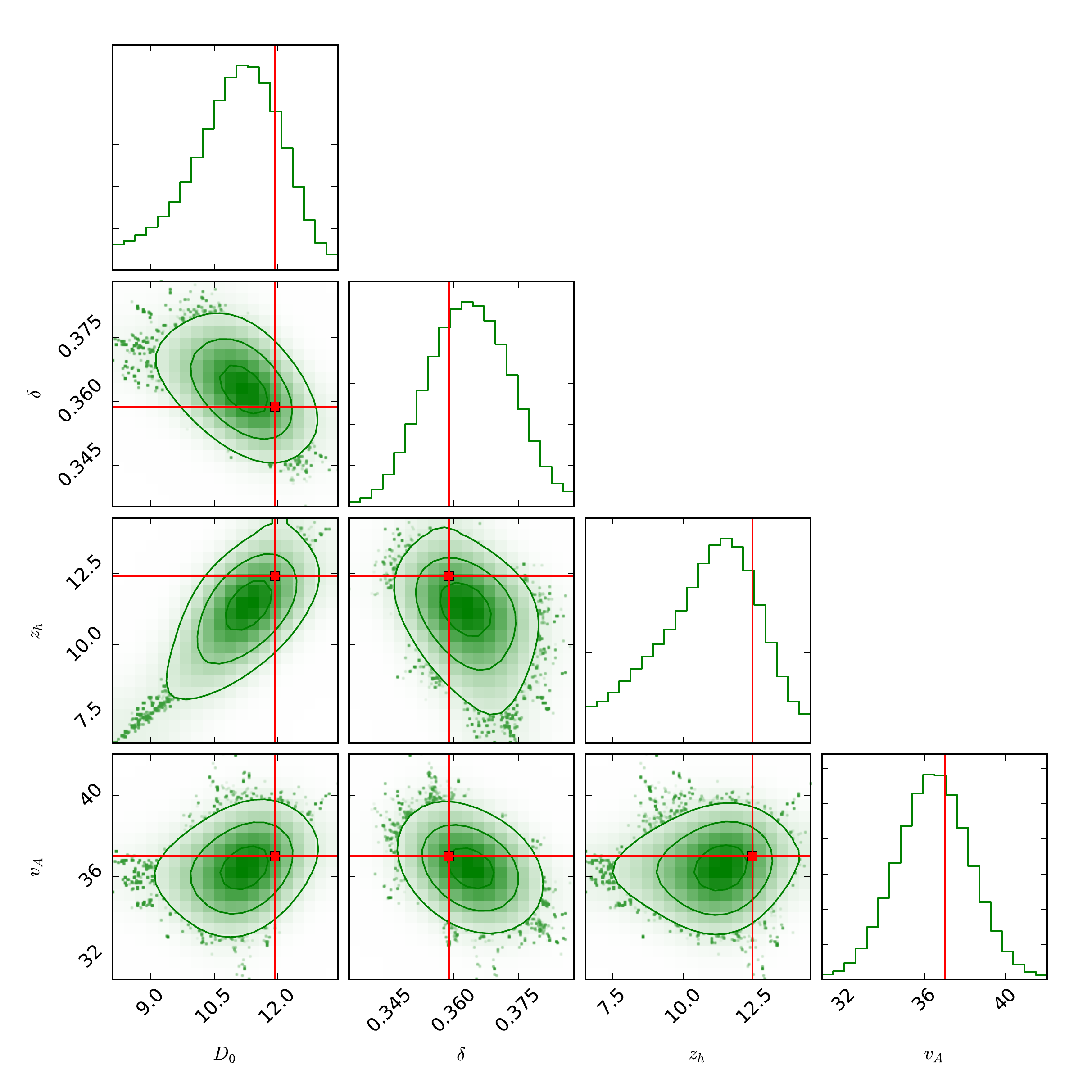}
\caption{Fitting 1D probability and 2D credible regions of posterior PDFs for the combinations of all propagation parameters from Scheme I. The regions enclosing $\sigma$, $2\sigma$ and $3\sigma$ CL are shown in step by step lighter green. The red cross lines and marks in each plot indicates the best-fit value (largest likelihood).}
\label{fig:dis_2d_1}
\end{figure}

\begin{figure}
\centering
\includegraphics[width=0.5\textwidth]{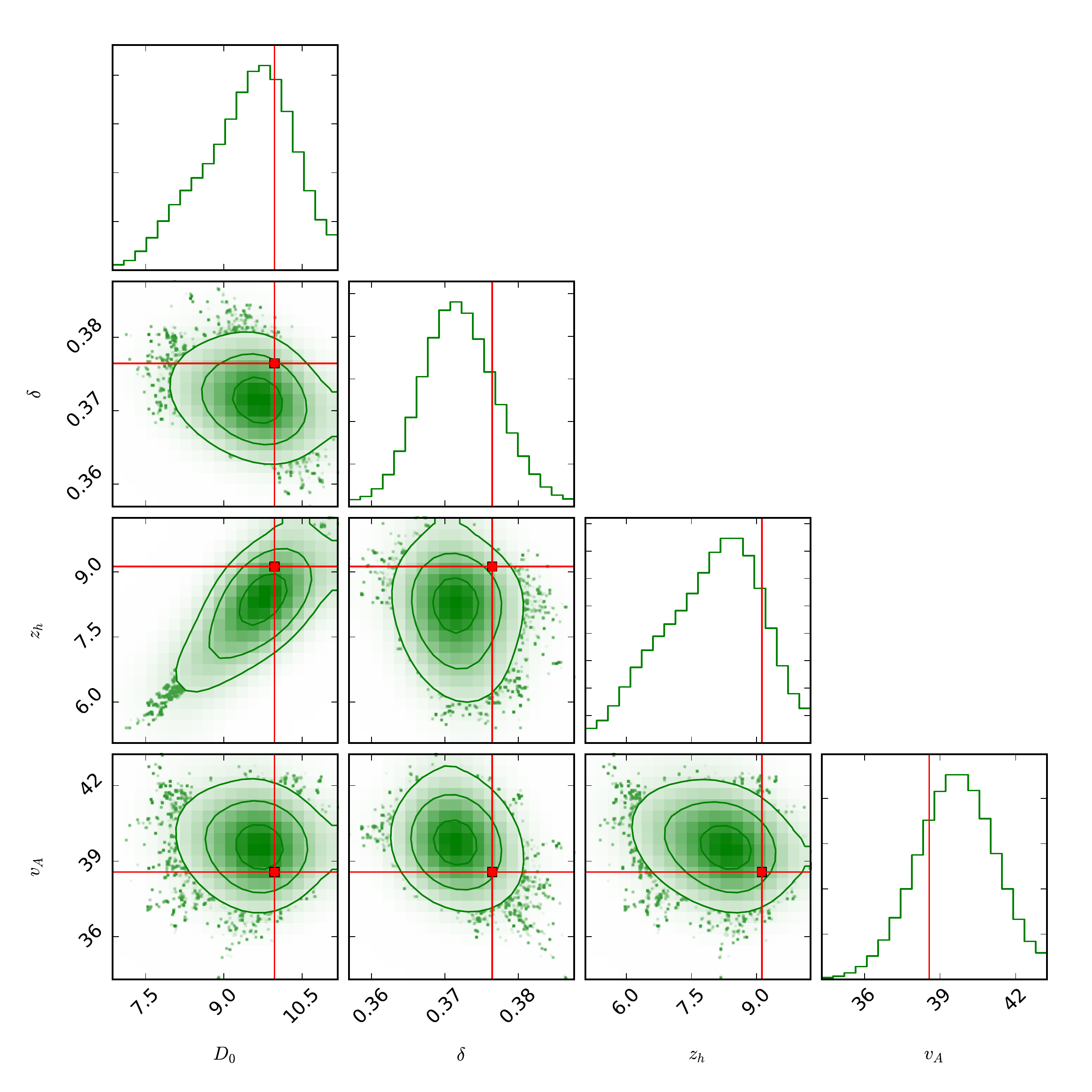}
\caption{Same as Fig. \ref{fig:dis_2d_1} but for Scheme II.}
\label{fig:dis_2d_2}
\end{figure}

\begin{figure}
\centering
\includegraphics[width=0.5\textwidth]{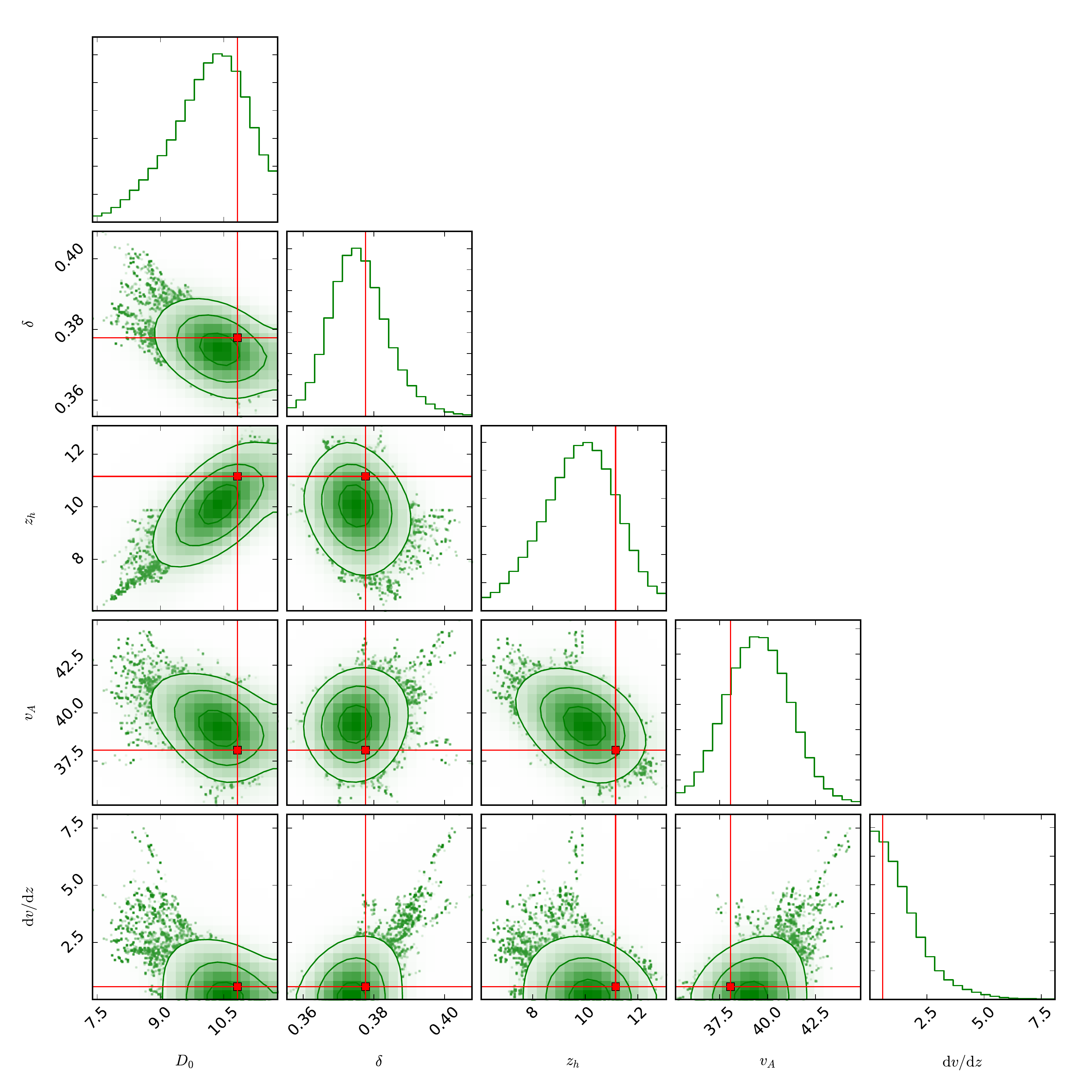}
\caption{Same as Fig. \ref{fig:dis_2d_1} but for Scheme III.}
\label{fig:dis_2d_3}
\end{figure}

The results of posterior probability distributions of the propagation parameters are show in  Fig. \ref{fig:dis_2d_1} (Scheme I), Fig. \ref{fig:dis_2d_2} (Scheme II) and Fig. \ref{fig:dis_2d_3} (Scheme III). 
In general, this data set (the new released AMS-02 B/C, proton and helium data) favors large values 
of $D_{0}$ and $z_{h}$  compared to some previous works, for examples, 
see Refs.~\citep{Trotta2011,Lin2015,Jin2015,Korsmeier2016,Johannesson2016,Yuan2017}.

In Fig. \ref{fig:dis_2d_1}, there is a clear degeneracy between $D_0$ and $z_h$. This is because the B/C data can only constrain $D_0/z_h$ effectively \citep{Maurin2001,Jin2015}.  From Table \ref{tab:propagation_params_I}, 
we can see that the data set of Scheme I (without $\pbarp$ data) gives us a similar result 
from \citet{Yuan2017}. Consequently, the $^{10}$Be/$^{9}$Be data in \citet{Yuan2017} is unnecessary because the AMS-02 B/C and proton data are precise enough to relieve the degeneracy of the correlation between $D_0$ and $z_h$ at that level \citep{Jin2015}.

 In Fig. \ref{fig:dis_2d_2}, although there still exists the degeneracy between $D_{0}$ and $z_{h}$, the $\pbarp$ data can relieve the degeneracy of this classical correlation more effectively and our results show a concrete improvement compared with previous works (see for e.g., \citep{Johannesson2016,Korsmeier2016,Yuan2017}).\footnote{The $\pbar$ here is entirely produced as the secondary product of proton and helium, other than some other primary component.} This improvement 
may arise from the high precision of the $\pbarp$ ratio data which reveal the high order products in propagation. 
 Note that the $\pbar$ flux arises from not only the primary proton and helium but also the secondary proton 
interacting with ISM. At the same time, the tertiary antiproton, which is included in our calculations, 
 may also contribute to this improvement.

In Fig. \ref{fig:dis_2d_3}, the constraints on $D_0$ and $z_h$ are relaxed by the additional parameter 
$\diff V_c/ \diff z$. But what is interesting is that the result favors a small value of $\diff V_c/ \diff z \sim 0.558 \km/\s)$, which is largely different from the result in \citet{Yuan2017} ($\diff V_c/ \diff z \sim 11.99 \km/\s)$). This difference may come from the bias of different experiment and large uncertainties of the $^{10}$Be/$^{9}$Be data 
 and the bias in $^{10}$Be production cross section \citep{Tomassetti2015prc}.  Therefore, 
this data set disfavors the DRC model.


\subsection{Primary source parameters}

\begin{figure*}
\centering
\includegraphics[height=0.7\textheight]{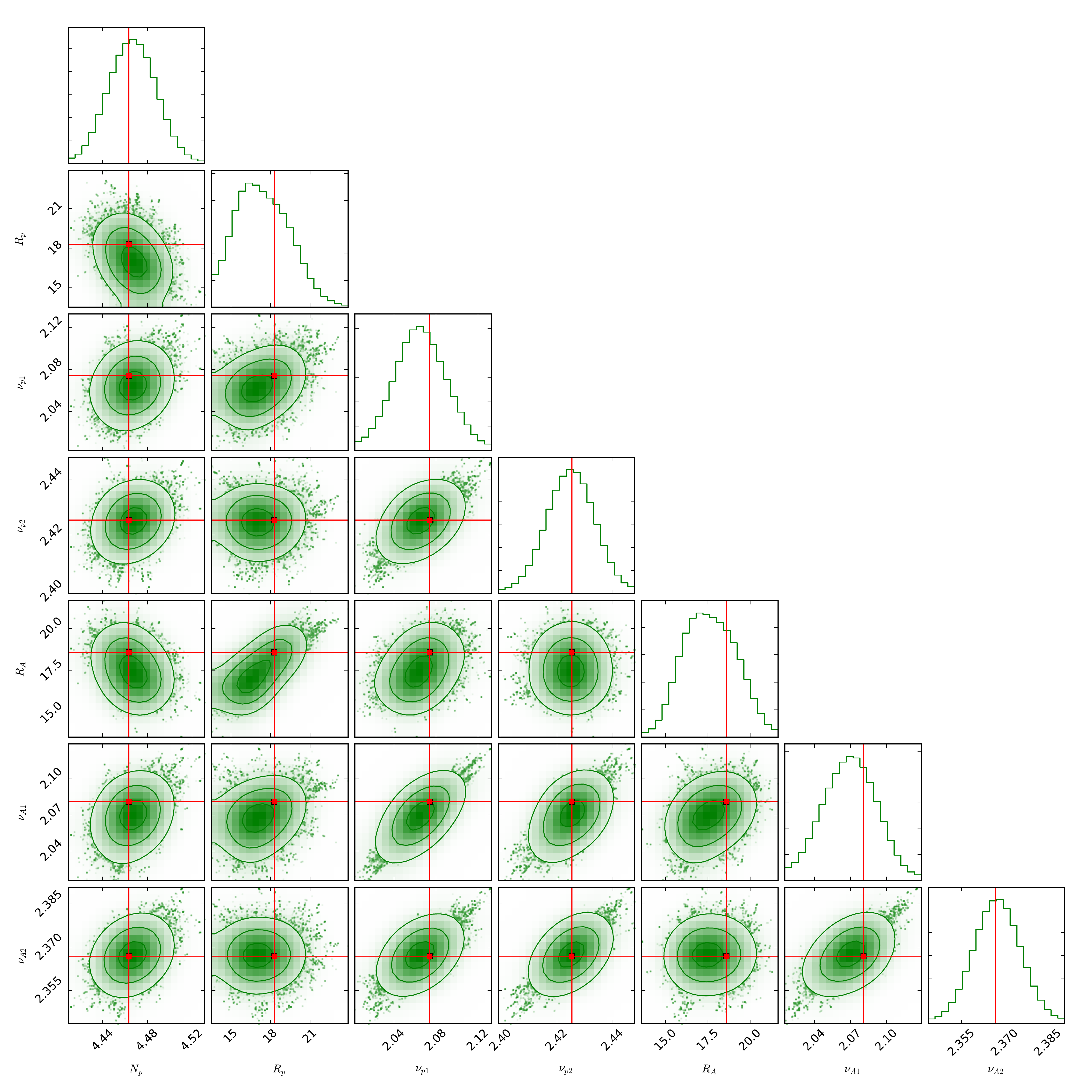}
\caption{Fitting 1D probability and 2D credible regions of posterior PDFs for the combinations of primary source parameters from Scheme I. The regions enclosing $\sigma$, $2\sigma$ and $3\sigma$ CL are shown in step by step lighter green. The red cross lines and marks in each plot indicates the best-fit value (largest likelihood).}
\label{fig:inj_spe_2d_1}
\end{figure*}

\begin{figure*}
\centering
\includegraphics[height=0.7\textheight]{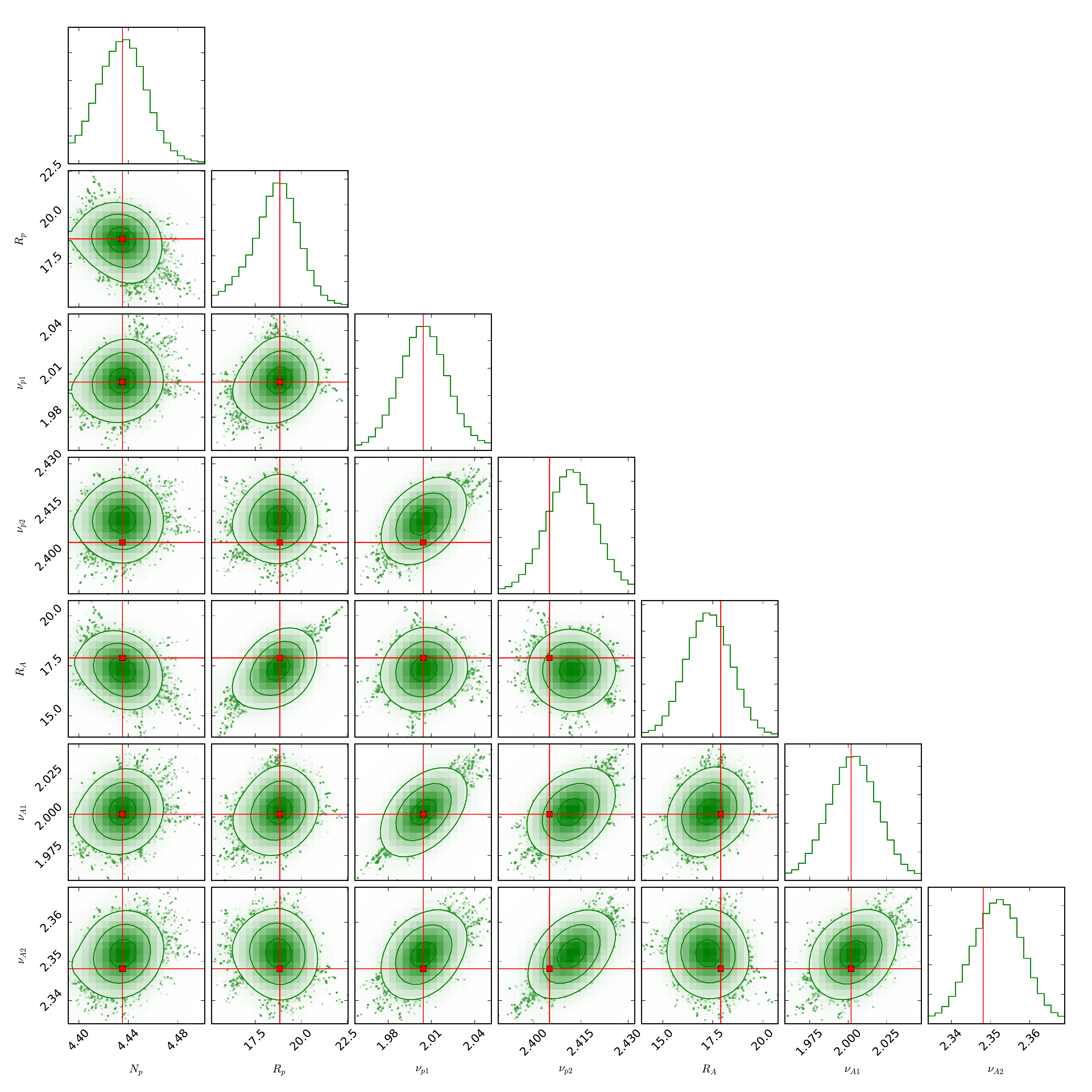}
\caption{Same as Fig. \ref{fig:inj_spe_2d_1} but for Scheme II.}
\label{fig:inj_spe_2d_2}
\end{figure*}

\begin{figure*}
\centering
\includegraphics[height=0.7\textheight]{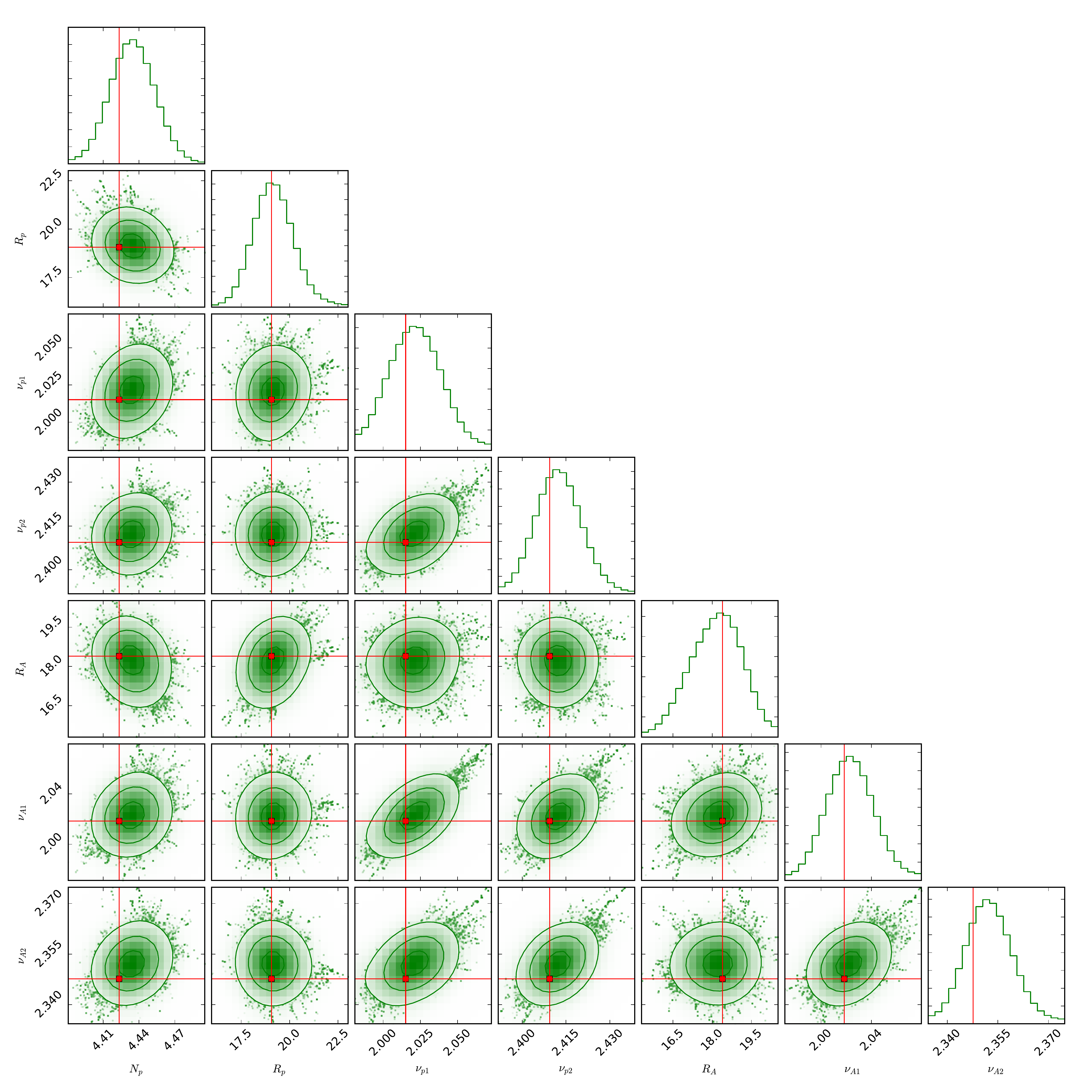}
\caption{Same as Fig. \ref{fig:inj_spe_2d_1} but for Scheme III.}
\label{fig:inj_spe_2d_3}
\end{figure*}

The results of posterior probability distributions of the primary source parameters are shown in 
 Figs. \ref{fig:inj_spe_2d_1},  \ref{fig:inj_spe_2d_2}, and \ref{fig:inj_spe_2d_3}
for Schemes I, II, and III, respectively.
Because we do not have obvious correlations in these figures, the posterior PDFs of these parameters are in a high confidence level and provide us the opportunity to study the CR physics behind them.

\begin{figure}
\centering
\includegraphics[width=0.5\textwidth]{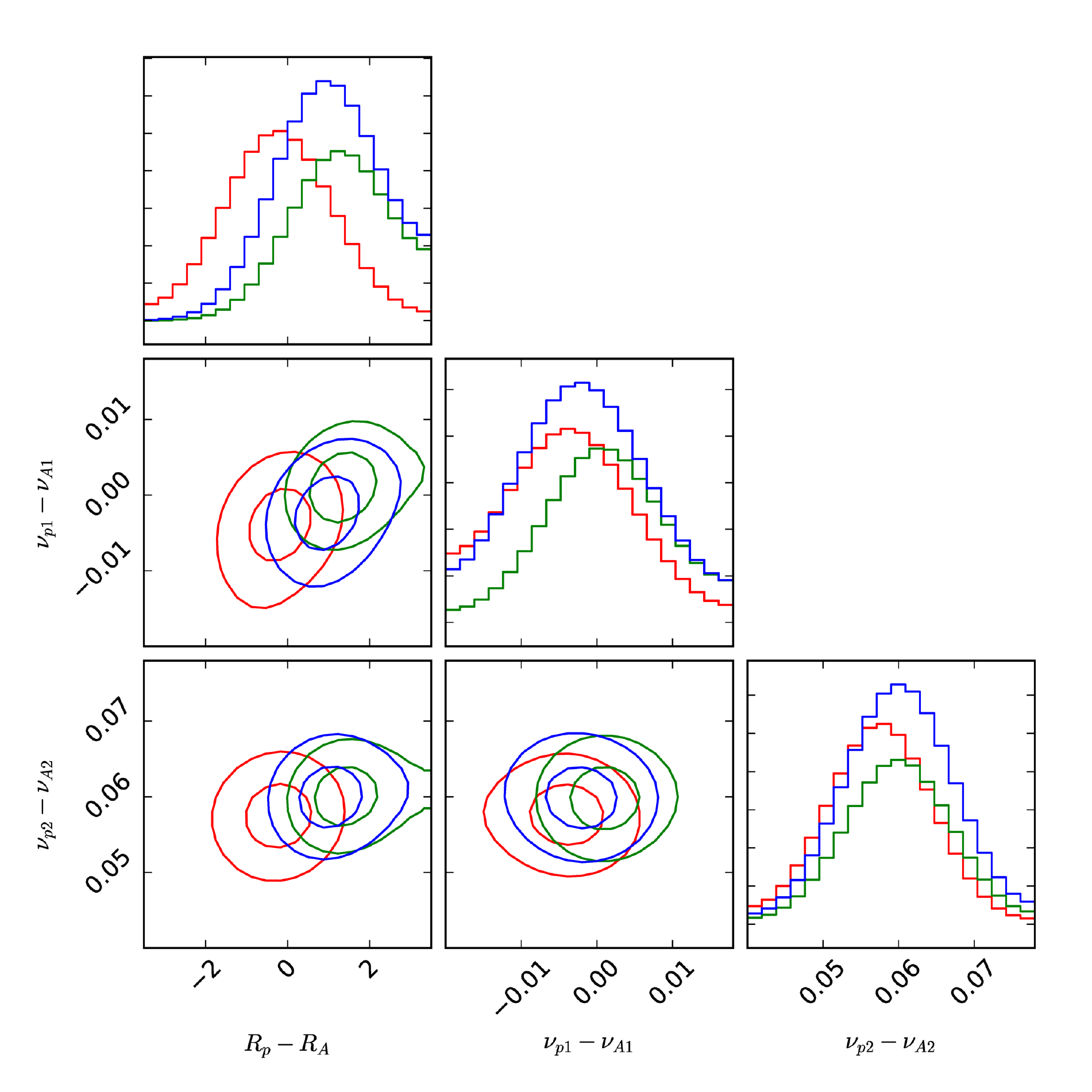}
\caption{Fitting 1D probability and 2D credible regions of posterior PDFs for the differences of primary source parameters (red for Scheme I, green for Scheme II, blue for Scheme III). The regions enclosing $\sigma$ and $2\sigma$ CL are indicated by the contours. }
\label{fig:inj_spe_para_2d}
\end{figure}

Benefited from the independent injection spectra for proton and other nuclei, we present the differences between rigidity breaks and slopes for proton and other nuclei species ($R_{\p}-R_{\A}$, $\nu_{\p1}-\nu_{\A1}$, $\nu_{\p2}-\nu_{\A2}$) in Fig. \ref{fig:inj_spe_para_2d} and Table \ref{tab:diff_source}.  In details, $R_{\p}-R_{\A}$ of Scheme I is largely different from those in Schemes II and III, which is influenced by the existence of $\pbarp$ ratio data in global fitting. $\nu_{\p1}-\nu_{\A1}$ and $\nu_{\p2}-\nu_{\A2}$ have  slightly different values for 3 Schemes, but has a relatively large overlap. For $\nu_{\p2}-\nu_{\A2}$, we have a high confidence level that the value is $\sim 0.06$.

\begin{table}[htb]
\begin{center}
\begin{tabular}{c| c c c}
  \hline\hline
ID  &Scheme I   &Scheme II   &Scheme III             \\
\hline
$R_{\p}-R_{\A}$ &-0.20 $\pm$ 0.81  &1.30 $\pm$ 0.65  &1.05 $\pm$ 0.81  \\

$\nu_{\p1}-\nu_{\A1}$ &-0.0040 $\pm$ 0.0068  &0.0014 $\pm$ 0.0063  &-0.0020 $\pm$ 0.0066  \\

$\nu_{\p2}-\nu_{\A2}$ &0.0575 $\pm$ 0.0040  &0.0600 $\pm$ 0.0041  &0.060 $\pm$ 0.0044  \\

\hline\hline

\end{tabular}
\end{center}
\caption{ The posterior mean and standard deviation of $R_{\p}-R_{\A}$, $\nu_{\p1}-\nu_{\A1}$, $\nu_{\p2}-\nu_{\A2}$ for the 3 schemes.}
\label{tab:diff_source}
\end{table}

\subsection{Nuisance parameters}

\begin{figure}
\centering
\includegraphics[width=0.5\textwidth]{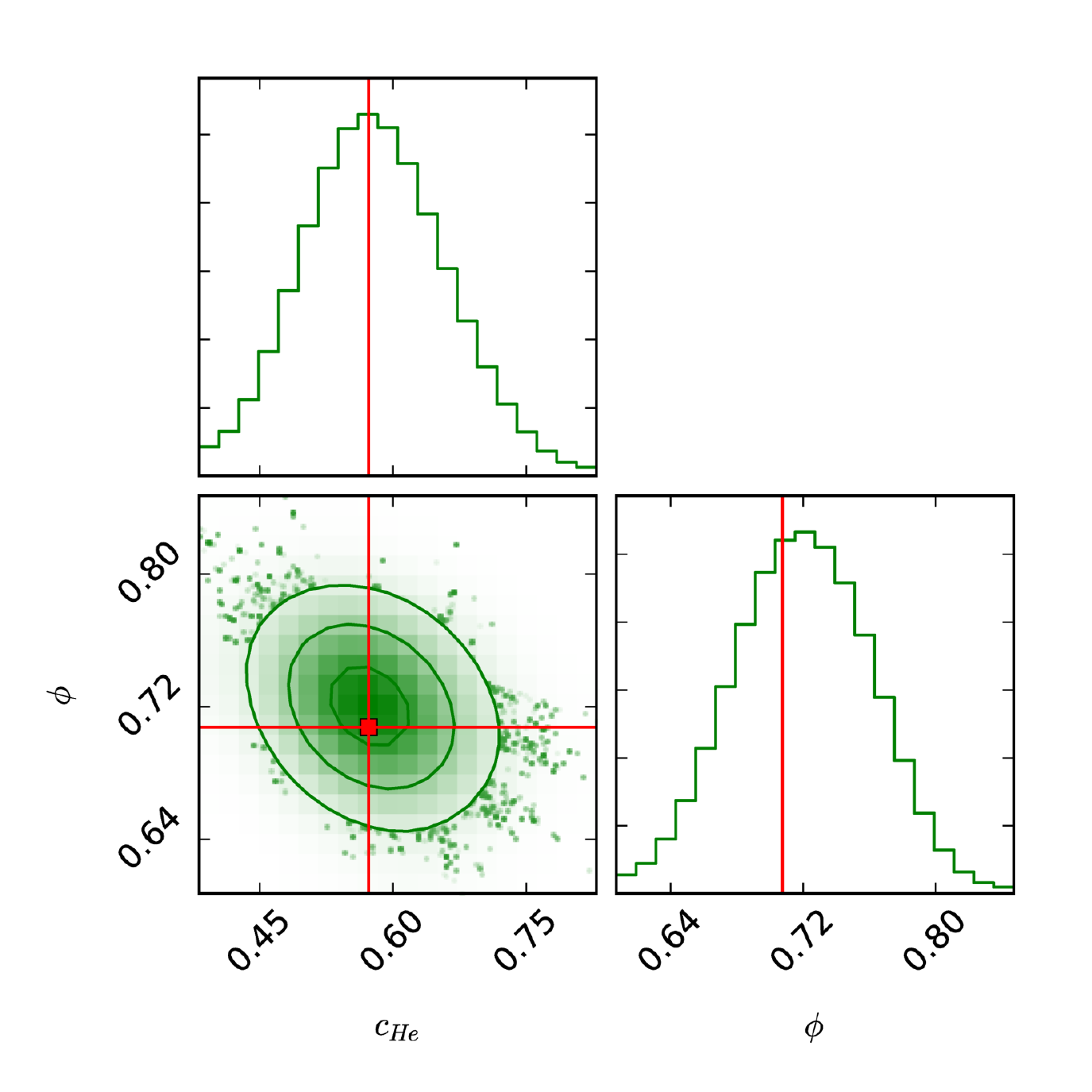}
\caption{Fitting 1D probability and 2D credible regions of posterior PDFs for the combinations of nuisance parameters from Scheme I. The regions enclosing $\sigma$, $2\sigma$ and $3\sigma$ CL are shown in step by step lighter green. The red cross lines and marks in each plot indicates the best-fit value (largest likelihood). }
\label{fig:nui_para_2d_1}
\end{figure}

\begin{figure}
\centering
\includegraphics[width=0.5\textwidth]{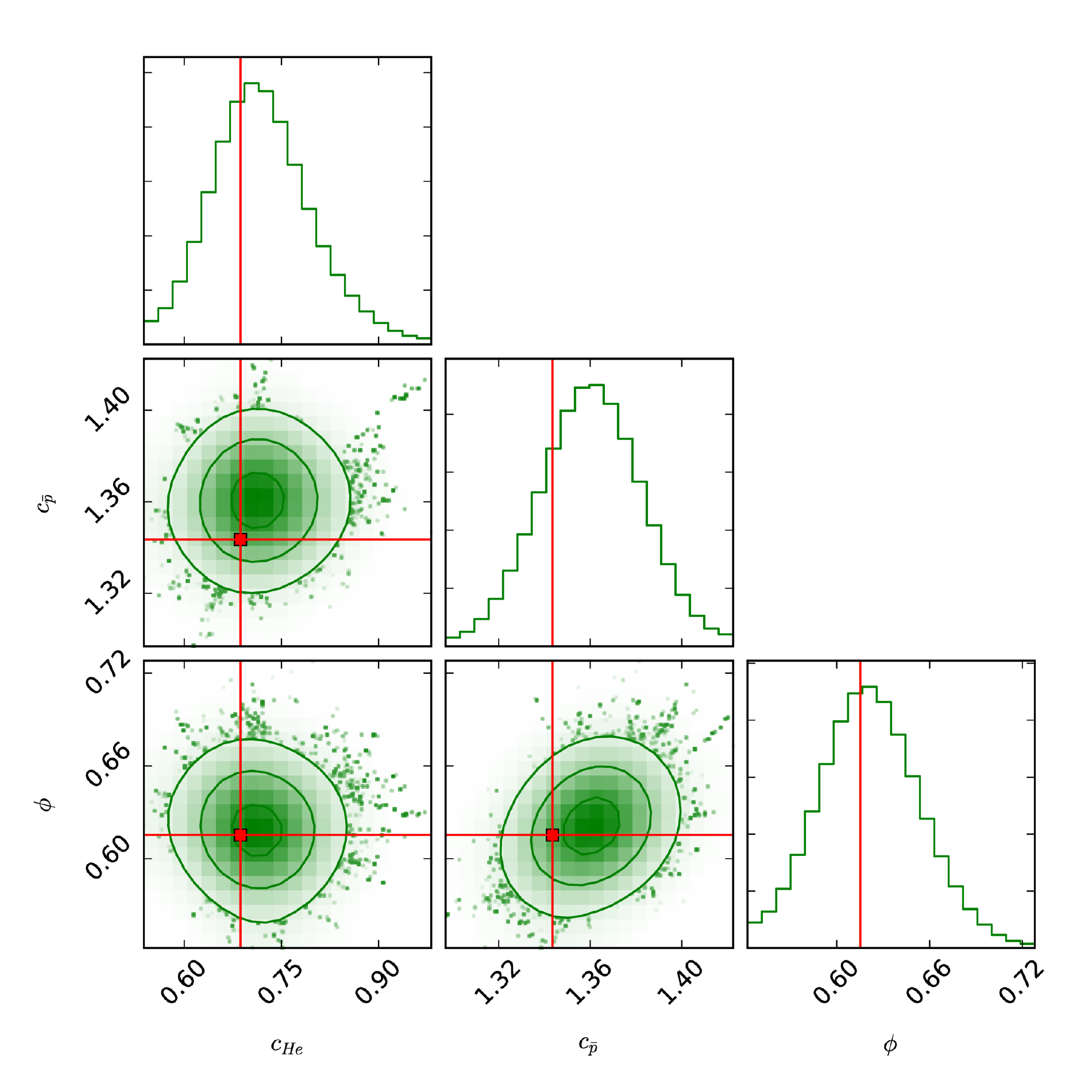}
\caption{Same as Fig. \ref{fig:nui_para_2d_1} but for Scheme II.}
\label{fig:nui_para_2d_2}
\end{figure}

\begin{figure}
\centering
\includegraphics[width=0.5\textwidth]{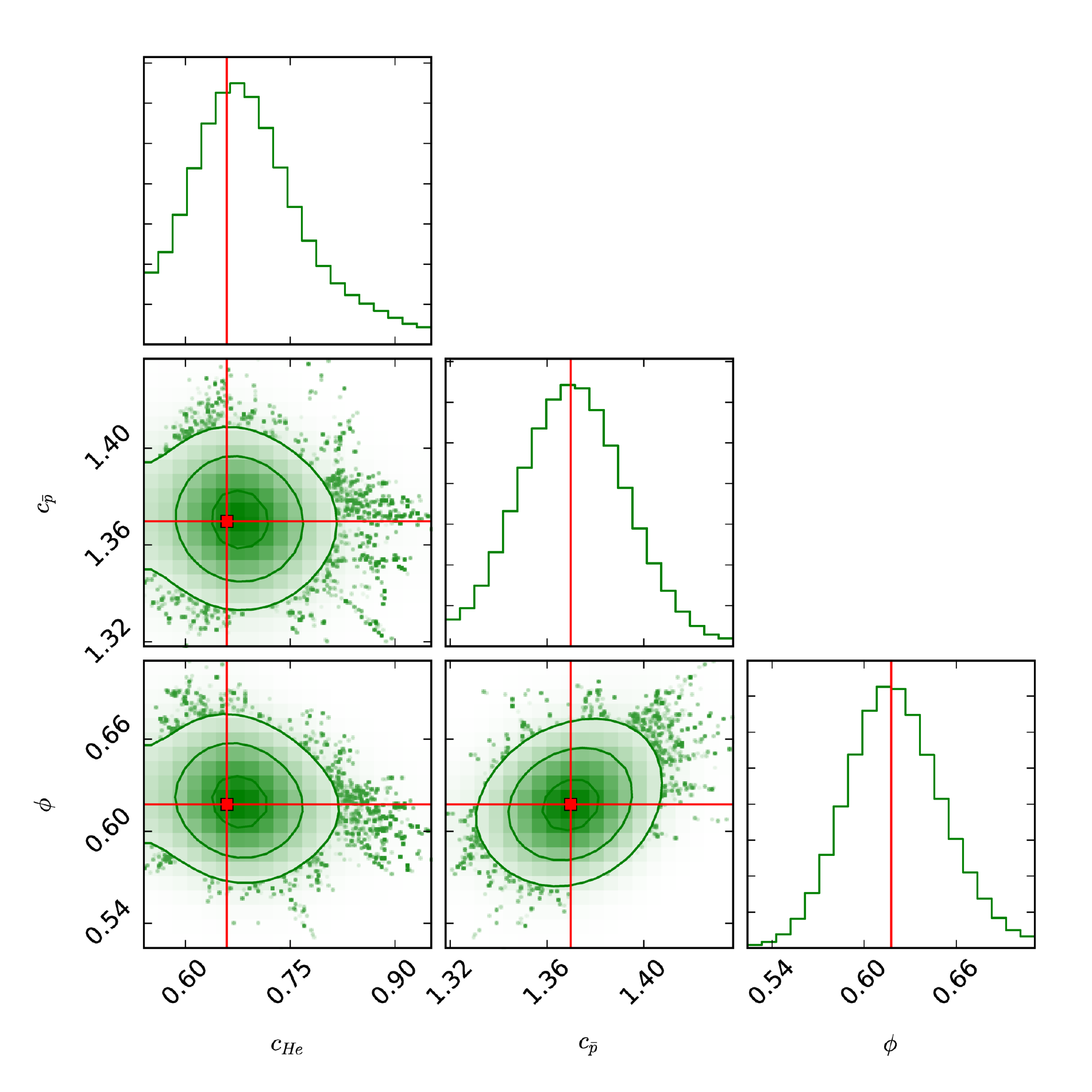}
\caption{Same as Fig. \ref{fig:nui_para_2d_1} but for Scheme III.}
\label{fig:nui_para_2d_3}
\end{figure}

 In Figs. \ref{fig:nui_para_2d_1}, \ref{fig:nui_para_2d_2}, and \ref{fig:nui_para_2d_3}, the results of posterior probability distributions  represent the necessity to introduce them in the global fitting.

In this work, we can see that if we want to fit the AMS-02 helium data in a self-consistent way, the helium-4 abundance should have a factor $\sim 0.68$ compared to the original value in {\sc galprop} ($7.199 \times 10^4$).

The uncertainties on the antiproton production cross sections could produce the relevant uncertainties in the antiproton flux \citep{diMauro2014,Lin2016}, and the employed energy (or rigidity) independent factor $c_{\pbar}$ can reproduce the AMS-02 antiproton flux result well except when $R \gtrsim 100 \GV$.

The solar modulation $\phi$ provide an relatively effective but not that precise fitting of the current data set. Benefited from the precise and self-consistent AMS-02 nuclei data set, the inefficient fitting in low-energy regions of force-field approximation is obviously represented in Fig.~\ref{fig:prop_results}.

\section{Discussions}

 In three schemes, we studied the widely used one break power law to describe the injection spectra for all kinds of nuclei (different breaks and slopes for proton and other nucleus species), and use the classical DR or DRC model with a uniform diffusion coefficient in the whole propagation region.  In Fig.~\ref{fig:prop_results}, we found 
a spectral break at $\sim 300 \GeV$ for proton and helium fluxes, which implies the deficiency of our schemes to fit the results in high energy region. Moreover, the underestimation of multi-TeV fluxes of proton and helium may cause an underestimation of sub-TeV fluxes of $\pbar$. For the purpose of this work not in this energy region and the lack of AMS-02 data and its relatively large uncertainties in high-energy region ($\gtrsim 1 \TV$), we did not consider more details on this problem. The proposed solutions to this problem include new break in high-energy region ($\sim 300 \GeV$) to the injection spectra (see, e.g., \cite{Korsmeier2016,Boschini2017}), as well as new break to the diffusion coefficient (see, e.g., \cite{Genolini2017}), and inhomogeneous diffusion (see, e.g., \cite{Blasi2012,Tomassetti2012,Tomassetti2015apjl01,Tomassetti2015prd,Guo2016,Feng2016}) or the superposition of local and distant sources (see for e.g. \cite{Vladimirov2012,Bernard2013,Thoudam2013,Tomassetti2015apjl02}). Based on our simplicity, the $\pbar$ excess in $100 \sim 300 \GeV$ might be interpreted as dark matter annihilation \citep{Cui2017,Cuoco2017}.

In low energy region, we find that the fitting is not that good. This may arise from (i) the published 
AMS-02 data on B/C, proton, helium and $\pbarp$ are collected during different periods 
(see Table \ref{tab:data_period}); (ii) the force-field approximation cannot deal with 
the charge-sign dependent solar modulation 
in reality~\footnote{ As the Scheme I (absence of $\pbar$ data) and Scheme II (presence of $\pbar$ data) 
give different $\phi$ values, the charge-sign dependent modulation is clearly supported here.}; (iii) there is the Sun's magnetic field reversal in early 2013 and it would bring the effects which cannot be described by a signal $\phi$ for all these data; (iv) the heavier elements suffer different diffusion coefficient from light ones which may arise from unaccounted inhomogeneity in CR diffusion (or in the medium) \citep{Johannesson2016}; (v) there may exist extra source which leads to the MeV excesses for some nucleus species.

\begin{figure*}[!htbp]
  \centering
  \includegraphics[width=0.3\textwidth]{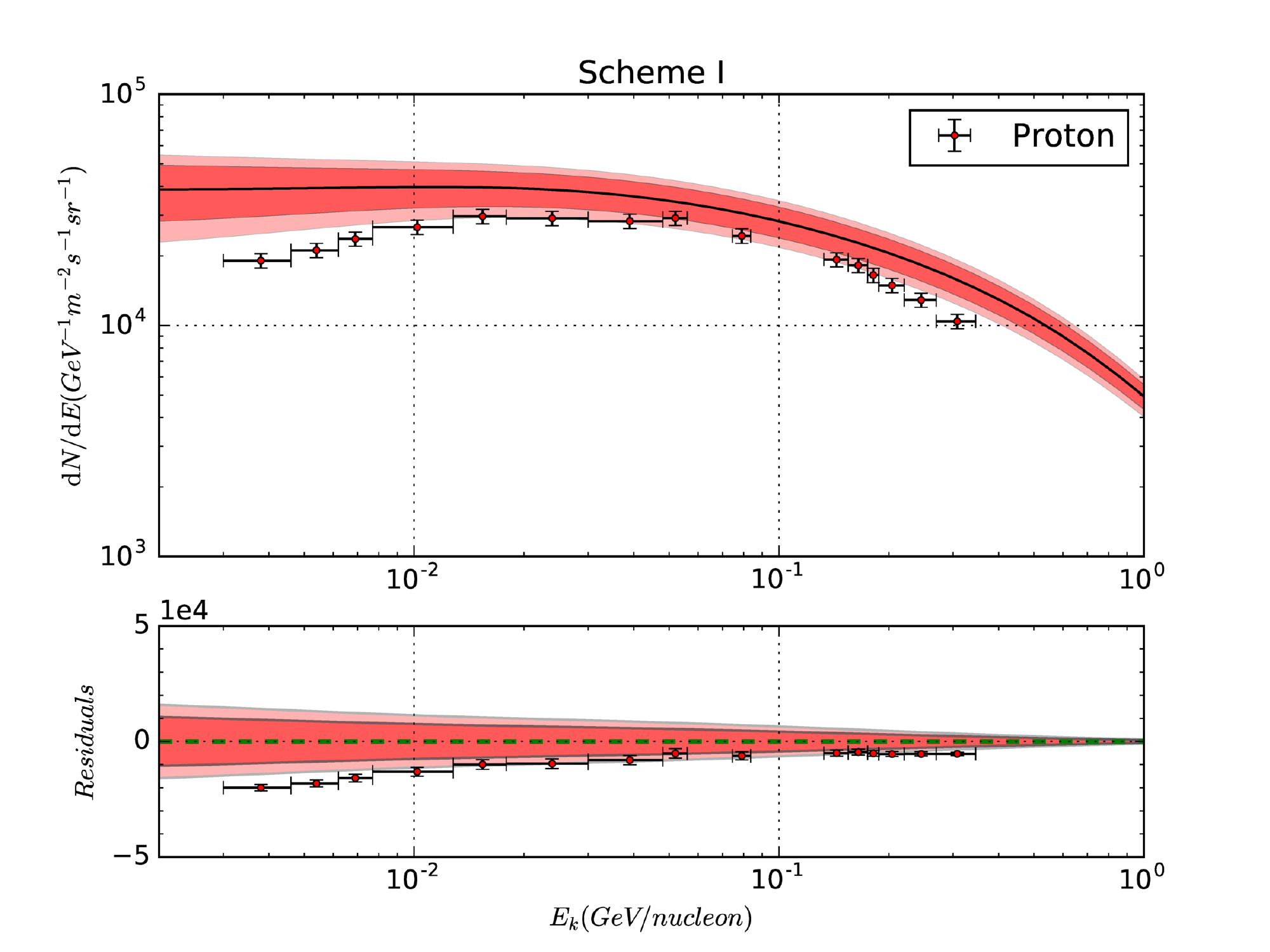}
  \includegraphics[width=0.3\textwidth]{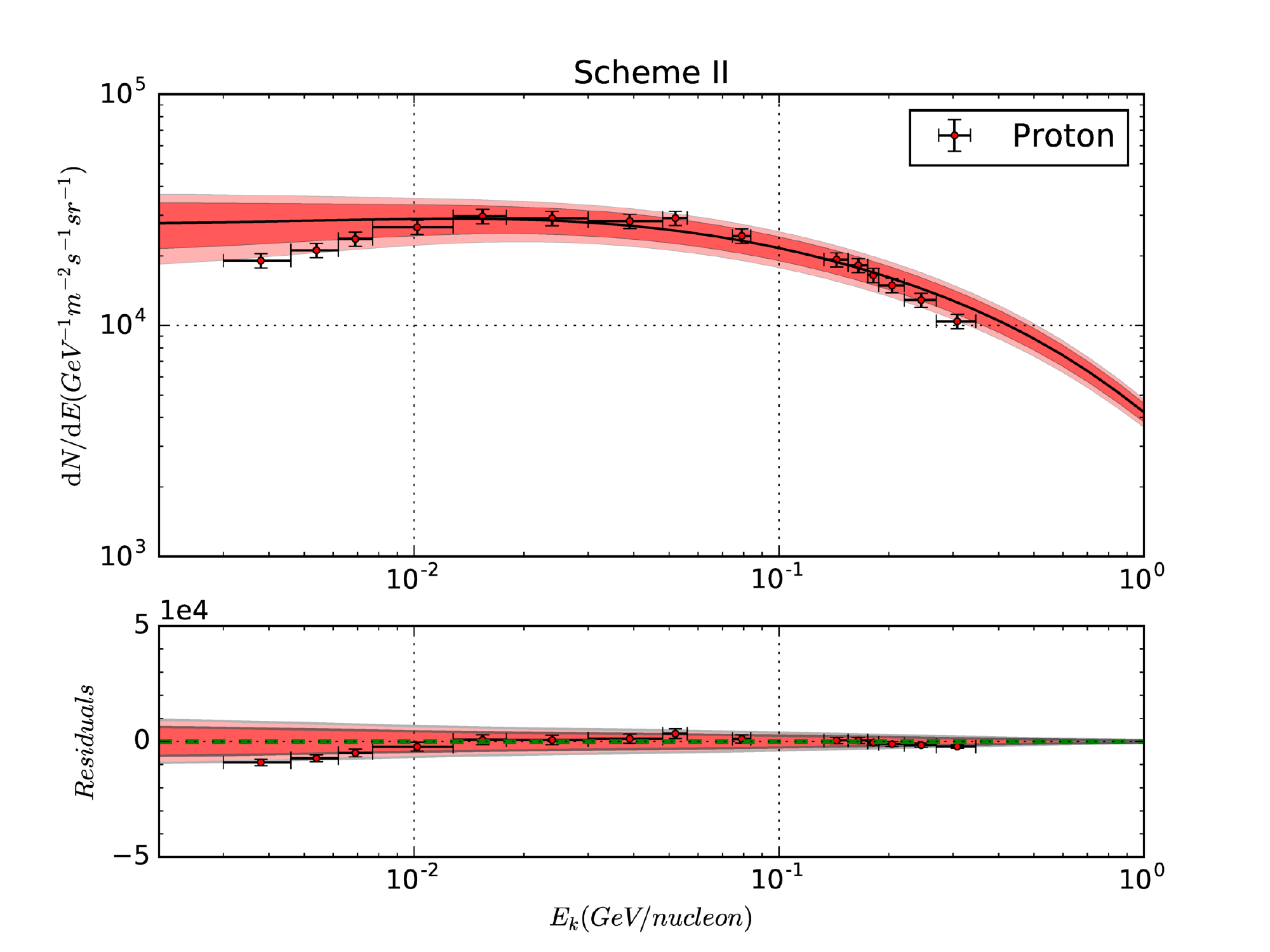}
  \includegraphics[width=0.3\textwidth]{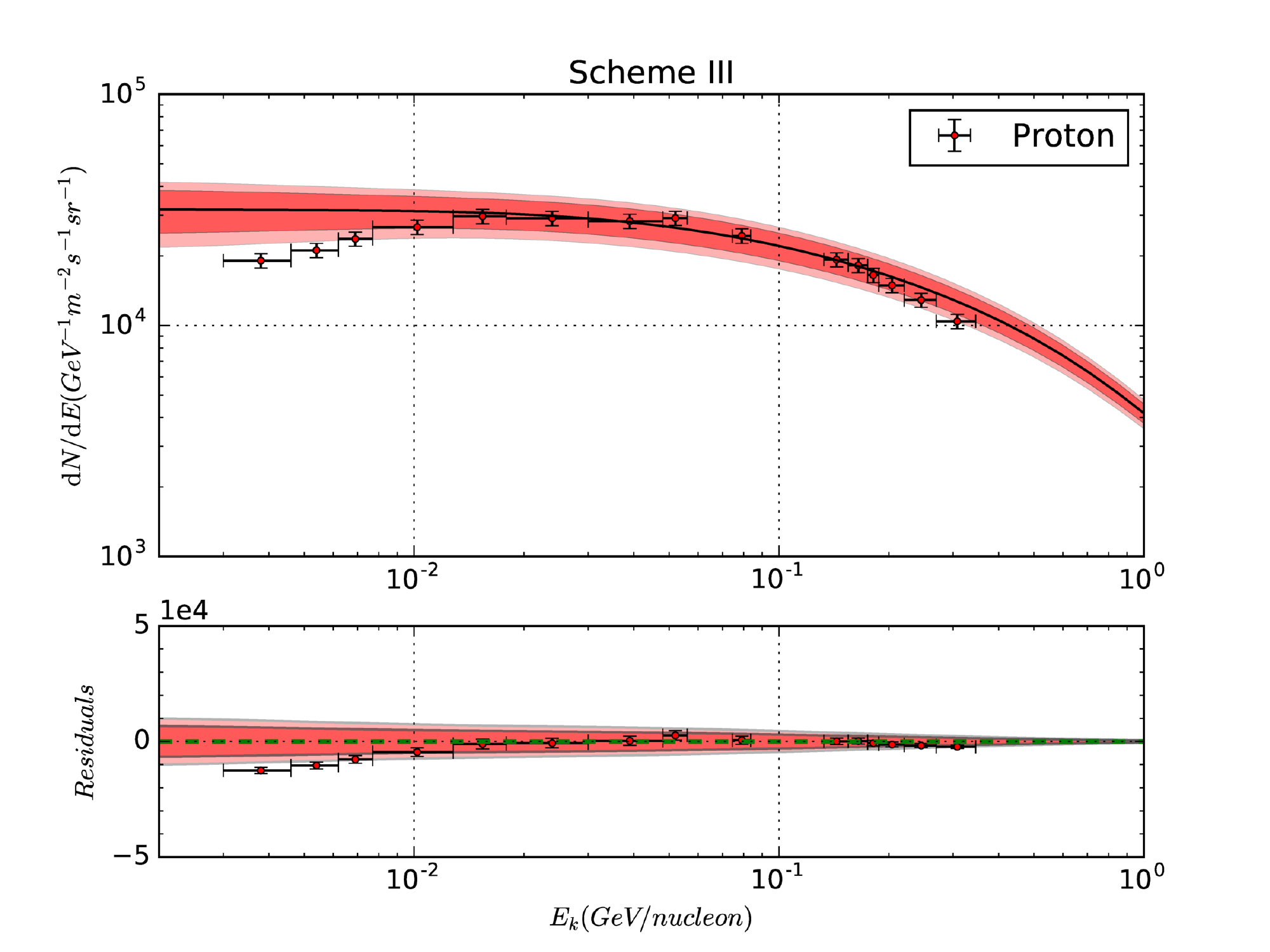}
  \includegraphics[width=0.3\textwidth]{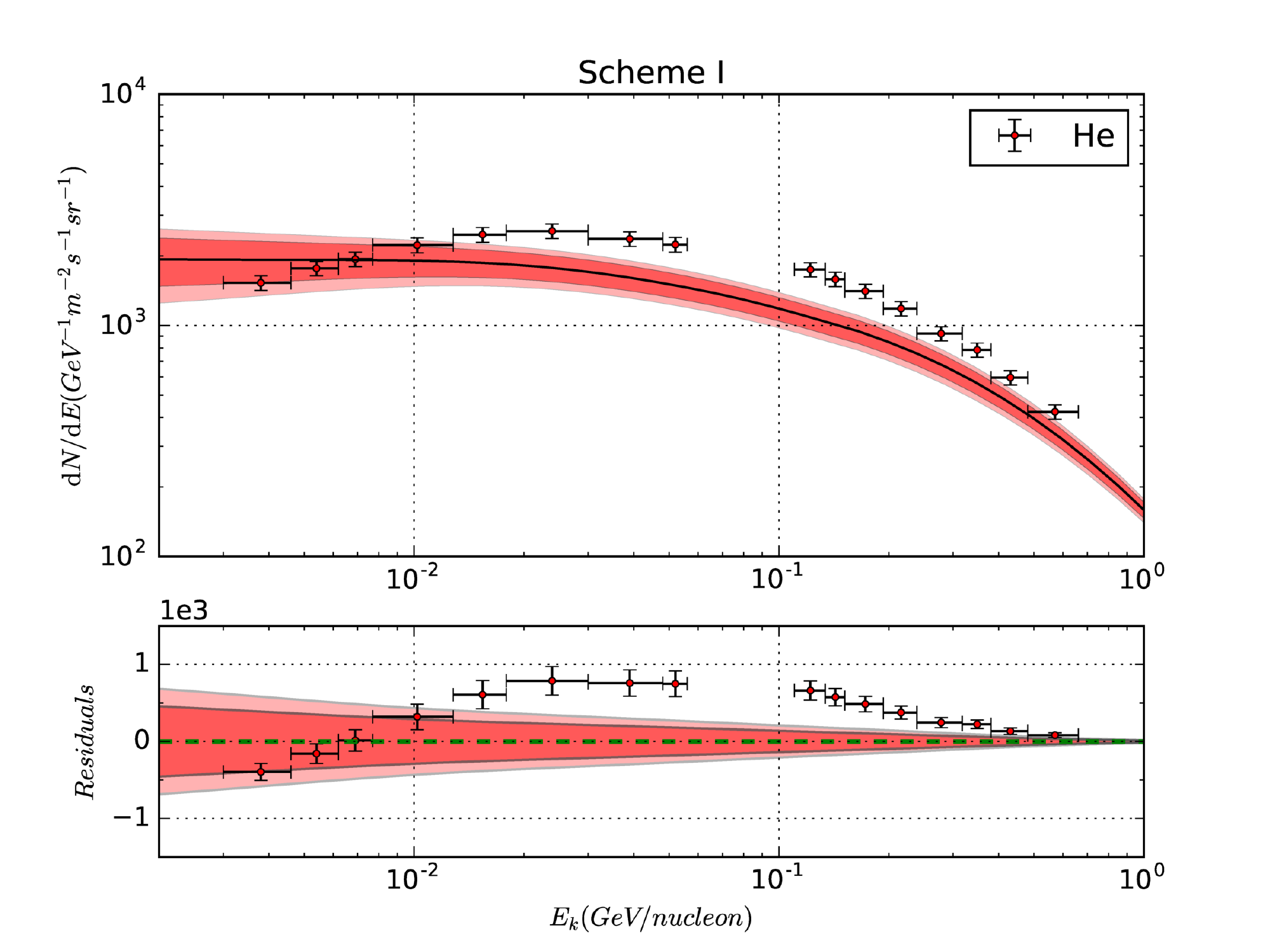}
  \includegraphics[width=0.3\textwidth]{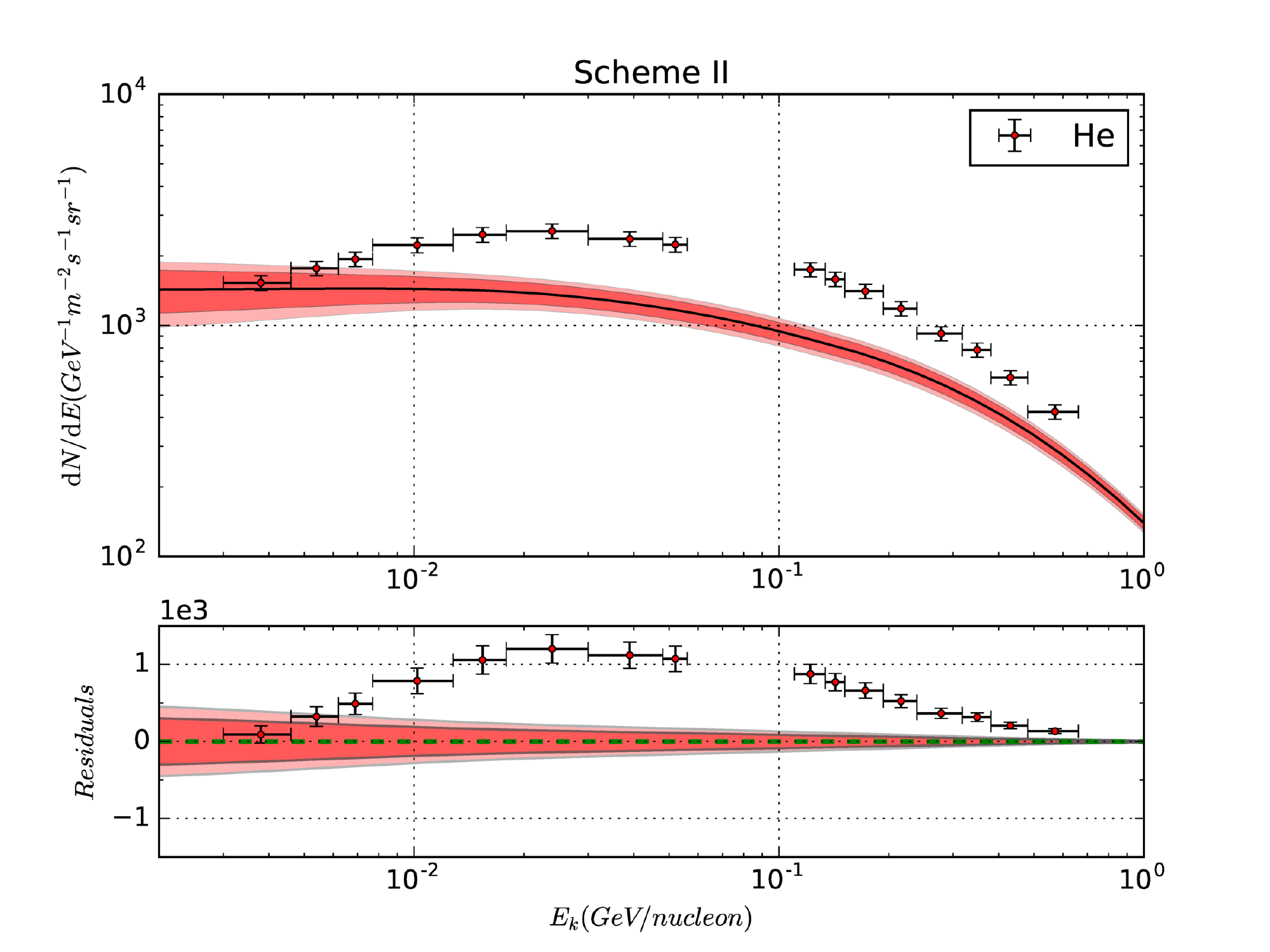}
  \includegraphics[width=0.3\textwidth]{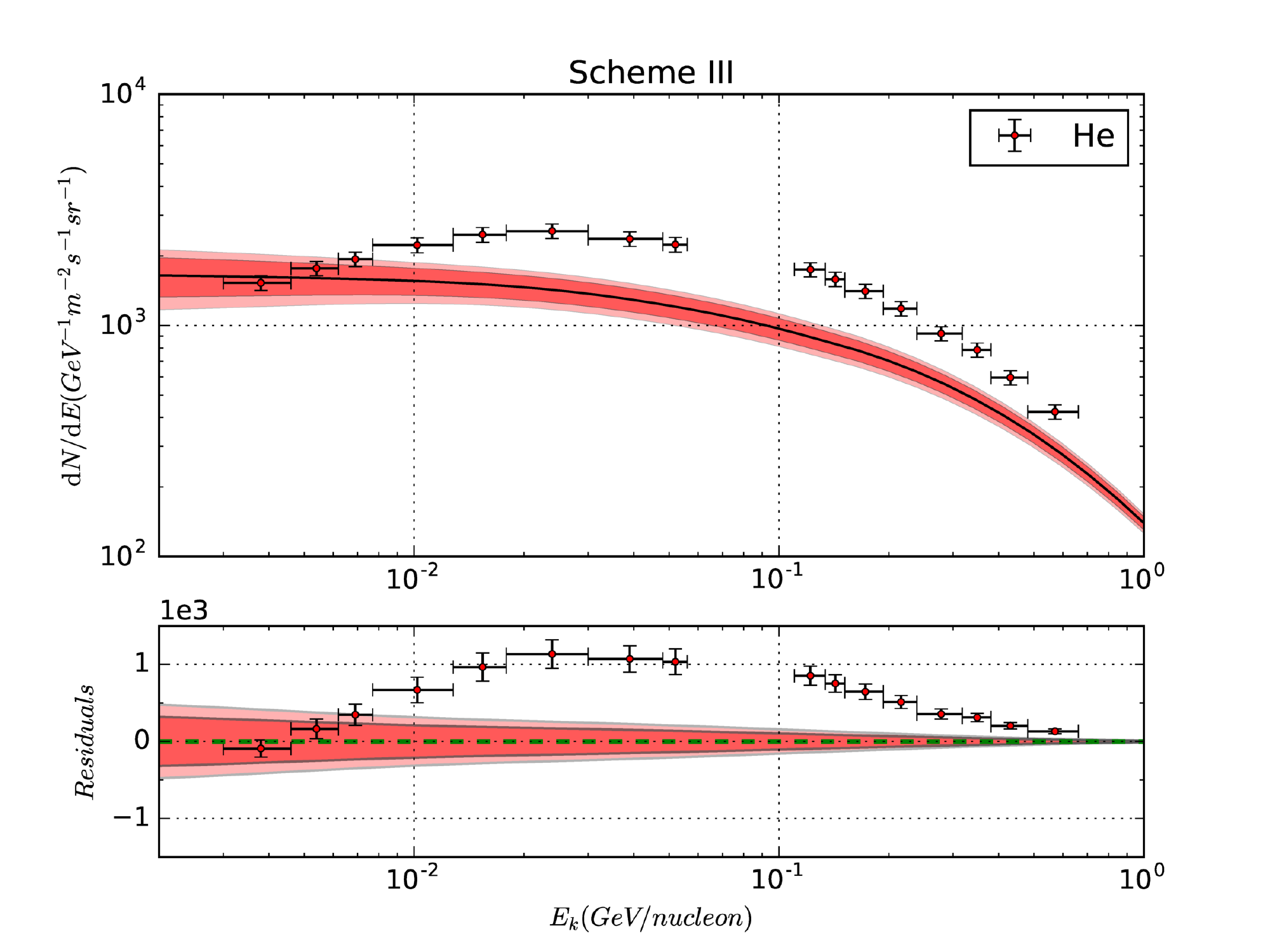}
  \includegraphics[width=0.3\textwidth]{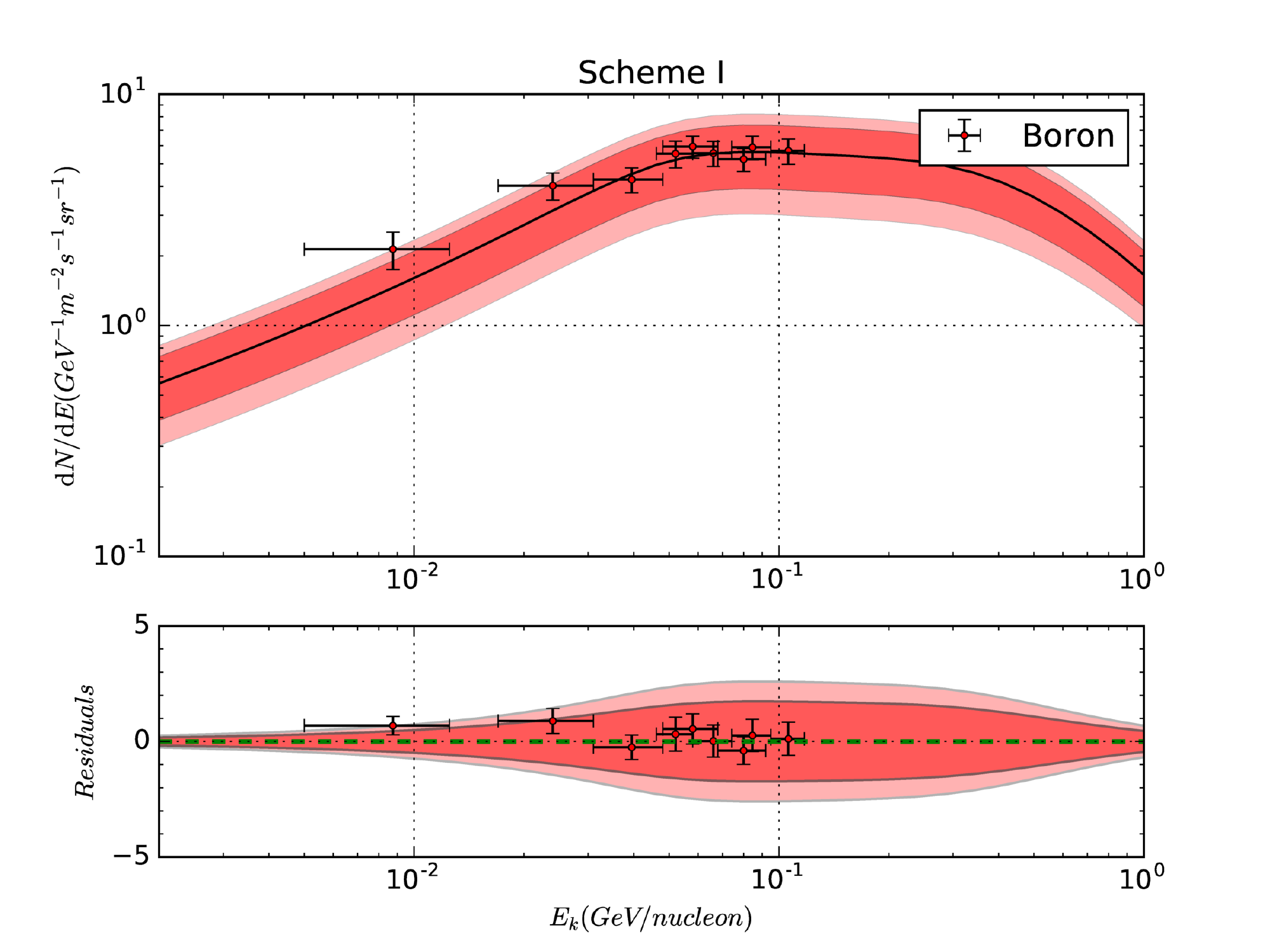}
  \includegraphics[width=0.3\textwidth]{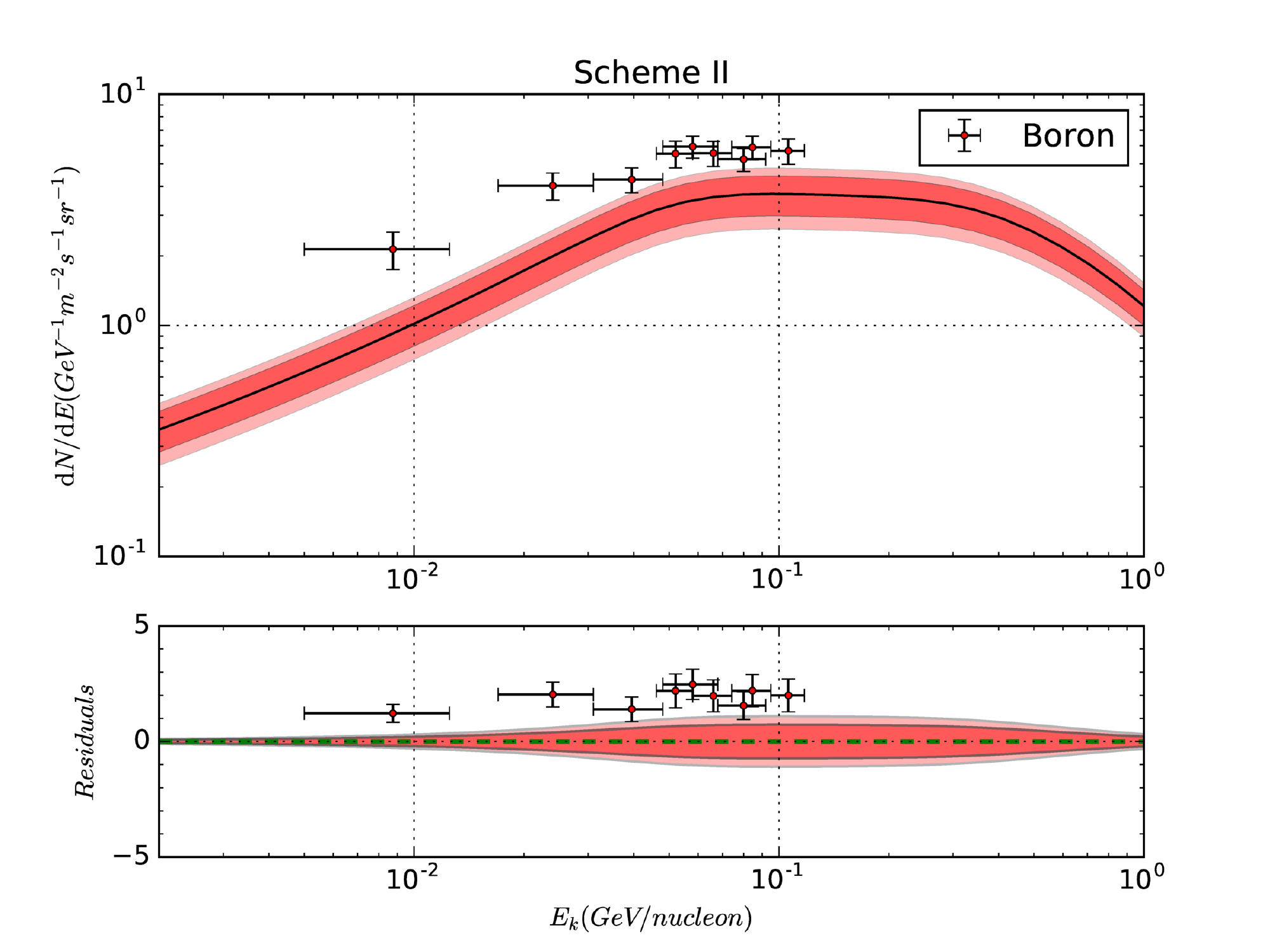}
  \includegraphics[width=0.3\textwidth]{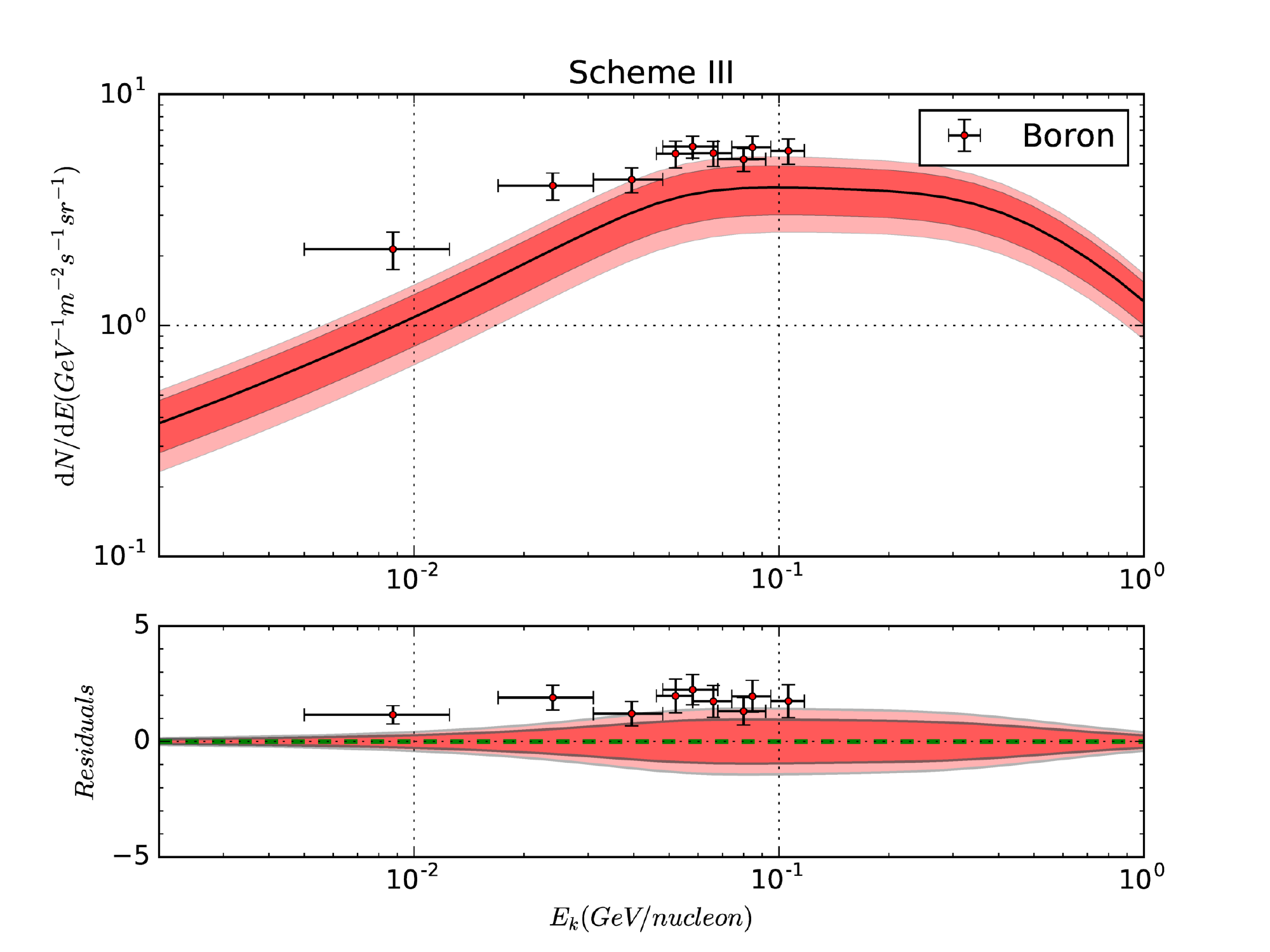}
  \includegraphics[width=0.3\textwidth]{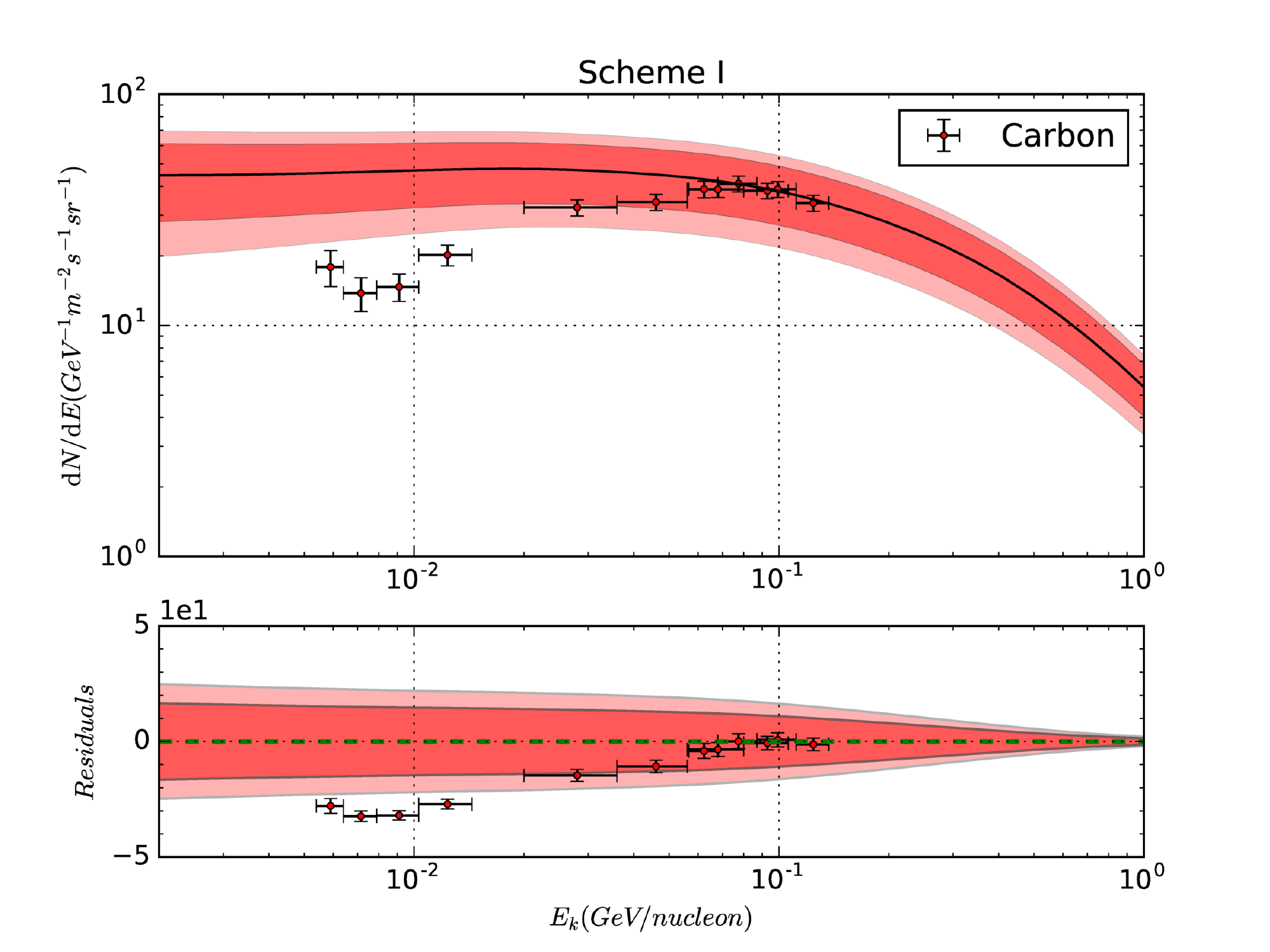}
  \includegraphics[width=0.3\textwidth]{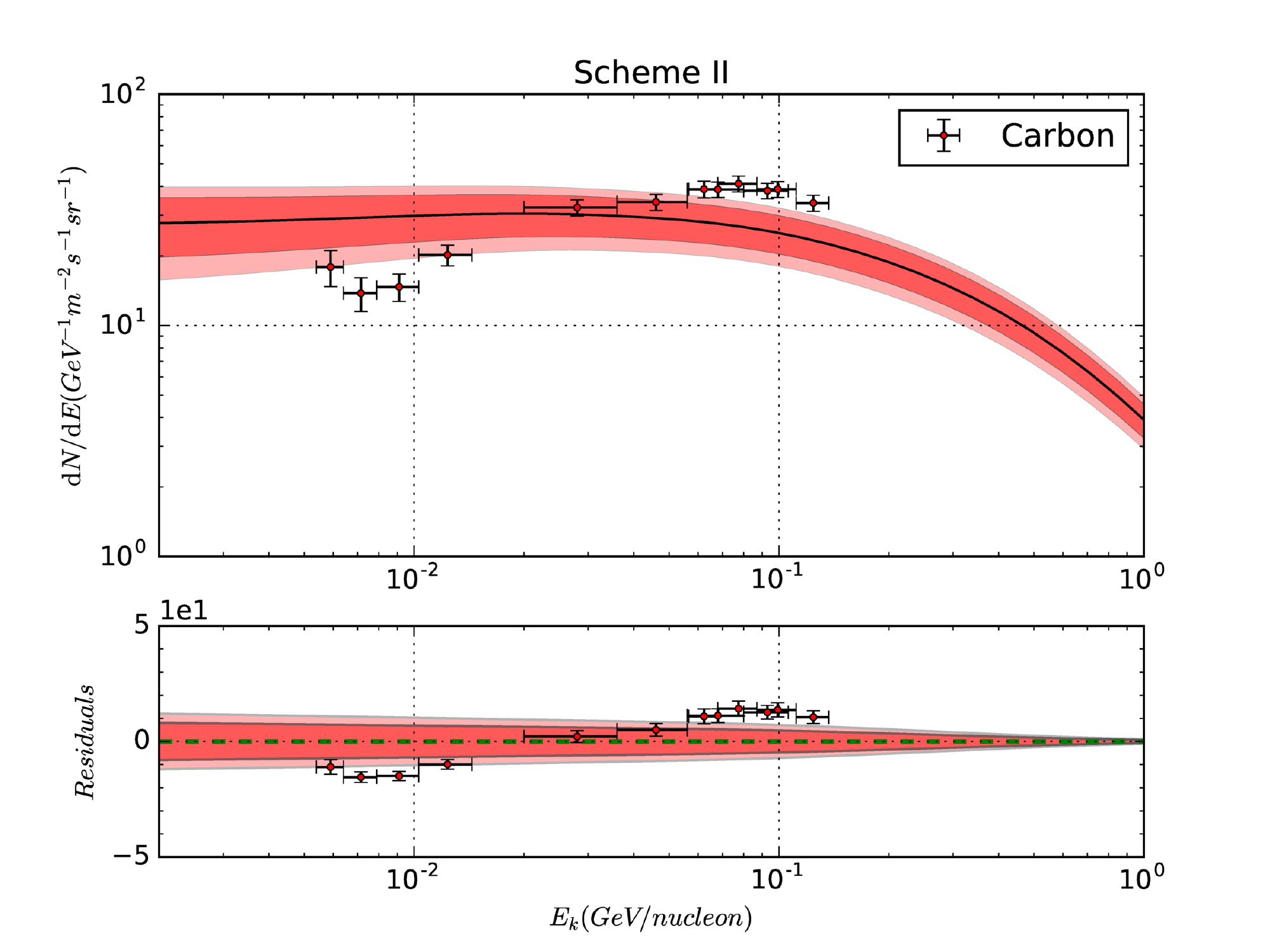}
  \includegraphics[width=0.3\textwidth]{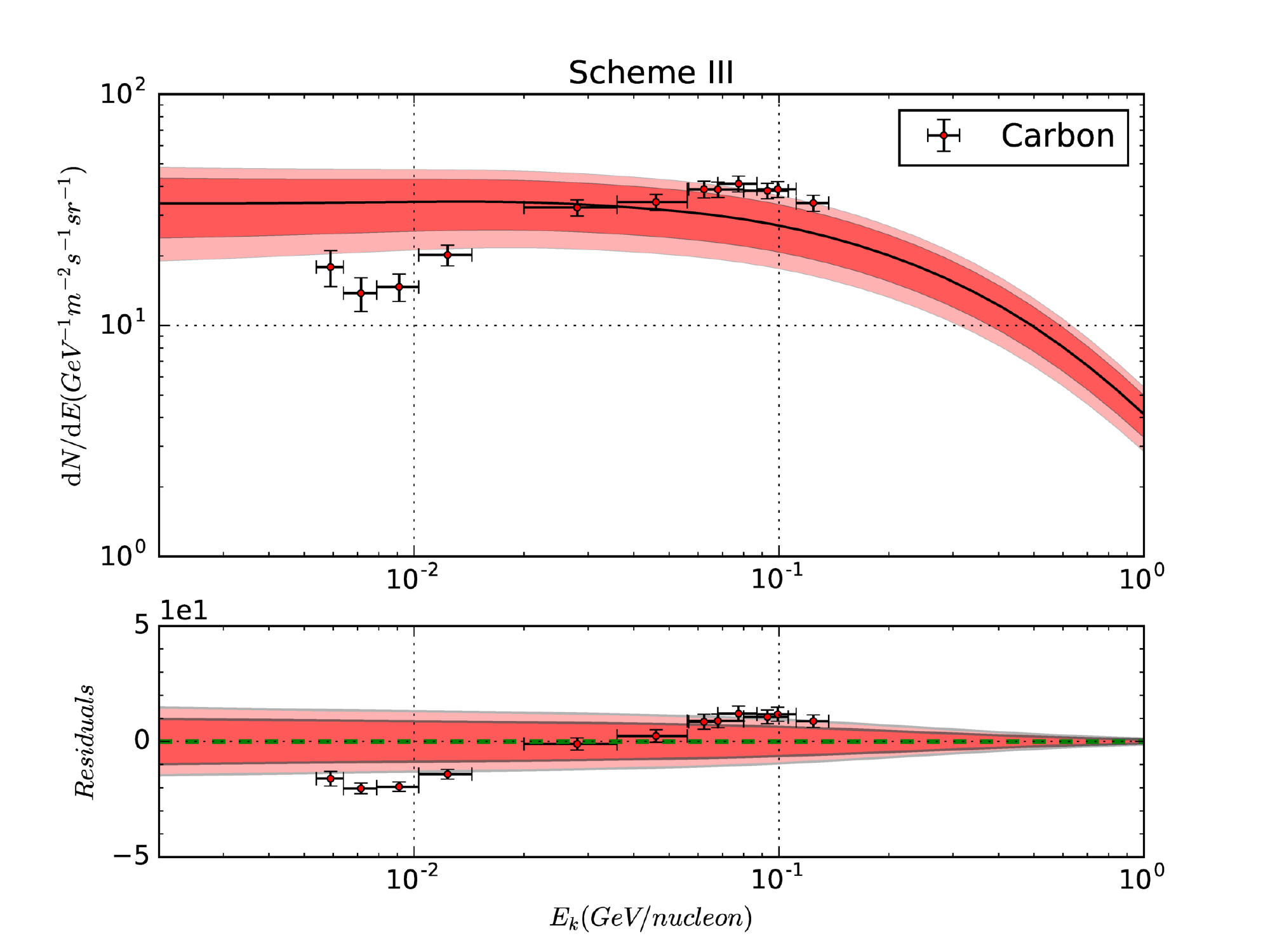}

  \caption{The comparison between the best fitting results (with $\phi = 0$) and the VOYAGER-1 data for Scheme I, II and III. The $2\sigma$ (deep red) and $3\sigma$ (light red) bound are also showed in the figures.}
  \label{fig:voyager}
\end{figure*}

  In order to study the details using the fitting results as far as possible, we take $\phi = 0$ and extrapolate 
the fitting results of the 3 Schemes to 1 MeV/nucleon -- 1 GeV/nucleon in Fig. \ref{fig:voyager}. The data   
in Fig. \ref{fig:voyager} from VOYAGER-1 \citep{Cummings2016}, which has been measured outside of the heliosphere, 
is considered as the local interstellar spectra (LIS) that was unaffected (or little affected) by solar modulation. The comparison between the LIS measured by VOYAGER-1 and the fitting results ($\phi = 0$) gives us more information about the CRs propagation in low energy region. In Fig. \ref{fig:voyager}, the trend of  proton flux and boron flux is well fitted but there exist overestimation for proton in Scheme I and underestimation for boron in Schemes II and III. 
For carbon, there exists fine structure in the spectrum which is mis-modeled. Considering the different collection periods of the AMS-02 data in global fitting and the above reasons (ii) and (iii), 
we do not focus on these features further more in this work.
  What is more interesting comes from the defective fitting of helium flux which is largely different with the result of proton flux. From Table \ref{tab:data_period}, we note that the collection periods of proton and helium fluxes, which are used for our global fitting, are the same. If all the configurations are right, the results for proton and helium fluxes in Fig. \ref{fig:voyager} should give a same or similar level of residuals.   The different levels of the fitting results between proton and helium reveal the different primary sources or propagation mechanisms between these two species in low energy region.

  Additionally, the results in Fig. \ref{fig:inj_spe_para_2d} reveal the differences between the injection spectra of proton and helium $\gtrsim 18 \GV$ ($\nu_{\p2}-\nu_{\A2} \sim 0.06$). This result is called p/He anomaly which is generally ascribed to particle-dependent acceleration mechanisms occurring in Galactic CR sources (see for e.g. \cite{Vladimirov2012}). And many specific mechanisms are proposed to interpret this anomaly (see for e.g. \cite{Erlykin2015,Malkov2012,Fisk2012,Ohira2011,Tomassetti2015apjl01}).

 Comparing with the slope difference for the observed spectra ($\sim 0.08$ at rigidity $R = 45-1800 \GV$ \citep{Tomassetti2015apjl01} ) from AMS-02, we can ascribe this difference ($\sim 0.06$ from injection and $\sim 0.08$ from propagated) to propagation effects, because helium particles interact with the ISM more than proton (see for e.g., \citep{Blasi2012,Tomassetti2012,Tomassetti2015prd,Aloisio2015}). 

 In the energy region $\lesssim 18 \GV$, we can conclude from Figs. \ref{fig:prop_results} and \ref{fig:voyager} 
that the fitting results for proton and helium are also obviously different. But if we consider the discrepancy 
from the fitting of helium flux in Fig. \ref{fig:voyager}, we can conclude that the helium propagation in this region 
is mis-modeled.  In any event, it seems that the primary sources ($\gtrsim 18 \GV$) and 
propagation mechanisms ($\lesssim 18 \GV$) between proton and helium are different, which need further
studies to reveal the physics behind it.

\begin{table}[htb]
\begin{center}
\begin{tabular}{c | c}
  \hline\hline
ID  & Periods             \\
\hline
B/C & 2011/05/19-00:00:00 -- 2016/05/26-00:00:00 \\

Proton & 2011/05/19-00:00:00 -- 2013/11/26-00:00:00 \\

Helium & 2011/05/19-00:00:00 -- 2013/11/26-00:00:00 \\

$\pbarp$ & 2011/05/19-00:00:00 -- 2015/05/26-00:00:00 \\
\hline\hline

\end{tabular}
\end{center}
\caption{ The periods of the relevant AMS-02 data collected.}
\label{tab:data_period}
\end{table}

\begin{figure*}[!htbp]
  \centering
  \includegraphics[width=0.4\textwidth]{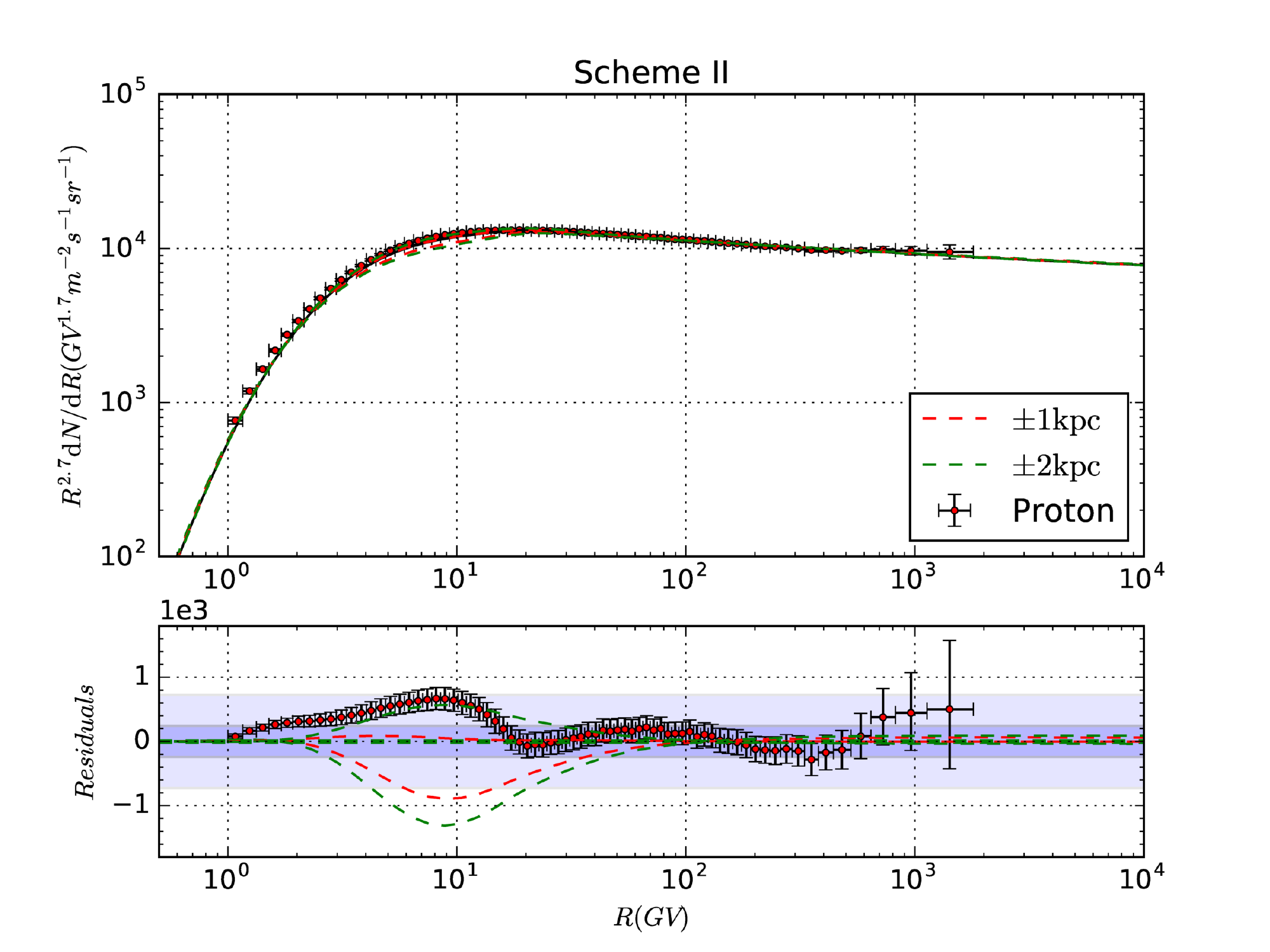}
  \includegraphics[width=0.4\textwidth]{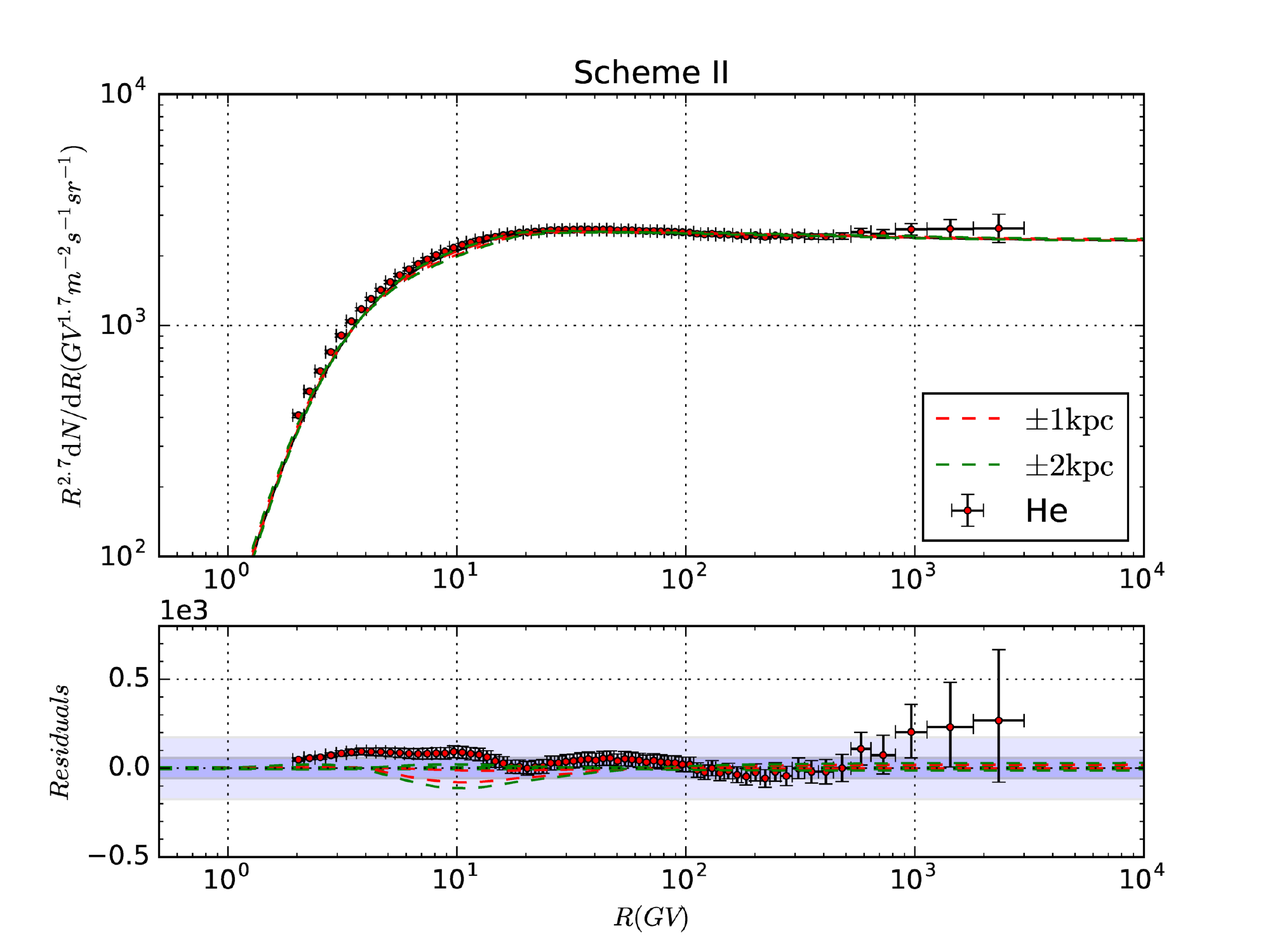}
  \includegraphics[width=0.4\textwidth]{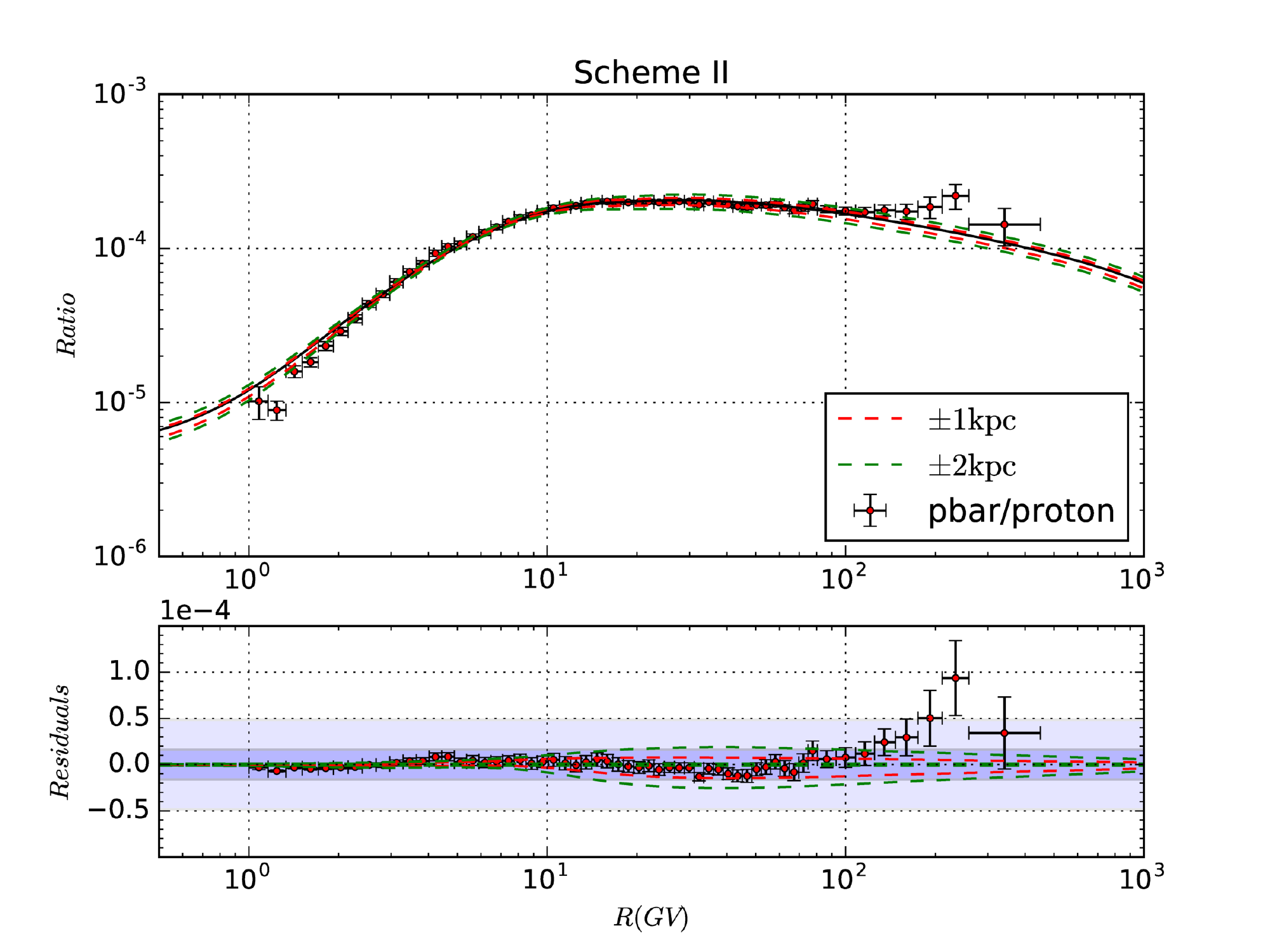}
  \includegraphics[width=0.4\textwidth]{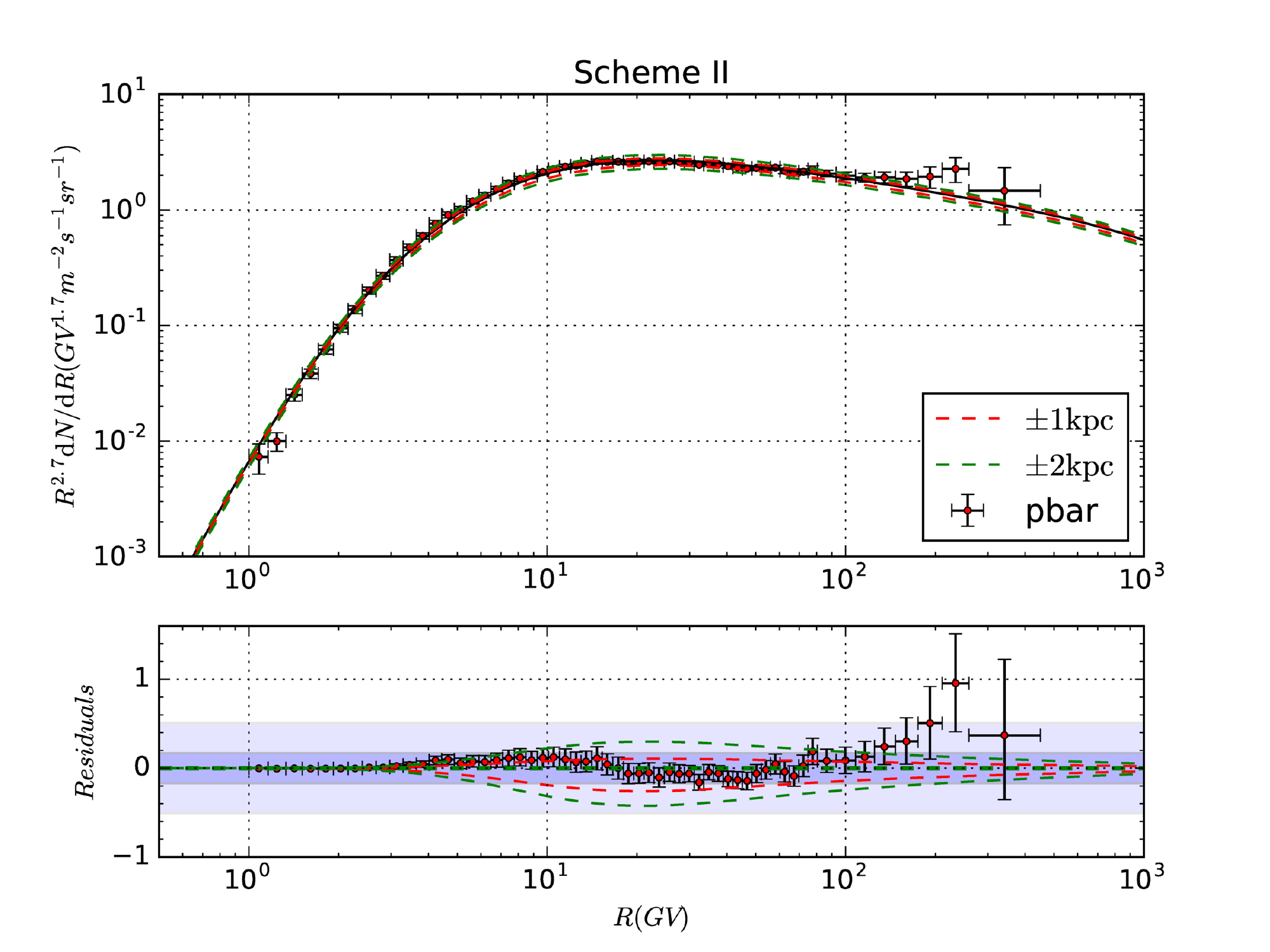}
  \includegraphics[width=0.4\textwidth]{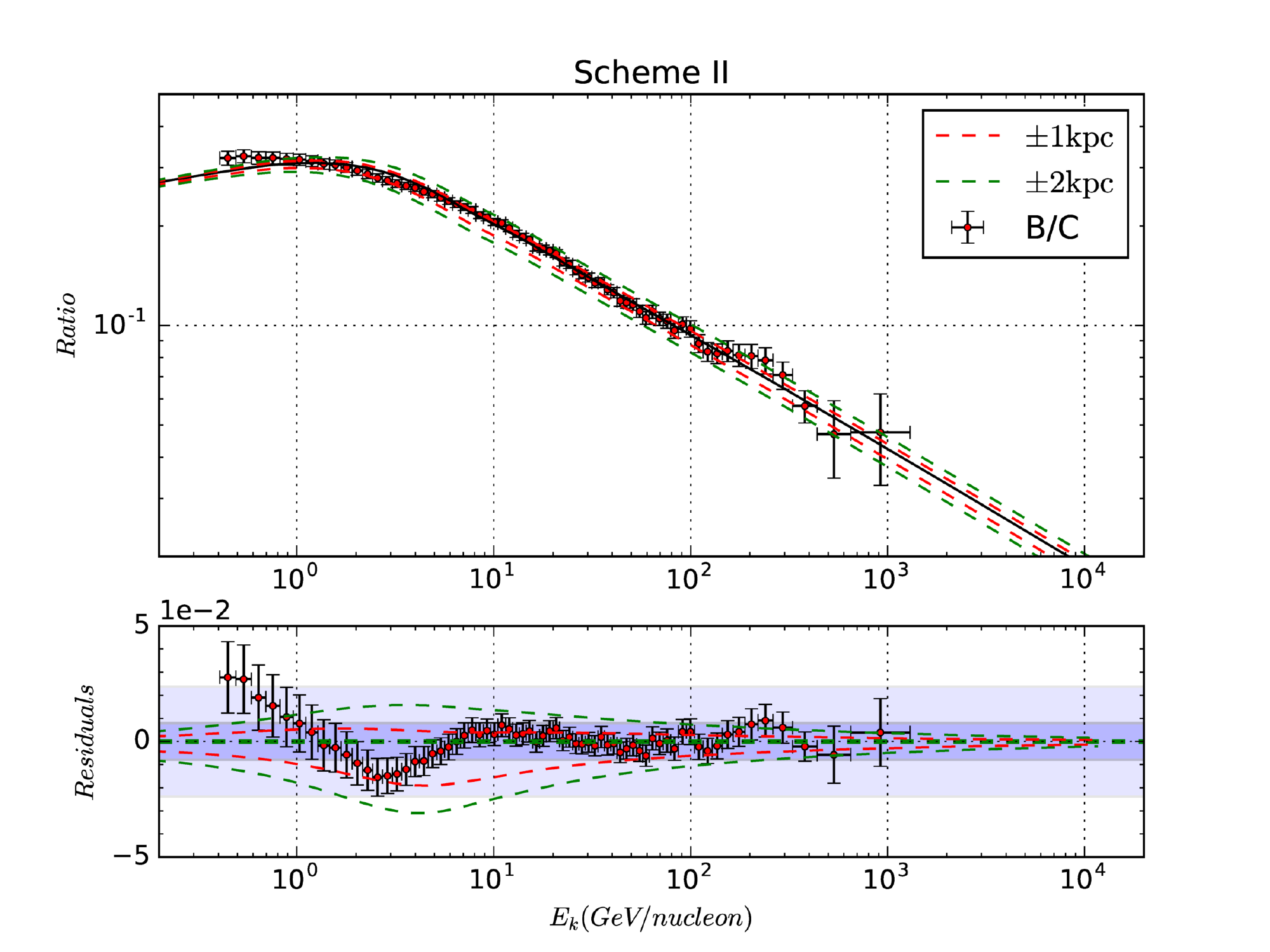}
\caption{The comparison between the best fit results and the perturbation results which use different $D_{0}$ and $z_{h}$ values (labeled by the perturbations on $z_{h}$) with fixed $D_{0}/z_{h}$ at best fit parameters from Scheme II. The $\pm 1 \kpc$ and $\pm 2 \kpc$ perturbation results are presented by red and green dash lines. The $\sigma$ and 3$\sigma$ of the fitting residuals are showed in deep blue and light blue in the lower panel of each sub-figures respectively. }
  \label{fig:results_diff_z}
\end{figure*}

 In order to see how the degeneracy between $D_{0}$ and $z_{h}$ is relieved by the attendance of $\pbar$ data, 
we pertubate the values of $D_{0}$ and $z_{h}$ at the same time and hold the  $D_{0}/z_{h}$ fixed at its best fit value 
based on the Scheme II (all the other parameters are also fixed in this case).  Figure \ref{fig:results_diff_z} shows 
the results when $z_{h}$ is perturbated by 1 and 2 kpc. From Fig. \ref{fig:results_diff_z}, we can find 
that the sensitivity regions of B/C, proton and helium are all $\lesssim 10 \GV$. In this energy region, 
the results are seriously influenced by solar modulation which is mismodeled by force field approximation. 
On the other hand, the sensitivity region of $\pbar$ data locate at $10-100 \GV$, where the influence of 
solar modulation can be ignored and the propagation effects 
(which is closely related to the $D_{0}$ and $z_{h}$ values) plays a main role. This is the visualized interpretation 
of the degeneracy's break. As a result, we cannot relieve this degeneracy more efficiently using the B/C, proton, helium and $^{10}$Be/$^{9}$Be data (which is always $< 10 \GeV$) before the solar modulation have been precisely modeled.

\section{Conclusions}

In this work, we use the newly released AMS-02  nuclei data (B/C, proton, helium and $\pbarp$) only, to study 
 3 Schemes with different data sets and different propagation models. In this scenario, the systematics between different experiments are avoided. Additionally,  we use separate primary spectra settings for proton and other nuclei 
 (all $Z >1$ nuclei have the same injection parameters) because of the observed significant difference in the slopes of proton and helium, which can reveal the sources' differences between them.

According to the fitting results and the posterior PDFs of different groups of the schemes' parameters, we present 
our main conclusion as follows.

\begin{itemize}
\item[(i)] The newly reported AMS-02 nuclei data set (B/C, proton, helium and $\pbarp$) can effectively relieve the degeneracy of the classical correlation between $D_0$ and $z_h$, and our results for the constraints 
on some parameters show a concrete improvement compared with previous works. Benefitted from 
the self-consistence of the new data set from AMS-02, the fitting results 
(see Fig.~\ref{fig:prop_results}) show a little bias (note the 2$\sigma$ and 3$\sigma$ bounds), and thus the disadvantages and limitations of the existed propagation models emerge.

\item[(ii)] Based on (i), the major discrepancy obviously comes from the fitting results lower 
than $\sim 10 \GeV$. This discrepancy shows that the force-field approximation cannot deal with the solar modulation in reality and more detailed treatments should be employed in this data level (see, e.g., \citep{DellaTorre2017,Boskini2017}).

\item[(iii)]  Also based on (i), there is an obvious excess for $\pbar$ flux and $\pbarp$ ratio data 
 from the corresponding fitting results in Fig. \ref{fig:prop_results} for Scheme I, which could not be 
explained by the standard propagation models  (one break for injection spectra and 
a uniform diffusion coefficient in the whole propagation region). This gives a concrete hint 
for new solutions, including dark matter (see, e.g., \citep{Cui2017,Cuoco2017}).

\item[(iv)] The difference of the second slopes between them $\nu_{\p2} - \nu_{\A2} \sim 0.06$ which has a high level of confidence and interpret that the primary source of proton is different from other nuclei when $R \gtrsim 18 \GV$ .
Additionally, the comparison between the best-fitting results with $\phi = 0$ and VOYAGER-1 data shows that 
the corresponding results of proton and helium fluxes after propagation ($\lesssim 1 \GeV$) are obviously different. 
Altogether, the primary sources or propagation mechanisms should be different between proton and helium (and other heavier nucleus species). These results do need proper explanation. 

\item[(v)]  If we want to fit the data set precisely, the helium-4 abundance should have a value of $\sim 4.895 \times 10^{4} = 0.6868 \times (7.199 \times 10^{4})$, the energy-independent rescaling factor $c_{\pbar}$ should have a value of $1.33 - 1.39$ within a  confidence level of $95 \%$ and the effective solar modulation $\phi \sim 0.62$. The physics behind $c_{\He}$ and $c_{\pbar}$ should be attended in further research.

\item[(vi)]  The new data set (B/C, proton, helium and $\pbarp$) favors a very small value 
($\diff V_{c} / \diff z \sim 0.558 \km / \s$) of convection (or disfavors the model with convection), 
which is different from some previous works (see, e.g., \citep{Yuan2017}), and needs further studies. 

\end{itemize}

Thanks to the precise measurements of CR data by AMS-02, with more and more precise data available, we are going into a precision-driven era and able to investigate the CR-related problems in great details. With the results of this work, it turns out that the problem seems to be more complicated than what we expected based on the rough measurements in the past (especially in the low-energy region). Thus, CR physics becomes a comprehensive discipline which now requires the improvement not only for itself, but also other disciplines like atomic physics and space physics.

\section*{ACKNOWLEDGMENTS}
We would like to thank Xiao-Jun Bi, Su-Jie Lin, and Qiang Yuan very much for helpful discussions,
 \citet{corner} to provide us the tool to visualize multidimensional samples using a scatterplot matrix,
 and \citet{Maurin2014} to collect database and associated online tools  for charged cosmic-ray measurements.
 Many thanks for the referees' valuable and detailed suggestions, which led to a great progress in this work.
This research was supported in part by the Projects 11475238 and 11647601 supported by National  Science Foundation of China,
and by Key Research Program of Frontier Sciences, CAS. 
The calculation in this paper are supported by HPC Cluster of SKLTP/ITP-CAS.


\begin{thebibliography}{80}%
\makeatletter
\providecommand \@ifxundefined [1]{%
 \@ifx{#1\undefined}
}%
\providecommand \@ifnum [1]{%
 \ifnum #1\expandafter \@firstoftwo
 \else \expandafter \@secondoftwo
 \fi
}%
\providecommand \@ifx [1]{%
 \ifx #1\expandafter \@firstoftwo
 \else \expandafter \@secondoftwo
 \fi
}%
\providecommand \natexlab [1]{#1}%
\providecommand \enquote  [1]{``#1''}%
\providecommand \bibnamefont  [1]{#1}%
\providecommand \bibfnamefont [1]{#1}%
\providecommand \citenamefont [1]{#1}%
\providecommand \href@noop [0]{\@secondoftwo}%
\providecommand \href [0]{\begingroup \@sanitize@url \@href}%
\providecommand \@href[1]{\@@startlink{#1}\@@href}%
\providecommand \@@href[1]{\endgroup#1\@@endlink}%
\providecommand \@sanitize@url [0]{\catcode `\\12\catcode `\$12\catcode
  `\&12\catcode `\#12\catcode `\^12\catcode `\_12\catcode `\%12\relax}%
\providecommand \@@startlink[1]{}%
\providecommand \@@endlink[0]{}%
\providecommand \url  [0]{\begingroup\@sanitize@url \@url }%
\providecommand \@url [1]{\endgroup\@href {#1}{\urlprefix }}%
\providecommand \urlprefix  [0]{URL }%
\providecommand \Eprint [0]{\href }%
\providecommand \doibase [0]{http://dx.doi.org/}%
\providecommand \selectlanguage [0]{\@gobble}%
\providecommand \bibinfo  [0]{\@secondoftwo}%
\providecommand \bibfield  [0]{\@secondoftwo}%
\providecommand \translation [1]{[#1]}%
\providecommand \BibitemOpen [0]{}%
\providecommand \bibitemStop [0]{}%
\providecommand \bibitemNoStop [0]{.\EOS\space}%
\providecommand \EOS [0]{\spacefactor3000\relax}%
\providecommand \BibitemShut  [1]{\csname bibitem#1\endcsname}%
\let\auto@bib@innerbib\@empty
\bibitem [{\citenamefont {{Strong}}\ \emph {et~al.}(2007)\citenamefont
  {{Strong}}, \citenamefont {{Moskalenko}},\ and\ \citenamefont
  {{Ptuskin}}}]{Strong2007}%
  \BibitemOpen
  \bibfield  {author} {\bibinfo {author} {\bibfnamefont {A.~W.}\ \bibnamefont
  {{Strong}}}, \bibinfo {author} {\bibfnamefont {I.~V.}\ \bibnamefont
  {{Moskalenko}}}, \ and\ \bibinfo {author} {\bibfnamefont {V.~S.}\
  \bibnamefont {{Ptuskin}}},\ }\href {\doibase
  10.1146/annurev.nucl.57.090506.123011} {\bibfield  {journal} {\bibinfo
  {journal} {Annual Review of Nuclear and Particle Science}\ }\textbf {\bibinfo
  {volume} {57}},\ \bibinfo {pages} {285} (\bibinfo {year} {2007})},\ \Eprint
  {http://arxiv.org/abs/astro-ph/0701517} {astro-ph/0701517} \BibitemShut
  {NoStop}%
\bibitem [{\citenamefont {{Webber}}\ \emph {et~al.}(1992)\citenamefont
  {{Webber}}, \citenamefont {{Lee}},\ and\ \citenamefont
  {{Gupta}}}]{Webber1992}%
  \BibitemOpen
  \bibfield  {author} {\bibinfo {author} {\bibfnamefont {W.~R.}\ \bibnamefont
  {{Webber}}}, \bibinfo {author} {\bibfnamefont {M.~A.}\ \bibnamefont {{Lee}}},
  \ and\ \bibinfo {author} {\bibfnamefont {M.}~\bibnamefont {{Gupta}}},\ }\href
  {\doibase 10.1086/171262} {\bibfield  {journal} {\bibinfo  {journal} {\apj}\
  }\textbf {\bibinfo {volume} {390}},\ \bibinfo {pages} {96} (\bibinfo {year}
  {1992})}\BibitemShut {NoStop}%
\bibitem [{\citenamefont {{Bloemen}}\ \emph {et~al.}(1993)\citenamefont
  {{Bloemen}}, \citenamefont {{Dogiel}}, \citenamefont {{Dorman}},\ and\
  \citenamefont {{Ptuskin}}}]{Bloemen1993}%
  \BibitemOpen
  \bibfield  {author} {\bibinfo {author} {\bibfnamefont {J.~B.~G.~M.}\
  \bibnamefont {{Bloemen}}}, \bibinfo {author} {\bibfnamefont {V.~A.}\
  \bibnamefont {{Dogiel}}}, \bibinfo {author} {\bibfnamefont {V.~L.}\
  \bibnamefont {{Dorman}}}, \ and\ \bibinfo {author} {\bibfnamefont {V.~S.}\
  \bibnamefont {{Ptuskin}}},\ }\href@noop {} {\bibfield  {journal} {\bibinfo
  {journal} {\aap}\ }\textbf {\bibinfo {volume} {267}},\ \bibinfo {pages} {372}
  (\bibinfo {year} {1993})}\BibitemShut {NoStop}%
\bibitem [{\citenamefont {{Maurin}}\ \emph
  {et~al.}(2002{\natexlab{a}})\citenamefont {{Maurin}}, \citenamefont
  {{Taillet}},\ and\ \citenamefont {{Donato}}}]{Maurin2002a}%
  \BibitemOpen
  \bibfield  {author} {\bibinfo {author} {\bibfnamefont {D.}~\bibnamefont
  {{Maurin}}}, \bibinfo {author} {\bibfnamefont {R.}~\bibnamefont {{Taillet}}},
  \ and\ \bibinfo {author} {\bibfnamefont {F.}~\bibnamefont {{Donato}}},\
  }\href {\doibase 10.1051/0004-6361:20021176} {\bibfield  {journal} {\bibinfo
  {journal} {\aap}\ }\textbf {\bibinfo {volume} {394}},\ \bibinfo {pages}
  {1039} (\bibinfo {year} {2002}{\natexlab{a}})},\ \Eprint
  {http://arxiv.org/abs/astro-ph/0206286} {astro-ph/0206286} \BibitemShut
  {NoStop}%
\bibitem [{\citenamefont {{Shibata}}\ \emph {et~al.}(2004)\citenamefont
  {{Shibata}}, \citenamefont {{Hareyama}}, \citenamefont {{Nakazawa}},\ and\
  \citenamefont {{Saito}}}]{Shibata2004}%
  \BibitemOpen
  \bibfield  {author} {\bibinfo {author} {\bibfnamefont {T.}~\bibnamefont
  {{Shibata}}}, \bibinfo {author} {\bibfnamefont {M.}~\bibnamefont
  {{Hareyama}}}, \bibinfo {author} {\bibfnamefont {M.}~\bibnamefont
  {{Nakazawa}}}, \ and\ \bibinfo {author} {\bibfnamefont {C.}~\bibnamefont
  {{Saito}}},\ }\href {\doibase 10.1086/422548} {\bibfield  {journal} {\bibinfo
   {journal} {\apj}\ }\textbf {\bibinfo {volume} {612}},\ \bibinfo {pages}
  {238} (\bibinfo {year} {2004})}\BibitemShut {NoStop}%
\bibitem [{\citenamefont {{Strong}}\ and\ \citenamefont
  {{Moskalenko}}(1998)}]{Strong1998}%
  \BibitemOpen
  \bibfield  {author} {\bibinfo {author} {\bibfnamefont {A.~W.}\ \bibnamefont
  {{Strong}}}\ and\ \bibinfo {author} {\bibfnamefont {I.~V.}\ \bibnamefont
  {{Moskalenko}}},\ }\href {\doibase 10.1086/306470} {\bibfield  {journal}
  {\bibinfo  {journal} {\apj}\ }\textbf {\bibinfo {volume} {509}},\ \bibinfo
  {pages} {212} (\bibinfo {year} {1998})},\ \Eprint
  {http://arxiv.org/abs/astro-ph/9807150} {astro-ph/9807150} \BibitemShut
  {NoStop}%
\bibitem [{\citenamefont {{Evoli}}\ \emph {et~al.}(2008)\citenamefont
  {{Evoli}}, \citenamefont {{Gaggero}}, \citenamefont {{Grasso}},\ and\
  \citenamefont {{Maccione}}}]{Evoli2008}%
  \BibitemOpen
  \bibfield  {author} {\bibinfo {author} {\bibfnamefont {C.}~\bibnamefont
  {{Evoli}}}, \bibinfo {author} {\bibfnamefont {D.}~\bibnamefont {{Gaggero}}},
  \bibinfo {author} {\bibfnamefont {D.}~\bibnamefont {{Grasso}}}, \ and\
  \bibinfo {author} {\bibfnamefont {L.}~\bibnamefont {{Maccione}}},\ }\href
  {\doibase 10.1088/1475-7516/2008/10/018} {\bibfield  {journal} {\bibinfo
  {journal} {\jcap}\ }\textbf {\bibinfo {volume} {10}},\ \bibinfo {eid} {018}
  (\bibinfo {year} {2008})},\ \Eprint {http://arxiv.org/abs/0807.4730}
  {arXiv:0807.4730} \BibitemShut {NoStop}%
\bibitem [{\citenamefont {{Kissmann}}(2014)}]{Kissmann2014}%
  \BibitemOpen
  \bibfield  {author} {\bibinfo {author} {\bibfnamefont {R.}~\bibnamefont
  {{Kissmann}}},\ }\href {\doibase 10.1016/j.astropartphys.2014.02.002}
  {\bibfield  {journal} {\bibinfo  {journal} {Astroparticle Physics}\ }\textbf
  {\bibinfo {volume} {55}},\ \bibinfo {pages} {37} (\bibinfo {year} {2014})},\
  \Eprint {http://arxiv.org/abs/1401.4035} {arXiv:1401.4035 [astro-ph.HE]}
  \BibitemShut {NoStop}%
\bibitem [{\citenamefont {{Trotta}}\ \emph {et~al.}(2011)\citenamefont
  {{Trotta}}, \citenamefont {{J{\'o}hannesson}}, \citenamefont {{Moskalenko}},
  \citenamefont {{Porter}}, \citenamefont {{Ruiz de Austri}},\ and\
  \citenamefont {{Strong}}}]{Trotta2011}%
  \BibitemOpen
  \bibfield  {author} {\bibinfo {author} {\bibfnamefont {R.}~\bibnamefont
  {{Trotta}}}, \bibinfo {author} {\bibfnamefont {G.}~\bibnamefont
  {{J{\'o}hannesson}}}, \bibinfo {author} {\bibfnamefont {I.~V.}\ \bibnamefont
  {{Moskalenko}}}, \bibinfo {author} {\bibfnamefont {T.~A.}\ \bibnamefont
  {{Porter}}}, \bibinfo {author} {\bibfnamefont {R.}~\bibnamefont {{Ruiz de
  Austri}}}, \ and\ \bibinfo {author} {\bibfnamefont {A.~W.}\ \bibnamefont
  {{Strong}}},\ }\href {\doibase 10.1088/0004-637X/729/2/106} {\bibfield
  {journal} {\bibinfo  {journal} {\apj}\ }\textbf {\bibinfo {volume} {729}},\
  \bibinfo {eid} {106} (\bibinfo {year} {2011})},\ \Eprint
  {http://arxiv.org/abs/1011.0037} {arXiv:1011.0037 [astro-ph.HE]} \BibitemShut
  {NoStop}%
\bibitem [{\citenamefont {{J{\'o}hannesson}}\ \emph {et~al.}(2016)\citenamefont
  {{J{\'o}hannesson}}, \citenamefont {{Ruiz de Austri}}, \citenamefont
  {{Vincent}}, \citenamefont {{Moskalenko}}, \citenamefont {{Orlando}},
  \citenamefont {{Porter}}, \citenamefont {{Strong}}, \citenamefont {{Trotta}},
  \citenamefont {{Feroz}}, \citenamefont {{Graff}},\ and\ \citenamefont
  {{Hobson}}}]{Johannesson2016}%
  \BibitemOpen
  \bibfield  {author} {\bibinfo {author} {\bibfnamefont {G.}~\bibnamefont
  {{J{\'o}hannesson}}}, \bibinfo {author} {\bibfnamefont {R.}~\bibnamefont
  {{Ruiz de Austri}}}, \bibinfo {author} {\bibfnamefont {A.~C.}\ \bibnamefont
  {{Vincent}}}, \bibinfo {author} {\bibfnamefont {I.~V.}\ \bibnamefont
  {{Moskalenko}}}, \bibinfo {author} {\bibfnamefont {E.}~\bibnamefont
  {{Orlando}}}, \bibinfo {author} {\bibfnamefont {T.~A.}\ \bibnamefont
  {{Porter}}}, \bibinfo {author} {\bibfnamefont {A.~W.}\ \bibnamefont
  {{Strong}}}, \bibinfo {author} {\bibfnamefont {R.}~\bibnamefont {{Trotta}}},
  \bibinfo {author} {\bibfnamefont {F.}~\bibnamefont {{Feroz}}}, \bibinfo
  {author} {\bibfnamefont {P.}~\bibnamefont {{Graff}}}, \ and\ \bibinfo
  {author} {\bibfnamefont {M.~P.}\ \bibnamefont {{Hobson}}},\ }\href {\doibase
  10.3847/0004-637X/824/1/16} {\bibfield  {journal} {\bibinfo  {journal}
  {\apj}\ }\textbf {\bibinfo {volume} {824}},\ \bibinfo {eid} {16} (\bibinfo
  {year} {2016})},\ \Eprint {http://arxiv.org/abs/1602.02243} {arXiv:1602.02243
  [astro-ph.HE]} \BibitemShut {NoStop}%
\bibitem [{\citenamefont {Lin}\ \emph {et~al.}(2015)\citenamefont {Lin},
  \citenamefont {Yuan},\ and\ \citenamefont {Bi}}]{Lin2015}%
  \BibitemOpen
  \bibfield  {author} {\bibinfo {author} {\bibfnamefont {S.-J.}\ \bibnamefont
  {Lin}}, \bibinfo {author} {\bibfnamefont {Q.}~\bibnamefont {Yuan}}, \ and\
  \bibinfo {author} {\bibfnamefont {X.-J.}\ \bibnamefont {Bi}},\ }\href
  {\doibase 10.1103/PhysRevD.91.063508} {\bibfield  {journal} {\bibinfo
  {journal} {Physical Review D}\ }\textbf {\bibinfo {volume} {91}},\ \bibinfo
  {pages} {063508} (\bibinfo {year} {2015})},\ \Eprint
  {http://arxiv.org/abs/1409.6248} {arXiv:1409.6248 [astro-ph.HE]} \BibitemShut
  {NoStop}%
\bibitem [{\citenamefont {{Yuan}}\ \emph {et~al.}(2017)\citenamefont {{Yuan}},
  \citenamefont {{Lin}}, \citenamefont {{Fang}},\ and\ \citenamefont
  {{Bi}}}]{Yuan2017}%
  \BibitemOpen
  \bibfield  {author} {\bibinfo {author} {\bibfnamefont {Q.}~\bibnamefont
  {{Yuan}}}, \bibinfo {author} {\bibfnamefont {S.-J.}\ \bibnamefont {{Lin}}},
  \bibinfo {author} {\bibfnamefont {K.}~\bibnamefont {{Fang}}}, \ and\ \bibinfo
  {author} {\bibfnamefont {X.-J.}\ \bibnamefont {{Bi}}},\ }\href@noop {}
  {\bibfield  {journal} {\bibinfo  {journal} {ArXiv e-prints}\ } (\bibinfo
  {year} {2017})},\ \Eprint {http://arxiv.org/abs/1701.06149} {arXiv:1701.06149
  [astro-ph.HE]} \BibitemShut {NoStop}%
\bibitem [{\citenamefont {{Jin}}\ \emph {et~al.}(2015)\citenamefont {{Jin}},
  \citenamefont {{Wu}},\ and\ \citenamefont {{Zhou}}}]{Jin2015}%
  \BibitemOpen
  \bibfield  {author} {\bibinfo {author} {\bibfnamefont {H.-B.}\ \bibnamefont
  {{Jin}}}, \bibinfo {author} {\bibfnamefont {Y.-L.}\ \bibnamefont {{Wu}}}, \
  and\ \bibinfo {author} {\bibfnamefont {Y.-F.}\ \bibnamefont {{Zhou}}},\
  }\href {\doibase 10.1088/1475-7516/2015/09/049} {\bibfield  {journal}
  {\bibinfo  {journal} {\jcap}\ }\textbf {\bibinfo {volume} {9}},\ \bibinfo
  {eid} {049} (\bibinfo {year} {2015})},\ \Eprint
  {http://arxiv.org/abs/1410.0171} {arXiv:1410.0171 [hep-ph]} \BibitemShut
  {NoStop}%
\bibitem [{\citenamefont {{AMS Collaboration}}\ \emph
  {et~al.}(2013)\citenamefont {{AMS Collaboration}}, \citenamefont {{Aguilar}},
  \citenamefont {{Alberti}}, \citenamefont {{Alpat}}, \citenamefont {{Alvino}},
  \citenamefont {{Ambrosi}}, \citenamefont {{Andeen}}, \citenamefont
  {{Anderhub}}, \citenamefont {{Arruda}}, \citenamefont {{Azzarello}},
  \citenamefont {{Bachlechner}},\ and\ \citenamefont {et~al.}}]{AMS2013}%
  \BibitemOpen
  \bibfield  {author} {\bibinfo {author} {\bibnamefont {{AMS Collaboration}}},
  \bibinfo {author} {\bibfnamefont {M.}~\bibnamefont {{Aguilar}}}, \bibinfo
  {author} {\bibfnamefont {G.}~\bibnamefont {{Alberti}}}, \bibinfo {author}
  {\bibfnamefont {B.}~\bibnamefont {{Alpat}}}, \bibinfo {author} {\bibfnamefont
  {A.}~\bibnamefont {{Alvino}}}, \bibinfo {author} {\bibfnamefont
  {G.}~\bibnamefont {{Ambrosi}}}, \bibinfo {author} {\bibfnamefont
  {K.}~\bibnamefont {{Andeen}}}, \bibinfo {author} {\bibfnamefont
  {H.}~\bibnamefont {{Anderhub}}}, \bibinfo {author} {\bibfnamefont
  {L.}~\bibnamefont {{Arruda}}}, \bibinfo {author} {\bibfnamefont
  {P.}~\bibnamefont {{Azzarello}}}, \bibinfo {author} {\bibfnamefont
  {A.}~\bibnamefont {{Bachlechner}}}, \ and\ \bibinfo {author} {\bibnamefont
  {et~al.}},\ }\href {\doibase 10.1103/PhysRevLett.110.141102} {\bibfield
  {journal} {\bibinfo  {journal} {Physical Review Letters}\ }\textbf {\bibinfo
  {volume} {110}},\ \bibinfo {eid} {141102} (\bibinfo {year}
  {2013})}\BibitemShut {NoStop}%
\bibitem [{\citenamefont {{Jin}}\ \emph {et~al.}(2013)\citenamefont {{Jin}},
  \citenamefont {{Wu}},\ and\ \citenamefont {{Zhou}}}]{Jin2013}%
  \BibitemOpen
  \bibfield  {author} {\bibinfo {author} {\bibfnamefont {H.-B.}\ \bibnamefont
  {{Jin}}}, \bibinfo {author} {\bibfnamefont {Y.-L.}\ \bibnamefont {{Wu}}}, \
  and\ \bibinfo {author} {\bibfnamefont {Y.-F.}\ \bibnamefont {{Zhou}}},\
  }\href {\doibase 10.1088/1475-7516/2013/11/026} {\bibfield  {journal}
  {\bibinfo  {journal} {\jcap}\ }\textbf {\bibinfo {volume} {11}},\ \bibinfo
  {eid} {026} (\bibinfo {year} {2013})},\ \Eprint
  {http://arxiv.org/abs/1304.1997} {arXiv:1304.1997 [hep-ph]} \BibitemShut
  {NoStop}%
\bibitem [{\citenamefont {{Feng}}\ \emph {et~al.}(2014)\citenamefont {{Feng}},
  \citenamefont {{Yang}}, \citenamefont {{He}}, \citenamefont {{Dong}},
  \citenamefont {{Fan}},\ and\ \citenamefont {{Chang}}}]{Feng2014}%
  \BibitemOpen
  \bibfield  {author} {\bibinfo {author} {\bibfnamefont {L.}~\bibnamefont
  {{Feng}}}, \bibinfo {author} {\bibfnamefont {R.-Z.}\ \bibnamefont {{Yang}}},
  \bibinfo {author} {\bibfnamefont {H.-N.}\ \bibnamefont {{He}}}, \bibinfo
  {author} {\bibfnamefont {T.-K.}\ \bibnamefont {{Dong}}}, \bibinfo {author}
  {\bibfnamefont {Y.-Z.}\ \bibnamefont {{Fan}}}, \ and\ \bibinfo {author}
  {\bibfnamefont {J.}~\bibnamefont {{Chang}}},\ }\href {\doibase
  10.1016/j.physletb.2013.12.012} {\bibfield  {journal} {\bibinfo  {journal}
  {Physics Letters B}\ }\textbf {\bibinfo {volume} {728}},\ \bibinfo {pages}
  {250} (\bibinfo {year} {2014})},\ \Eprint {http://arxiv.org/abs/1303.0530}
  {arXiv:1303.0530 [astro-ph.HE]} \BibitemShut {NoStop}%
\bibitem [{\citenamefont {{Di Mauro}}\ \emph {et~al.}(2014)\citenamefont {{Di
  Mauro}}, \citenamefont {{Donato}}, \citenamefont {{Fornengo}}, \citenamefont
  {{Lineros}},\ and\ \citenamefont {{Vittino}}}]{Mauro2014}%
  \BibitemOpen
  \bibfield  {author} {\bibinfo {author} {\bibfnamefont {M.}~\bibnamefont {{Di
  Mauro}}}, \bibinfo {author} {\bibfnamefont {F.}~\bibnamefont {{Donato}}},
  \bibinfo {author} {\bibfnamefont {N.}~\bibnamefont {{Fornengo}}}, \bibinfo
  {author} {\bibfnamefont {R.}~\bibnamefont {{Lineros}}}, \ and\ \bibinfo
  {author} {\bibfnamefont {A.}~\bibnamefont {{Vittino}}},\ }\href {\doibase
  10.1088/1475-7516/2014/04/006} {\bibfield  {journal} {\bibinfo  {journal}
  {\jcap}\ }\textbf {\bibinfo {volume} {4}},\ \bibinfo {eid} {006} (\bibinfo
  {year} {2014})},\ \Eprint {http://arxiv.org/abs/1402.0321} {arXiv:1402.0321
  [astro-ph.HE]} \BibitemShut {NoStop}%
\bibitem [{\citenamefont {{Yuan}}\ and\ \citenamefont {{Bi}}(2015)}]{Yuan2015}%
  \BibitemOpen
  \bibfield  {author} {\bibinfo {author} {\bibfnamefont {Q.}~\bibnamefont
  {{Yuan}}}\ and\ \bibinfo {author} {\bibfnamefont {X.-J.}\ \bibnamefont
  {{Bi}}},\ }\href {\doibase 10.1088/1475-7516/2015/03/033} {\bibfield
  {journal} {\bibinfo  {journal} {\jcap}\ }\textbf {\bibinfo {volume} {3}},\
  \bibinfo {eid} {033} (\bibinfo {year} {2015})},\ \Eprint
  {http://arxiv.org/abs/1408.2424} {arXiv:1408.2424 [astro-ph.HE]} \BibitemShut
  {NoStop}%
\bibitem [{\citenamefont {{AMS Collaboration}}\ \emph
  {et~al.}(2015{\natexlab{a}})\citenamefont {{AMS Collaboration}},
  \citenamefont {{Aguilar}}, \citenamefont {{Aisa}}, \citenamefont {{Alpat}},
  \citenamefont {{Alvino}}, \citenamefont {{Ambrosi}}, \citenamefont
  {{Andeen}}, \citenamefont {{Arruda}}, \citenamefont {{Attig}}, \citenamefont
  {{Azzarello}}, \citenamefont {{Bachlechner}},\ and\ \citenamefont
  {et~al.}}]{AMS02_proton}%
  \BibitemOpen
  \bibfield  {author} {\bibinfo {author} {\bibnamefont {{AMS Collaboration}}},
  \bibinfo {author} {\bibfnamefont {M.}~\bibnamefont {{Aguilar}}}, \bibinfo
  {author} {\bibfnamefont {D.}~\bibnamefont {{Aisa}}}, \bibinfo {author}
  {\bibfnamefont {B.}~\bibnamefont {{Alpat}}}, \bibinfo {author} {\bibfnamefont
  {A.}~\bibnamefont {{Alvino}}}, \bibinfo {author} {\bibfnamefont
  {G.}~\bibnamefont {{Ambrosi}}}, \bibinfo {author} {\bibfnamefont
  {K.}~\bibnamefont {{Andeen}}}, \bibinfo {author} {\bibfnamefont
  {L.}~\bibnamefont {{Arruda}}}, \bibinfo {author} {\bibfnamefont
  {N.}~\bibnamefont {{Attig}}}, \bibinfo {author} {\bibfnamefont
  {P.}~\bibnamefont {{Azzarello}}}, \bibinfo {author} {\bibfnamefont
  {A.}~\bibnamefont {{Bachlechner}}}, \ and\ \bibinfo {author} {\bibnamefont
  {et~al.}},\ }\href {\doibase 10.1103/PhysRevLett.114.171103} {\bibfield
  {journal} {\bibinfo  {journal} {Physical Review Letters}\ }\textbf {\bibinfo
  {volume} {114}},\ \bibinfo {eid} {171103} (\bibinfo {year}
  {2015}{\natexlab{a}})}\BibitemShut {NoStop}%
\bibitem [{\citenamefont {{AMS Collaboration}}\ \emph
  {et~al.}(2015{\natexlab{b}})\citenamefont {{AMS Collaboration}},
  \citenamefont {{Aguilar}}, \citenamefont {{Aisa}}, \citenamefont {{Alpat}},
  \citenamefont {{Alvino}}, \citenamefont {{Ambrosi}}, \citenamefont
  {{Andeen}}, \citenamefont {{Arruda}}, \citenamefont {{Attig}}, \citenamefont
  {{Azzarello}}, \citenamefont {{Bachlechner}},\ and\ \citenamefont
  {et~al.}}]{AMS02_helium}%
  \BibitemOpen
  \bibfield  {author} {\bibinfo {author} {\bibnamefont {{AMS Collaboration}}},
  \bibinfo {author} {\bibfnamefont {M.}~\bibnamefont {{Aguilar}}}, \bibinfo
  {author} {\bibfnamefont {D.}~\bibnamefont {{Aisa}}}, \bibinfo {author}
  {\bibfnamefont {B.}~\bibnamefont {{Alpat}}}, \bibinfo {author} {\bibfnamefont
  {A.}~\bibnamefont {{Alvino}}}, \bibinfo {author} {\bibfnamefont
  {G.}~\bibnamefont {{Ambrosi}}}, \bibinfo {author} {\bibfnamefont
  {K.}~\bibnamefont {{Andeen}}}, \bibinfo {author} {\bibfnamefont
  {L.}~\bibnamefont {{Arruda}}}, \bibinfo {author} {\bibfnamefont
  {N.}~\bibnamefont {{Attig}}}, \bibinfo {author} {\bibfnamefont
  {P.}~\bibnamefont {{Azzarello}}}, \bibinfo {author} {\bibfnamefont
  {A.}~\bibnamefont {{Bachlechner}}}, \ and\ \bibinfo {author} {\bibnamefont
  {et~al.}},\ }\href {\doibase 10.1103/PhysRevLett.115.211101} {\bibfield
  {journal} {\bibinfo  {journal} {Physical Review Letters}\ }\textbf {\bibinfo
  {volume} {115}},\ \bibinfo {eid} {211101} (\bibinfo {year}
  {2015}{\natexlab{b}})}\BibitemShut {NoStop}%
\bibitem [{\citenamefont {{AMS Collaboration}}\ \emph
  {et~al.}(2016{\natexlab{a}})\citenamefont {{AMS Collaboration}},
  \citenamefont {Aguilar}, \citenamefont {Ali~Cavasonza}, \citenamefont
  {Ambrosi}, \citenamefont {Arruda}, \citenamefont {Attig}, \citenamefont
  {Aupetit}, \citenamefont {Azzarello}, \citenamefont {Bachlechner},
  \citenamefont {Barao}, \citenamefont {Barrau},\ and\ \citenamefont
  {et~al.}}]{AMS02_b_c}%
  \BibitemOpen
  \bibfield  {author} {\bibinfo {author} {\bibnamefont {{AMS Collaboration}}},
  \bibinfo {author} {\bibfnamefont {M.}~\bibnamefont {Aguilar}}, \bibinfo
  {author} {\bibfnamefont {L.}~\bibnamefont {Ali~Cavasonza}}, \bibinfo {author}
  {\bibfnamefont {G.}~\bibnamefont {Ambrosi}}, \bibinfo {author} {\bibfnamefont
  {L.}~\bibnamefont {Arruda}}, \bibinfo {author} {\bibfnamefont
  {N.}~\bibnamefont {Attig}}, \bibinfo {author} {\bibfnamefont
  {S.}~\bibnamefont {Aupetit}}, \bibinfo {author} {\bibfnamefont
  {P.}~\bibnamefont {Azzarello}}, \bibinfo {author} {\bibfnamefont
  {A.}~\bibnamefont {Bachlechner}}, \bibinfo {author} {\bibfnamefont
  {F.}~\bibnamefont {Barao}}, \bibinfo {author} {\bibfnamefont
  {A.}~\bibnamefont {Barrau}}, \ and\ \bibinfo {author} {\bibnamefont
  {et~al.}},\ }\href {\doibase 10.1103/PhysRevLett.117.231102} {\bibfield
  {journal} {\bibinfo  {journal} {Phys. Rev. Lett.}\ }\textbf {\bibinfo
  {volume} {117}},\ \bibinfo {pages} {231102} (\bibinfo {year}
  {2016}{\natexlab{a}})}\BibitemShut {NoStop}%
\bibitem [{\citenamefont {{AMS Collaboration}}\ \emph
  {et~al.}(2016{\natexlab{b}})\citenamefont {{AMS Collaboration}},
  \citenamefont {{Aguilar}}, \citenamefont {{Ali Cavasonza}}, \citenamefont
  {{Alpat}}, \citenamefont {{Ambrosi}}, \citenamefont {{Arruda}}, \citenamefont
  {{Attig}}, \citenamefont {{Aupetit}}, \citenamefont {{Azzarello}},
  \citenamefont {{Bachlechner}}, \citenamefont {{Barao}},\ and\ \citenamefont
  {et~al.}}]{AMS02_pbar_proton}%
  \BibitemOpen
  \bibfield  {author} {\bibinfo {author} {\bibnamefont {{AMS Collaboration}}},
  \bibinfo {author} {\bibfnamefont {M.}~\bibnamefont {{Aguilar}}}, \bibinfo
  {author} {\bibfnamefont {L.}~\bibnamefont {{Ali Cavasonza}}}, \bibinfo
  {author} {\bibfnamefont {B.}~\bibnamefont {{Alpat}}}, \bibinfo {author}
  {\bibfnamefont {G.}~\bibnamefont {{Ambrosi}}}, \bibinfo {author}
  {\bibfnamefont {L.}~\bibnamefont {{Arruda}}}, \bibinfo {author}
  {\bibfnamefont {N.}~\bibnamefont {{Attig}}}, \bibinfo {author} {\bibfnamefont
  {S.}~\bibnamefont {{Aupetit}}}, \bibinfo {author} {\bibfnamefont
  {P.}~\bibnamefont {{Azzarello}}}, \bibinfo {author} {\bibfnamefont
  {A.}~\bibnamefont {{Bachlechner}}}, \bibinfo {author} {\bibfnamefont
  {F.}~\bibnamefont {{Barao}}}, \ and\ \bibinfo {author} {\bibnamefont
  {et~al.}},\ }\href {\doibase 10.1103/PhysRevLett.117.091103} {\bibfield
  {journal} {\bibinfo  {journal} {Physical Review Letters}\ }\textbf {\bibinfo
  {volume} {117}},\ \bibinfo {eid} {091103} (\bibinfo {year}
  {2016}{\natexlab{b}})}\BibitemShut {NoStop}%
\bibitem [{\citenamefont {{Lewis}}\ and\ \citenamefont
  {{Bridle}}(2002)}]{Lewis2002}%
  \BibitemOpen
  \bibfield  {author} {\bibinfo {author} {\bibfnamefont {A.}~\bibnamefont
  {{Lewis}}}\ and\ \bibinfo {author} {\bibfnamefont {S.}~\bibnamefont
  {{Bridle}}},\ }\href {\doibase 10.1103/PhysRevD.66.103511} {\bibfield
  {journal} {\bibinfo  {journal} {\prd}\ }\textbf {\bibinfo {volume} {66}},\
  \bibinfo {eid} {103511} (\bibinfo {year} {2002})},\ \Eprint
  {http://arxiv.org/abs/astro-ph/0205436} {astro-ph/0205436} \BibitemShut
  {NoStop}%
\bibitem [{\citenamefont {{Liu}}\ \emph {et~al.}(2010)\citenamefont {{Liu}},
  \citenamefont {{Yuan}}, \citenamefont {{Bi}}, \citenamefont {{Li}},\ and\
  \citenamefont {{Zhang}}}]{Liu2010}%
  \BibitemOpen
  \bibfield  {author} {\bibinfo {author} {\bibfnamefont {J.}~\bibnamefont
  {{Liu}}}, \bibinfo {author} {\bibfnamefont {Q.}~\bibnamefont {{Yuan}}},
  \bibinfo {author} {\bibfnamefont {X.}~\bibnamefont {{Bi}}}, \bibinfo {author}
  {\bibfnamefont {H.}~\bibnamefont {{Li}}}, \ and\ \bibinfo {author}
  {\bibfnamefont {X.}~\bibnamefont {{Zhang}}},\ }\href {\doibase
  10.1103/PhysRevD.81.023516} {\bibfield  {journal} {\bibinfo  {journal}
  {\prd}\ }\textbf {\bibinfo {volume} {81}},\ \bibinfo {eid} {023516} (\bibinfo
  {year} {2010})},\ \Eprint {http://arxiv.org/abs/0906.3858} {arXiv:0906.3858
  [astro-ph.CO]} \BibitemShut {NoStop}%
\bibitem [{\citenamefont {{PAMELA Collaboration}}\ \emph
  {et~al.}(2011)\citenamefont {{PAMELA Collaboration}}, \citenamefont
  {{Adriani}}, \citenamefont {{Barbarino}}, \citenamefont {{Bazilevskaya}},
  \citenamefont {{Bellotti}}, \citenamefont {{Boezio}}, \citenamefont
  {{Bogomolov}}, \citenamefont {{Bonechi}}, \citenamefont {{Bongi}},
  \citenamefont {{Bonvicini}}, \citenamefont {{Borisov}},\ and\ \citenamefont
  {el~al.}}]{Adriani2011}%
  \BibitemOpen
  \bibfield  {author} {\bibinfo {author} {\bibnamefont {{PAMELA
  Collaboration}}}, \bibinfo {author} {\bibfnamefont {O.}~\bibnamefont
  {{Adriani}}}, \bibinfo {author} {\bibfnamefont {G.~C.}\ \bibnamefont
  {{Barbarino}}}, \bibinfo {author} {\bibfnamefont {G.~A.}\ \bibnamefont
  {{Bazilevskaya}}}, \bibinfo {author} {\bibfnamefont {R.}~\bibnamefont
  {{Bellotti}}}, \bibinfo {author} {\bibfnamefont {M.}~\bibnamefont
  {{Boezio}}}, \bibinfo {author} {\bibfnamefont {E.~A.}\ \bibnamefont
  {{Bogomolov}}}, \bibinfo {author} {\bibfnamefont {L.}~\bibnamefont
  {{Bonechi}}}, \bibinfo {author} {\bibfnamefont {M.}~\bibnamefont {{Bongi}}},
  \bibinfo {author} {\bibfnamefont {V.}~\bibnamefont {{Bonvicini}}}, \bibinfo
  {author} {\bibfnamefont {S.}~\bibnamefont {{Borisov}}}, \ and\ \bibinfo
  {author} {\bibnamefont {el~al.}},\ }\href {\doibase 10.1126/science.1199172}
  {\bibfield  {journal} {\bibinfo  {journal} {Science}\ }\textbf {\bibinfo
  {volume} {332}},\ \bibinfo {pages} {69} (\bibinfo {year} {2011})},\ \Eprint
  {http://arxiv.org/abs/1103.4055} {arXiv:1103.4055 [astro-ph.HE]} \BibitemShut
  {NoStop}%
\bibitem [{\citenamefont {{Seo}}\ and\ \citenamefont
  {{Ptuskin}}(1994)}]{Seo1994}%
  \BibitemOpen
  \bibfield  {author} {\bibinfo {author} {\bibfnamefont {E.~S.}\ \bibnamefont
  {{Seo}}}\ and\ \bibinfo {author} {\bibfnamefont {V.~S.}\ \bibnamefont
  {{Ptuskin}}},\ }\href {\doibase 10.1086/174520} {\bibfield  {journal}
  {\bibinfo  {journal} {\apj}\ }\textbf {\bibinfo {volume} {431}},\ \bibinfo
  {pages} {705} (\bibinfo {year} {1994})}\BibitemShut {NoStop}%
\bibitem [{\citenamefont {{Case}}\ and\ \citenamefont
  {{Bhattacharya}}(1996)}]{Case1996}%
  \BibitemOpen
  \bibfield  {author} {\bibinfo {author} {\bibfnamefont {G.}~\bibnamefont
  {{Case}}}\ and\ \bibinfo {author} {\bibfnamefont {D.}~\bibnamefont
  {{Bhattacharya}}},\ }\href@noop {} {\bibfield  {journal} {\bibinfo  {journal}
  {\aaps}\ }\textbf {\bibinfo {volume} {120}},\ \bibinfo {pages} {437}
  (\bibinfo {year} {1996})}\BibitemShut {NoStop}%
\bibitem [{\citenamefont {{Fermi-LAT Collaboration}}\ \emph
  {et~al.}(2009)\citenamefont {{Fermi-LAT Collaboration}}, \citenamefont
  {{Tibaldo}},\ and\ \citenamefont {{Grenier}}}]{Tibaldo2009}%
  \BibitemOpen
  \bibfield  {author} {\bibinfo {author} {\bibnamefont {{Fermi-LAT
  Collaboration}}}, \bibinfo {author} {\bibfnamefont {L.}~\bibnamefont
  {{Tibaldo}}}, \ and\ \bibinfo {author} {\bibfnamefont {I.~A.}\ \bibnamefont
  {{Grenier}}},\ }\href@noop {} {\bibfield  {journal} {\bibinfo  {journal}
  {ArXiv e-prints}\ } (\bibinfo {year} {2009})},\ \Eprint
  {http://arxiv.org/abs/0907.0312} {arXiv:0907.0312 [astro-ph.HE]} \BibitemShut
  {NoStop}%
\bibitem [{\citenamefont {{Tan}}\ and\ \citenamefont {{Ng}}(1983)}]{Tan1983}%
  \BibitemOpen
  \bibfield  {author} {\bibinfo {author} {\bibfnamefont {L.~C.}\ \bibnamefont
  {{Tan}}}\ and\ \bibinfo {author} {\bibfnamefont {L.~K.}\ \bibnamefont
  {{Ng}}},\ }\href {\doibase 10.1088/0305-4616/9/10/015} {\bibfield  {journal}
  {\bibinfo  {journal} {Journal of Physics G Nuclear Physics}\ }\textbf
  {\bibinfo {volume} {9}},\ \bibinfo {pages} {1289} (\bibinfo {year}
  {1983})}\BibitemShut {NoStop}%
\bibitem [{\citenamefont {{Duperray}}\ \emph {et~al.}(2003)\citenamefont
  {{Duperray}}, \citenamefont {{Huang}}, \citenamefont {{Protasov}},\ and\
  \citenamefont {{Bu{\'e}nerd}}}]{Duperray2003}%
  \BibitemOpen
  \bibfield  {author} {\bibinfo {author} {\bibfnamefont {R.~P.}\ \bibnamefont
  {{Duperray}}}, \bibinfo {author} {\bibfnamefont {C.-Y.}\ \bibnamefont
  {{Huang}}}, \bibinfo {author} {\bibfnamefont {K.~V.}\ \bibnamefont
  {{Protasov}}}, \ and\ \bibinfo {author} {\bibfnamefont {M.}~\bibnamefont
  {{Bu{\'e}nerd}}},\ }\href {\doibase 10.1103/PhysRevD.68.094017} {\bibfield
  {journal} {\bibinfo  {journal} {\prd}\ }\textbf {\bibinfo {volume} {68}},\
  \bibinfo {eid} {094017} (\bibinfo {year} {2003})},\ \Eprint
  {http://arxiv.org/abs/astro-ph/0305274} {astro-ph/0305274} \BibitemShut
  {NoStop}%
\bibitem [{\citenamefont {{Kappl}}\ and\ \citenamefont
  {{Winkler}}(2014)}]{Kappl2014}%
  \BibitemOpen
  \bibfield  {author} {\bibinfo {author} {\bibfnamefont {R.}~\bibnamefont
  {{Kappl}}}\ and\ \bibinfo {author} {\bibfnamefont {M.~W.}\ \bibnamefont
  {{Winkler}}},\ }\href {\doibase 10.1088/1475-7516/2014/09/051} {\bibfield
  {journal} {\bibinfo  {journal} {\jcap}\ }\textbf {\bibinfo {volume} {9}},\
  \bibinfo {eid} {051} (\bibinfo {year} {2014})},\ \Eprint
  {http://arxiv.org/abs/1408.0299} {arXiv:1408.0299 [hep-ph]} \BibitemShut
  {NoStop}%
\bibitem [{\citenamefont {{di Mauro}}\ \emph {et~al.}(2014)\citenamefont {{di
  Mauro}}, \citenamefont {{Donato}}, \citenamefont {{Goudelis}},\ and\
  \citenamefont {{Serpico}}}]{diMauro2014}%
  \BibitemOpen
  \bibfield  {author} {\bibinfo {author} {\bibfnamefont {M.}~\bibnamefont {{di
  Mauro}}}, \bibinfo {author} {\bibfnamefont {F.}~\bibnamefont {{Donato}}},
  \bibinfo {author} {\bibfnamefont {A.}~\bibnamefont {{Goudelis}}}, \ and\
  \bibinfo {author} {\bibfnamefont {P.~D.}\ \bibnamefont {{Serpico}}},\ }\href
  {\doibase 10.1103/PhysRevD.90.085017} {\bibfield  {journal} {\bibinfo
  {journal} {\prd}\ }\textbf {\bibinfo {volume} {90}},\ \bibinfo {eid} {085017}
  (\bibinfo {year} {2014})},\ \Eprint {http://arxiv.org/abs/1408.0288}
  {arXiv:1408.0288 [hep-ph]} \BibitemShut {NoStop}%
\bibitem [{\citenamefont {{Gleeson}}\ and\ \citenamefont
  {{Axford}}(1968)}]{Gleeson1968}%
  \BibitemOpen
  \bibfield  {author} {\bibinfo {author} {\bibfnamefont {L.~J.}\ \bibnamefont
  {{Gleeson}}}\ and\ \bibinfo {author} {\bibfnamefont {W.~I.}\ \bibnamefont
  {{Axford}}},\ }\href {\doibase 10.1086/149822} {\bibfield  {journal}
  {\bibinfo  {journal} {\apj}\ }\textbf {\bibinfo {volume} {154}},\ \bibinfo
  {pages} {1011} (\bibinfo {year} {1968})}\BibitemShut {NoStop}%
\bibitem [{\citenamefont {{Moskalenko}}\ \emph {et~al.}(2002)\citenamefont
  {{Moskalenko}}, \citenamefont {{Strong}}, \citenamefont {{Ormes}},\ and\
  \citenamefont {{Potgieter}}}]{Moskalenko2002}%
  \BibitemOpen
  \bibfield  {author} {\bibinfo {author} {\bibfnamefont {I.~V.}\ \bibnamefont
  {{Moskalenko}}}, \bibinfo {author} {\bibfnamefont {A.~W.}\ \bibnamefont
  {{Strong}}}, \bibinfo {author} {\bibfnamefont {J.~F.}\ \bibnamefont
  {{Ormes}}}, \ and\ \bibinfo {author} {\bibfnamefont {M.~S.}\ \bibnamefont
  {{Potgieter}}},\ }\href {\doibase 10.1086/324402} {\bibfield  {journal}
  {\bibinfo  {journal} {\apj}\ }\textbf {\bibinfo {volume} {565}},\ \bibinfo
  {pages} {280} (\bibinfo {year} {2002})},\ \Eprint
  {http://arxiv.org/abs/astro-ph/0106567} {astro-ph/0106567} \BibitemShut
  {NoStop}%
\bibitem [{\citenamefont {{Strong}}\ and\ \citenamefont
  {{Moskalenko}}(2001)}]{Strong2001}%
  \BibitemOpen
  \bibfield  {author} {\bibinfo {author} {\bibfnamefont {A.~W.}\ \bibnamefont
  {{Strong}}}\ and\ \bibinfo {author} {\bibfnamefont {I.~V.}\ \bibnamefont
  {{Moskalenko}}},\ }\href {\doibase 10.1016/S0273-1177(01)00112-0} {\bibfield
  {journal} {\bibinfo  {journal} {Advances in Space Research}\ }\textbf
  {\bibinfo {volume} {27}},\ \bibinfo {pages} {717} (\bibinfo {year} {2001})},\
  \Eprint {http://arxiv.org/abs/astro-ph/0101068} {astro-ph/0101068}
  \BibitemShut {NoStop}%
\bibitem [{\citenamefont {{Moskalenko}}\ \emph {et~al.}(2003)\citenamefont
  {{Moskalenko}}, \citenamefont {{Strong}}, \citenamefont {{Mashnik}},\ and\
  \citenamefont {{Ormes}}}]{Moskalenko2003}%
  \BibitemOpen
  \bibfield  {author} {\bibinfo {author} {\bibfnamefont {I.~V.}\ \bibnamefont
  {{Moskalenko}}}, \bibinfo {author} {\bibfnamefont {A.~W.}\ \bibnamefont
  {{Strong}}}, \bibinfo {author} {\bibfnamefont {S.~G.}\ \bibnamefont
  {{Mashnik}}}, \ and\ \bibinfo {author} {\bibfnamefont {J.~F.}\ \bibnamefont
  {{Ormes}}},\ }\href {\doibase 10.1086/367697} {\bibfield  {journal} {\bibinfo
   {journal} {\apj}\ }\textbf {\bibinfo {volume} {586}},\ \bibinfo {pages}
  {1050} (\bibinfo {year} {2003})},\ \Eprint
  {http://arxiv.org/abs/astro-ph/0210480} {astro-ph/0210480} \BibitemShut
  {NoStop}%
\bibitem [{\citenamefont {{Ptuskin}}\ \emph {et~al.}(2006)\citenamefont
  {{Ptuskin}}, \citenamefont {{Moskalenko}}, \citenamefont {{Jones}},
  \citenamefont {{Strong}},\ and\ \citenamefont
  {{Zirakashvili}}}]{Ptuskin2006}%
  \BibitemOpen
  \bibfield  {author} {\bibinfo {author} {\bibfnamefont {V.~S.}\ \bibnamefont
  {{Ptuskin}}}, \bibinfo {author} {\bibfnamefont {I.~V.}\ \bibnamefont
  {{Moskalenko}}}, \bibinfo {author} {\bibfnamefont {F.~C.}\ \bibnamefont
  {{Jones}}}, \bibinfo {author} {\bibfnamefont {A.~W.}\ \bibnamefont
  {{Strong}}}, \ and\ \bibinfo {author} {\bibfnamefont {V.~N.}\ \bibnamefont
  {{Zirakashvili}}},\ }\href {\doibase 10.1086/501117} {\bibfield  {journal}
  {\bibinfo  {journal} {\apj}\ }\textbf {\bibinfo {volume} {642}},\ \bibinfo
  {pages} {902} (\bibinfo {year} {2006})},\ \Eprint
  {http://arxiv.org/abs/astro-ph/0510335} {astro-ph/0510335} \BibitemShut
  {NoStop}%
\bibitem [{\citenamefont {{Donato}}\ \emph {et~al.}(2001)\citenamefont
  {{Donato}}, \citenamefont {{Maurin}}, \citenamefont {{Salati}}, \citenamefont
  {{Barrau}}, \citenamefont {{Boudoul}},\ and\ \citenamefont
  {{Taillet}}}]{Donato2001}%
  \BibitemOpen
  \bibfield  {author} {\bibinfo {author} {\bibfnamefont {F.}~\bibnamefont
  {{Donato}}}, \bibinfo {author} {\bibfnamefont {D.}~\bibnamefont {{Maurin}}},
  \bibinfo {author} {\bibfnamefont {P.}~\bibnamefont {{Salati}}}, \bibinfo
  {author} {\bibfnamefont {A.}~\bibnamefont {{Barrau}}}, \bibinfo {author}
  {\bibfnamefont {G.}~\bibnamefont {{Boudoul}}}, \ and\ \bibinfo {author}
  {\bibfnamefont {R.}~\bibnamefont {{Taillet}}},\ }\href {\doibase
  10.1086/323684} {\bibfield  {journal} {\bibinfo  {journal} {\apj}\ }\textbf
  {\bibinfo {volume} {563}},\ \bibinfo {pages} {172} (\bibinfo {year}
  {2001})},\ \Eprint {http://arxiv.org/abs/astro-ph/0103150} {astro-ph/0103150}
  \BibitemShut {NoStop}%
\bibitem [{\citenamefont {{Maurin}}\ \emph
  {et~al.}(2002{\natexlab{b}})\citenamefont {{Maurin}}, \citenamefont
  {{Taillet}}, \citenamefont {{Donato}}, \citenamefont {{Salati}},
  \citenamefont {{Barrau}},\ and\ \citenamefont {{Boudoul}}}]{Maurin2002b}%
  \BibitemOpen
  \bibfield  {author} {\bibinfo {author} {\bibfnamefont {D.}~\bibnamefont
  {{Maurin}}}, \bibinfo {author} {\bibfnamefont {R.}~\bibnamefont {{Taillet}}},
  \bibinfo {author} {\bibfnamefont {F.}~\bibnamefont {{Donato}}}, \bibinfo
  {author} {\bibfnamefont {P.}~\bibnamefont {{Salati}}}, \bibinfo {author}
  {\bibfnamefont {A.}~\bibnamefont {{Barrau}}}, \ and\ \bibinfo {author}
  {\bibfnamefont {G.}~\bibnamefont {{Boudoul}}},\ }\href@noop {} {\bibfield
  {journal} {\bibinfo  {journal} {ArXiv Astrophysics e-prints}\ } (\bibinfo
  {year} {2002}{\natexlab{b}})},\ \Eprint
  {http://arxiv.org/abs/astro-ph/0212111} {astro-ph/0212111} \BibitemShut
  {NoStop}%
\bibitem [{\citenamefont {{Donato}}\ \emph {et~al.}(2004)\citenamefont
  {{Donato}}, \citenamefont {{Fornengo}}, \citenamefont {{Maurin}},
  \citenamefont {{Salati}},\ and\ \citenamefont {{Taillet}}}]{Donato2004}%
  \BibitemOpen
  \bibfield  {author} {\bibinfo {author} {\bibfnamefont {F.}~\bibnamefont
  {{Donato}}}, \bibinfo {author} {\bibfnamefont {N.}~\bibnamefont
  {{Fornengo}}}, \bibinfo {author} {\bibfnamefont {D.}~\bibnamefont
  {{Maurin}}}, \bibinfo {author} {\bibfnamefont {P.}~\bibnamefont {{Salati}}},
  \ and\ \bibinfo {author} {\bibfnamefont {R.}~\bibnamefont {{Taillet}}},\
  }\href {\doibase 10.1103/PhysRevD.69.063501} {\bibfield  {journal} {\bibinfo
  {journal} {\prd}\ }\textbf {\bibinfo {volume} {69}},\ \bibinfo {eid} {063501}
  (\bibinfo {year} {2004})},\ \Eprint {http://arxiv.org/abs/astro-ph/0306207}
  {astro-ph/0306207} \BibitemShut {NoStop}%
\bibitem [{\citenamefont {{Putze}}\ \emph {et~al.}(2010)\citenamefont
  {{Putze}}, \citenamefont {{Derome}},\ and\ \citenamefont
  {{Maurin}}}]{Putze2010}%
  \BibitemOpen
  \bibfield  {author} {\bibinfo {author} {\bibfnamefont {A.}~\bibnamefont
  {{Putze}}}, \bibinfo {author} {\bibfnamefont {L.}~\bibnamefont {{Derome}}}, \
  and\ \bibinfo {author} {\bibfnamefont {D.}~\bibnamefont {{Maurin}}},\ }\href
  {\doibase 10.1051/0004-6361/201014010} {\bibfield  {journal} {\bibinfo
  {journal} {\aap}\ }\textbf {\bibinfo {volume} {516}},\ \bibinfo {eid} {A66}
  (\bibinfo {year} {2010})},\ \Eprint {http://arxiv.org/abs/1001.0551}
  {arXiv:1001.0551 [astro-ph.HE]} \BibitemShut {NoStop}%
\bibitem [{\citenamefont {{Cirelli}}\ \emph {et~al.}(2011)\citenamefont
  {{Cirelli}}, \citenamefont {{Corcella}}, \citenamefont {{Hektor}},
  \citenamefont {{H{\"u}tsi}}, \citenamefont {{Kadastik}}, \citenamefont
  {{Panci}}, \citenamefont {{Raidal}}, \citenamefont {{Sala}},\ and\
  \citenamefont {{Strumia}}}]{Cirelli2011}%
  \BibitemOpen
  \bibfield  {author} {\bibinfo {author} {\bibfnamefont {M.}~\bibnamefont
  {{Cirelli}}}, \bibinfo {author} {\bibfnamefont {G.}~\bibnamefont
  {{Corcella}}}, \bibinfo {author} {\bibfnamefont {A.}~\bibnamefont
  {{Hektor}}}, \bibinfo {author} {\bibfnamefont {G.}~\bibnamefont
  {{H{\"u}tsi}}}, \bibinfo {author} {\bibfnamefont {M.}~\bibnamefont
  {{Kadastik}}}, \bibinfo {author} {\bibfnamefont {P.}~\bibnamefont {{Panci}}},
  \bibinfo {author} {\bibfnamefont {M.}~\bibnamefont {{Raidal}}}, \bibinfo
  {author} {\bibfnamefont {F.}~\bibnamefont {{Sala}}}, \ and\ \bibinfo {author}
  {\bibfnamefont {A.}~\bibnamefont {{Strumia}}},\ }\href {\doibase
  10.1088/1475-7516/2011/03/051} {\bibfield  {journal} {\bibinfo  {journal}
  {\jcap}\ }\textbf {\bibinfo {volume} {3}},\ \bibinfo {eid} {051} (\bibinfo
  {year} {2011})},\ \Eprint {http://arxiv.org/abs/1012.4515} {arXiv:1012.4515
  [hep-ph]} \BibitemShut {NoStop}%
\bibitem [{\citenamefont {{Wiedenbeck}}\ \emph {et~al.}(2001)\citenamefont
  {{Wiedenbeck}}, \citenamefont {{Yanasak}}, \citenamefont {{Cummings}},
  \citenamefont {{Davis}}, \citenamefont {{George}}, \citenamefont {{Leske}},
  \citenamefont {{Mewaldt}}, \citenamefont {{Stone}}, \citenamefont {{Hink}},
  \citenamefont {{Israel}}, \citenamefont {{Lijowski}}, \citenamefont
  {{Christian}},\ and\ \citenamefont {{von Rosenvinge}}}]{Wiedenbeck2001}%
  \BibitemOpen
  \bibfield  {author} {\bibinfo {author} {\bibfnamefont {M.~E.}\ \bibnamefont
  {{Wiedenbeck}}}, \bibinfo {author} {\bibfnamefont {N.~E.}\ \bibnamefont
  {{Yanasak}}}, \bibinfo {author} {\bibfnamefont {A.~C.}\ \bibnamefont
  {{Cummings}}}, \bibinfo {author} {\bibfnamefont {A.~J.}\ \bibnamefont
  {{Davis}}}, \bibinfo {author} {\bibfnamefont {J.~S.}\ \bibnamefont
  {{George}}}, \bibinfo {author} {\bibfnamefont {R.~A.}\ \bibnamefont
  {{Leske}}}, \bibinfo {author} {\bibfnamefont {R.~A.}\ \bibnamefont
  {{Mewaldt}}}, \bibinfo {author} {\bibfnamefont {E.~C.}\ \bibnamefont
  {{Stone}}}, \bibinfo {author} {\bibfnamefont {P.~L.}\ \bibnamefont {{Hink}}},
  \bibinfo {author} {\bibfnamefont {M.~H.}\ \bibnamefont {{Israel}}}, \bibinfo
  {author} {\bibfnamefont {M.}~\bibnamefont {{Lijowski}}}, \bibinfo {author}
  {\bibfnamefont {E.~R.}\ \bibnamefont {{Christian}}}, \ and\ \bibinfo {author}
  {\bibfnamefont {T.~T.}\ \bibnamefont {{von Rosenvinge}}},\ }\href@noop {}
  {\bibfield  {journal} {\bibinfo  {journal} {\ssr}\ }\textbf {\bibinfo
  {volume} {99}},\ \bibinfo {pages} {15} (\bibinfo {year} {2001})}\BibitemShut
  {NoStop}%
\bibitem [{\citenamefont {{Wiedenbeck}}\ \emph {et~al.}(2008)\citenamefont
  {{Wiedenbeck}}, \citenamefont {{Binns}}, \citenamefont {{Cummings}},
  \citenamefont {{de Nolfo}}, \citenamefont {{Israel}}, \citenamefont
  {{Leske}}, \citenamefont {{Mewaldt}}, \citenamefont {{Ogliore}},
  \citenamefont {{Stone}},\ and\ \citenamefont {{von
  Rosenvinge}}}]{Wiedenbeck2008}%
  \BibitemOpen
  \bibfield  {author} {\bibinfo {author} {\bibfnamefont {M.~E.}\ \bibnamefont
  {{Wiedenbeck}}}, \bibinfo {author} {\bibfnamefont {W.~R.}\ \bibnamefont
  {{Binns}}}, \bibinfo {author} {\bibfnamefont {A.~C.}\ \bibnamefont
  {{Cummings}}}, \bibinfo {author} {\bibfnamefont {G.~A.}\ \bibnamefont {{de
  Nolfo}}}, \bibinfo {author} {\bibfnamefont {M.~H.}\ \bibnamefont {{Israel}}},
  \bibinfo {author} {\bibfnamefont {R.~A.}\ \bibnamefont {{Leske}}}, \bibinfo
  {author} {\bibfnamefont {R.~A.}\ \bibnamefont {{Mewaldt}}}, \bibinfo {author}
  {\bibfnamefont {R.~C.}\ \bibnamefont {{Ogliore}}}, \bibinfo {author}
  {\bibfnamefont {E.~C.}\ \bibnamefont {{Stone}}}, \ and\ \bibinfo {author}
  {\bibfnamefont {T.~T.}\ \bibnamefont {{von Rosenvinge}}},\ }\href@noop {}
  {\bibfield  {journal} {\bibinfo  {journal} {International Cosmic Ray
  Conference}\ }\textbf {\bibinfo {volume} {2}},\ \bibinfo {pages} {149}
  (\bibinfo {year} {2008})}\BibitemShut {NoStop}%
\bibitem [{\citenamefont {{Goodman}}\ and\ \citenamefont
  {{Weare}}(2010)}]{Goodman2010}%
  \BibitemOpen
  \bibfield  {author} {\bibinfo {author} {\bibfnamefont {J.}~\bibnamefont
  {{Goodman}}}\ and\ \bibinfo {author} {\bibfnamefont {J.}~\bibnamefont
  {{Weare}}},\ }\href {\doibase 10.2140/camcos.2010.5.65} {\bibfield  {journal}
  {\bibinfo  {journal} {Communications in Applied Mathematics and Computational
  Science}\ }\textbf {\bibinfo {volume} {5}},\ \bibinfo {pages} {65} (\bibinfo
  {year} {2010})}\BibitemShut {NoStop}%
\bibitem [{\citenamefont {{Foreman-Mackey}}\ \emph {et~al.}(2013)\citenamefont
  {{Foreman-Mackey}}, \citenamefont {{Hogg}}, \citenamefont {{Lang}},\ and\
  \citenamefont {{Goodman}}}]{Mackey2013}%
  \BibitemOpen
  \bibfield  {author} {\bibinfo {author} {\bibfnamefont {D.}~\bibnamefont
  {{Foreman-Mackey}}}, \bibinfo {author} {\bibfnamefont {D.~W.}\ \bibnamefont
  {{Hogg}}}, \bibinfo {author} {\bibfnamefont {D.}~\bibnamefont {{Lang}}}, \
  and\ \bibinfo {author} {\bibfnamefont {J.}~\bibnamefont {{Goodman}}},\ }\href
  {\doibase 10.1086/670067} {\bibfield  {journal} {\bibinfo  {journal} {\pasp}\
  }\textbf {\bibinfo {volume} {125}},\ \bibinfo {pages} {306} (\bibinfo {year}
  {2013})},\ \Eprint {http://arxiv.org/abs/1202.3665} {arXiv:1202.3665
  [astro-ph.IM]} \BibitemShut {NoStop}%
\bibitem [{\citenamefont {{Korsmeier}}\ and\ \citenamefont
  {{Cuoco}}(2016)}]{Korsmeier2016}%
  \BibitemOpen
  \bibfield  {author} {\bibinfo {author} {\bibfnamefont {M.}~\bibnamefont
  {{Korsmeier}}}\ and\ \bibinfo {author} {\bibfnamefont {A.}~\bibnamefont
  {{Cuoco}}},\ }\href@noop {} {\bibfield  {journal} {\bibinfo  {journal} {ArXiv
  e-prints}\ } (\bibinfo {year} {2016})},\ \Eprint
  {http://arxiv.org/abs/1607.06093} {arXiv:1607.06093 [astro-ph.HE]}
  \BibitemShut {NoStop}%
\bibitem [{\citenamefont {Cui}\ \emph {et~al.}(2017)\citenamefont {Cui},
  \citenamefont {Yuan}, \citenamefont {Tsai},\ and\ \citenamefont
  {Fan}}]{Cui2017}%
  \BibitemOpen
  \bibfield  {author} {\bibinfo {author} {\bibfnamefont {M.-Y.}\ \bibnamefont
  {Cui}}, \bibinfo {author} {\bibfnamefont {Q.}~\bibnamefont {Yuan}}, \bibinfo
  {author} {\bibfnamefont {Y.-L.~S.}\ \bibnamefont {Tsai}}, \ and\ \bibinfo
  {author} {\bibfnamefont {Y.-Z.}\ \bibnamefont {Fan}},\ }\href {\doibase
  10.1103/PhysRevLett.118.191101} {\bibfield  {journal} {\bibinfo  {journal}
  {Phys. Rev. Lett.}\ }\textbf {\bibinfo {volume} {118}},\ \bibinfo {pages}
  {191101} (\bibinfo {year} {2017})}\BibitemShut {NoStop}%
\bibitem [{\citenamefont {{Connell}}(1998)}]{Ulysses}%
  \BibitemOpen
  \bibfield  {author} {\bibinfo {author} {\bibfnamefont {J.~J.}\ \bibnamefont
  {{Connell}}},\ }\href {\doibase 10.1086/311437} {\bibfield  {journal}
  {\bibinfo  {journal} {\apjl}\ }\textbf {\bibinfo {volume} {501}},\ \bibinfo
  {pages} {L59} (\bibinfo {year} {1998})}\BibitemShut {NoStop}%
\bibitem [{\citenamefont {{Yanasak}}\ \emph {et~al.}(2001)\citenamefont
  {{Yanasak}}, \citenamefont {{Wiedenbeck}}, \citenamefont {{Mewaldt}},
  \citenamefont {{Davis}}, \citenamefont {{Cummings}}, \citenamefont
  {{George}}, \citenamefont {{Leske}}, \citenamefont {{Stone}}, \citenamefont
  {{Christian}}, \citenamefont {{von Rosenvinge}}, \citenamefont {{Binns}},
  \citenamefont {{Hink}},\ and\ \citenamefont {{Israel}}}]{ACE}%
  \BibitemOpen
  \bibfield  {author} {\bibinfo {author} {\bibfnamefont {N.~E.}\ \bibnamefont
  {{Yanasak}}}, \bibinfo {author} {\bibfnamefont {M.~E.}\ \bibnamefont
  {{Wiedenbeck}}}, \bibinfo {author} {\bibfnamefont {R.~A.}\ \bibnamefont
  {{Mewaldt}}}, \bibinfo {author} {\bibfnamefont {A.~J.}\ \bibnamefont
  {{Davis}}}, \bibinfo {author} {\bibfnamefont {A.~C.}\ \bibnamefont
  {{Cummings}}}, \bibinfo {author} {\bibfnamefont {J.~S.}\ \bibnamefont
  {{George}}}, \bibinfo {author} {\bibfnamefont {R.~A.}\ \bibnamefont
  {{Leske}}}, \bibinfo {author} {\bibfnamefont {E.~C.}\ \bibnamefont
  {{Stone}}}, \bibinfo {author} {\bibfnamefont {E.~R.}\ \bibnamefont
  {{Christian}}}, \bibinfo {author} {\bibfnamefont {T.~T.}\ \bibnamefont {{von
  Rosenvinge}}}, \bibinfo {author} {\bibfnamefont {W.~R.}\ \bibnamefont
  {{Binns}}}, \bibinfo {author} {\bibfnamefont {P.~L.}\ \bibnamefont {{Hink}}},
  \ and\ \bibinfo {author} {\bibfnamefont {M.~H.}\ \bibnamefont {{Israel}}},\
  }\href {\doibase 10.1086/323842} {\bibfield  {journal} {\bibinfo  {journal}
  {\apj}\ }\textbf {\bibinfo {volume} {563}},\ \bibinfo {pages} {768} (\bibinfo
  {year} {2001})}\BibitemShut {NoStop}%
\bibitem [{\citenamefont {{Lukasiak}}(1999)}]{Voyager}%
  \BibitemOpen
  \bibfield  {author} {\bibinfo {author} {\bibfnamefont {A.}~\bibnamefont
  {{Lukasiak}}},\ }\href@noop {} {\bibfield  {journal} {\bibinfo  {journal}
  {International Cosmic Ray Conference}\ }\textbf {\bibinfo {volume} {3}},\
  \bibinfo {pages} {41} (\bibinfo {year} {1999})}\BibitemShut {NoStop}%
\bibitem [{\citenamefont {{Simpson}}\ and\ \citenamefont
  {{Garcia-Munoz}}(1988)}]{IMP_ISEE3}%
  \BibitemOpen
  \bibfield  {author} {\bibinfo {author} {\bibfnamefont {J.~A.}\ \bibnamefont
  {{Simpson}}}\ and\ \bibinfo {author} {\bibfnamefont {M.}~\bibnamefont
  {{Garcia-Munoz}}},\ }\href {\doibase 10.1007/BF00212240} {\bibfield
  {journal} {\bibinfo  {journal} {\ssr}\ }\textbf {\bibinfo {volume} {46}},\
  \bibinfo {pages} {205} (\bibinfo {year} {1988})}\BibitemShut {NoStop}%
\bibitem [{\citenamefont {{Hams}}\ \emph {et~al.}(2004)\citenamefont {{Hams}},
  \citenamefont {{Barbier}}, \citenamefont {{Bremerich}}, \citenamefont
  {{Christian}}, \citenamefont {{de Nolfo}}, \citenamefont {{Geier}},
  \citenamefont {{G{\"o}bel}}, \citenamefont {{Gupta}}, \citenamefont {{Hof}},
  \citenamefont {{Menn}}, \citenamefont {{Mewaldt}}, \citenamefont
  {{Mitchell}}, \citenamefont {{Schindler}}, \citenamefont {{Simon}},\ and\
  \citenamefont {{Streitmatter}}}]{ISOMAX}%
  \BibitemOpen
  \bibfield  {author} {\bibinfo {author} {\bibfnamefont {T.}~\bibnamefont
  {{Hams}}}, \bibinfo {author} {\bibfnamefont {L.~M.}\ \bibnamefont
  {{Barbier}}}, \bibinfo {author} {\bibfnamefont {M.}~\bibnamefont
  {{Bremerich}}}, \bibinfo {author} {\bibfnamefont {E.~R.}\ \bibnamefont
  {{Christian}}}, \bibinfo {author} {\bibfnamefont {G.~A.}\ \bibnamefont {{de
  Nolfo}}}, \bibinfo {author} {\bibfnamefont {S.}~\bibnamefont {{Geier}}},
  \bibinfo {author} {\bibfnamefont {H.}~\bibnamefont {{G{\"o}bel}}}, \bibinfo
  {author} {\bibfnamefont {S.~K.}\ \bibnamefont {{Gupta}}}, \bibinfo {author}
  {\bibfnamefont {M.}~\bibnamefont {{Hof}}}, \bibinfo {author} {\bibfnamefont
  {W.}~\bibnamefont {{Menn}}}, \bibinfo {author} {\bibfnamefont {R.~A.}\
  \bibnamefont {{Mewaldt}}}, \bibinfo {author} {\bibfnamefont {J.~W.}\
  \bibnamefont {{Mitchell}}}, \bibinfo {author} {\bibfnamefont {S.~M.}\
  \bibnamefont {{Schindler}}}, \bibinfo {author} {\bibfnamefont
  {M.}~\bibnamefont {{Simon}}}, \ and\ \bibinfo {author} {\bibfnamefont
  {R.~E.}\ \bibnamefont {{Streitmatter}}},\ }\href {\doibase 10.1086/422384}
  {\bibfield  {journal} {\bibinfo  {journal} {\apj}\ }\textbf {\bibinfo
  {volume} {611}},\ \bibinfo {pages} {892} (\bibinfo {year}
  {2004})}\BibitemShut {NoStop}%
\bibitem [{\citenamefont {{Adriani}}\ \emph {et~al.}(2013)\citenamefont
  {{Adriani}}, \citenamefont {{Barbarino}}, \citenamefont {{Bazilevskaya}},
  \citenamefont {{Bellotti}}, \citenamefont {{Boezio}}, \citenamefont
  {{Bogomolov}}, \citenamefont {{Bongi}}, \citenamefont {{Bonvicini}},
  \citenamefont {{Borisov}}, \citenamefont {{Bottai}}, \citenamefont {{Bruno}},
  \citenamefont {{Cafagna}}, \citenamefont {{Campana}}, \citenamefont
  {{Carbone}}, \citenamefont {{Carlson}}, \citenamefont {{Casolino}},
  \citenamefont {{Castellini}}, \citenamefont {{De Pascale}}, \citenamefont
  {{De Santis}}, \citenamefont {{De Simone}}, \citenamefont {{Di Felice}},
  \citenamefont {{Formato}}, \citenamefont {{Galper}}, \citenamefont
  {{Grishantseva}}, \citenamefont {{Karelin}}, \citenamefont {{Koldashov}},
  \citenamefont {{Koldobskiy}}, \citenamefont {{Krutkov}}, \citenamefont
  {{Kvashnin}}, \citenamefont {{Leonov}}, \citenamefont {{Malakhov}},
  \citenamefont {{Marcelli}}, \citenamefont {{Mayorov}}, \citenamefont
  {{Menn}}, \citenamefont {{Mikhailov}}, \citenamefont {{Mocchiutti}},
  \citenamefont {{Monaco}}, \citenamefont {{Mori}}, \citenamefont {{Nikonov}},
  \citenamefont {{Osteria}}, \citenamefont {{Palma}}, \citenamefont {{Papini}},
  \citenamefont {{Pearce}}, \citenamefont {{Picozza}}, \citenamefont
  {{Pizzolotto}}, \citenamefont {{Ricci}}, \citenamefont {{Ricciarini}},
  \citenamefont {{Rossetto}}, \citenamefont {{Sarkar}}, \citenamefont
  {{Simon}}, \citenamefont {{Sparvoli}}, \citenamefont {{Spillantini}},
  \citenamefont {{Stozhkov}}, \citenamefont {{Vacchi}}, \citenamefont
  {{Vannuccini}}, \citenamefont {{Vasilyev}}, \citenamefont {{Voronov}},
  \citenamefont {{Yurkin}}, \citenamefont {{Wu}}, \citenamefont {{Zampa}},
  \citenamefont {{Zampa}}, \citenamefont {{Zverev}}, \citenamefont
  {{Potgieter}},\ and\ \citenamefont {{Vos}}}]{PAMELA2013}%
  \BibitemOpen
  \bibfield  {author} {\bibinfo {author} {\bibfnamefont {O.}~\bibnamefont
  {{Adriani}}}, \bibinfo {author} {\bibfnamefont {G.~C.}\ \bibnamefont
  {{Barbarino}}}, \bibinfo {author} {\bibfnamefont {G.~A.}\ \bibnamefont
  {{Bazilevskaya}}}, \bibinfo {author} {\bibfnamefont {R.}~\bibnamefont
  {{Bellotti}}}, \bibinfo {author} {\bibfnamefont {M.}~\bibnamefont
  {{Boezio}}}, \bibinfo {author} {\bibfnamefont {E.~A.}\ \bibnamefont
  {{Bogomolov}}}, \bibinfo {author} {\bibfnamefont {M.}~\bibnamefont
  {{Bongi}}}, \bibinfo {author} {\bibfnamefont {V.}~\bibnamefont
  {{Bonvicini}}}, \bibinfo {author} {\bibfnamefont {S.}~\bibnamefont
  {{Borisov}}}, \bibinfo {author} {\bibfnamefont {S.}~\bibnamefont {{Bottai}}},
  \bibinfo {author} {\bibfnamefont {A.}~\bibnamefont {{Bruno}}}, \bibinfo
  {author} {\bibfnamefont {F.}~\bibnamefont {{Cafagna}}}, \bibinfo {author}
  {\bibfnamefont {D.}~\bibnamefont {{Campana}}}, \bibinfo {author}
  {\bibfnamefont {R.}~\bibnamefont {{Carbone}}}, \bibinfo {author}
  {\bibfnamefont {P.}~\bibnamefont {{Carlson}}}, \bibinfo {author}
  {\bibfnamefont {M.}~\bibnamefont {{Casolino}}}, \bibinfo {author}
  {\bibfnamefont {G.}~\bibnamefont {{Castellini}}}, \bibinfo {author}
  {\bibfnamefont {M.~P.}\ \bibnamefont {{De Pascale}}}, \bibinfo {author}
  {\bibfnamefont {C.}~\bibnamefont {{De Santis}}}, \bibinfo {author}
  {\bibfnamefont {N.}~\bibnamefont {{De Simone}}}, \bibinfo {author}
  {\bibfnamefont {V.}~\bibnamefont {{Di Felice}}}, \bibinfo {author}
  {\bibfnamefont {V.}~\bibnamefont {{Formato}}}, \bibinfo {author}
  {\bibfnamefont {A.~M.}\ \bibnamefont {{Galper}}}, \bibinfo {author}
  {\bibfnamefont {L.}~\bibnamefont {{Grishantseva}}}, \bibinfo {author}
  {\bibfnamefont {A.~V.}\ \bibnamefont {{Karelin}}}, \bibinfo {author}
  {\bibfnamefont {S.~V.}\ \bibnamefont {{Koldashov}}}, \bibinfo {author}
  {\bibfnamefont {S.}~\bibnamefont {{Koldobskiy}}}, \bibinfo {author}
  {\bibfnamefont {S.~Y.}\ \bibnamefont {{Krutkov}}}, \bibinfo {author}
  {\bibfnamefont {A.~N.}\ \bibnamefont {{Kvashnin}}}, \bibinfo {author}
  {\bibfnamefont {A.}~\bibnamefont {{Leonov}}}, \bibinfo {author}
  {\bibfnamefont {V.}~\bibnamefont {{Malakhov}}}, \bibinfo {author}
  {\bibfnamefont {L.}~\bibnamefont {{Marcelli}}}, \bibinfo {author}
  {\bibfnamefont {A.~G.}\ \bibnamefont {{Mayorov}}}, \bibinfo {author}
  {\bibfnamefont {W.}~\bibnamefont {{Menn}}}, \bibinfo {author} {\bibfnamefont
  {V.~V.}\ \bibnamefont {{Mikhailov}}}, \bibinfo {author} {\bibfnamefont
  {E.}~\bibnamefont {{Mocchiutti}}}, \bibinfo {author} {\bibfnamefont
  {A.}~\bibnamefont {{Monaco}}}, \bibinfo {author} {\bibfnamefont
  {N.}~\bibnamefont {{Mori}}}, \bibinfo {author} {\bibfnamefont
  {N.}~\bibnamefont {{Nikonov}}}, \bibinfo {author} {\bibfnamefont
  {G.}~\bibnamefont {{Osteria}}}, \bibinfo {author} {\bibfnamefont
  {F.}~\bibnamefont {{Palma}}}, \bibinfo {author} {\bibfnamefont
  {P.}~\bibnamefont {{Papini}}}, \bibinfo {author} {\bibfnamefont
  {M.}~\bibnamefont {{Pearce}}}, \bibinfo {author} {\bibfnamefont
  {P.}~\bibnamefont {{Picozza}}}, \bibinfo {author} {\bibfnamefont
  {C.}~\bibnamefont {{Pizzolotto}}}, \bibinfo {author} {\bibfnamefont
  {M.}~\bibnamefont {{Ricci}}}, \bibinfo {author} {\bibfnamefont {S.~B.}\
  \bibnamefont {{Ricciarini}}}, \bibinfo {author} {\bibfnamefont
  {L.}~\bibnamefont {{Rossetto}}}, \bibinfo {author} {\bibfnamefont
  {R.}~\bibnamefont {{Sarkar}}}, \bibinfo {author} {\bibfnamefont
  {M.}~\bibnamefont {{Simon}}}, \bibinfo {author} {\bibfnamefont
  {R.}~\bibnamefont {{Sparvoli}}}, \bibinfo {author} {\bibfnamefont
  {P.}~\bibnamefont {{Spillantini}}}, \bibinfo {author} {\bibfnamefont {Y.~I.}\
  \bibnamefont {{Stozhkov}}}, \bibinfo {author} {\bibfnamefont
  {A.}~\bibnamefont {{Vacchi}}}, \bibinfo {author} {\bibfnamefont
  {E.}~\bibnamefont {{Vannuccini}}}, \bibinfo {author} {\bibfnamefont
  {G.}~\bibnamefont {{Vasilyev}}}, \bibinfo {author} {\bibfnamefont {S.~A.}\
  \bibnamefont {{Voronov}}}, \bibinfo {author} {\bibfnamefont {Y.~T.}\
  \bibnamefont {{Yurkin}}}, \bibinfo {author} {\bibfnamefont {J.}~\bibnamefont
  {{Wu}}}, \bibinfo {author} {\bibfnamefont {G.}~\bibnamefont {{Zampa}}},
  \bibinfo {author} {\bibfnamefont {N.}~\bibnamefont {{Zampa}}}, \bibinfo
  {author} {\bibfnamefont {V.~G.}\ \bibnamefont {{Zverev}}}, \bibinfo {author}
  {\bibfnamefont {M.~S.}\ \bibnamefont {{Potgieter}}}, \ and\ \bibinfo {author}
  {\bibfnamefont {E.~E.}\ \bibnamefont {{Vos}}},\ }\href {\doibase
  10.1088/0004-637X/765/2/91} {\bibfield  {journal} {\bibinfo  {journal}
  {\apj}\ }\textbf {\bibinfo {volume} {765}},\ \bibinfo {eid} {91} (\bibinfo
  {year} {2013})},\ \Eprint {http://arxiv.org/abs/1301.4108} {arXiv:1301.4108
  [astro-ph.HE]} \BibitemShut {NoStop}%
\bibitem [{\citenamefont {{Maurin}}\ \emph {et~al.}(2001)\citenamefont
  {{Maurin}}, \citenamefont {{Donato}}, \citenamefont {{Taillet}},\ and\
  \citenamefont {{Salati}}}]{Maurin2001}%
  \BibitemOpen
  \bibfield  {author} {\bibinfo {author} {\bibfnamefont {D.}~\bibnamefont
  {{Maurin}}}, \bibinfo {author} {\bibfnamefont {F.}~\bibnamefont {{Donato}}},
  \bibinfo {author} {\bibfnamefont {R.}~\bibnamefont {{Taillet}}}, \ and\
  \bibinfo {author} {\bibfnamefont {P.}~\bibnamefont {{Salati}}},\ }\href
  {\doibase 10.1086/321496} {\bibfield  {journal} {\bibinfo  {journal} {\apj}\
  }\textbf {\bibinfo {volume} {555}},\ \bibinfo {pages} {585} (\bibinfo {year}
  {2001})},\ \Eprint {http://arxiv.org/abs/astro-ph/0101231} {astro-ph/0101231}
  \BibitemShut {NoStop}%
\bibitem [{\citenamefont
  {{Tomassetti}}(2015{\natexlab{a}})}]{Tomassetti2015prc}%
  \BibitemOpen
  \bibfield  {author} {\bibinfo {author} {\bibfnamefont {N.}~\bibnamefont
  {{Tomassetti}}},\ }\href {\doibase 10.1103/PhysRevC.92.045808} {\bibfield
  {journal} {\bibinfo  {journal} {\prc}\ }\textbf {\bibinfo {volume} {92}},\
  \bibinfo {eid} {045808} (\bibinfo {year} {2015}{\natexlab{a}})},\ \Eprint
  {http://arxiv.org/abs/1509.05776} {arXiv:1509.05776 [astro-ph.HE]}
  \BibitemShut {NoStop}%
\bibitem [{\citenamefont {{Lin}}\ \emph {et~al.}(2016)\citenamefont {{Lin}},
  \citenamefont {{Bi}}, \citenamefont {{Feng}}, \citenamefont {{Yin}},\ and\
  \citenamefont {{Yu}}}]{Lin2016}%
  \BibitemOpen
  \bibfield  {author} {\bibinfo {author} {\bibfnamefont {S.-J.}\ \bibnamefont
  {{Lin}}}, \bibinfo {author} {\bibfnamefont {X.-J.}\ \bibnamefont {{Bi}}},
  \bibinfo {author} {\bibfnamefont {J.}~\bibnamefont {{Feng}}}, \bibinfo
  {author} {\bibfnamefont {P.-F.}\ \bibnamefont {{Yin}}}, \ and\ \bibinfo
  {author} {\bibfnamefont {Z.-H.}\ \bibnamefont {{Yu}}},\ }\href@noop {}
  {\bibfield  {journal} {\bibinfo  {journal} {ArXiv e-prints}\ } (\bibinfo
  {year} {2016})},\ \Eprint {http://arxiv.org/abs/1612.04001} {arXiv:1612.04001
  [astro-ph.HE]} \BibitemShut {NoStop}%
\bibitem [{\citenamefont {{Boschini}}\ \emph
  {et~al.}(2017{\natexlab{a}})\citenamefont {{Boschini}}, \citenamefont {{Della
  Torre}}, \citenamefont {{Gervasi}}, \citenamefont {{Grandi}}, \citenamefont
  {{J{\'o}hannesson}}, \citenamefont {{Kachelriess}}, \citenamefont {{La
  Vacca}}, \citenamefont {{Masi}}, \citenamefont {{Moskalenko}}, \citenamefont
  {{Orlando}}, \citenamefont {{Ostapchenko}}, \citenamefont {{Pensotti}},
  \citenamefont {{Porter}}, \citenamefont {{Quadrani}}, \citenamefont
  {{Rancoita}}, \citenamefont {{Rozza}},\ and\ \citenamefont
  {{Tacconi}}}]{Boschini2017}%
  \BibitemOpen
  \bibfield  {author} {\bibinfo {author} {\bibfnamefont {M.~J.}\ \bibnamefont
  {{Boschini}}}, \bibinfo {author} {\bibfnamefont {S.}~\bibnamefont {{Della
  Torre}}}, \bibinfo {author} {\bibfnamefont {M.}~\bibnamefont {{Gervasi}}},
  \bibinfo {author} {\bibfnamefont {D.}~\bibnamefont {{Grandi}}}, \bibinfo
  {author} {\bibfnamefont {G.}~\bibnamefont {{J{\'o}hannesson}}}, \bibinfo
  {author} {\bibfnamefont {M.}~\bibnamefont {{Kachelriess}}}, \bibinfo {author}
  {\bibfnamefont {G.}~\bibnamefont {{La Vacca}}}, \bibinfo {author}
  {\bibfnamefont {N.}~\bibnamefont {{Masi}}}, \bibinfo {author} {\bibfnamefont
  {I.~V.}\ \bibnamefont {{Moskalenko}}}, \bibinfo {author} {\bibfnamefont
  {E.}~\bibnamefont {{Orlando}}}, \bibinfo {author} {\bibfnamefont {S.~S.}\
  \bibnamefont {{Ostapchenko}}}, \bibinfo {author} {\bibfnamefont
  {S.}~\bibnamefont {{Pensotti}}}, \bibinfo {author} {\bibfnamefont {T.~A.}\
  \bibnamefont {{Porter}}}, \bibinfo {author} {\bibfnamefont {L.}~\bibnamefont
  {{Quadrani}}}, \bibinfo {author} {\bibfnamefont {P.~G.}\ \bibnamefont
  {{Rancoita}}}, \bibinfo {author} {\bibfnamefont {D.}~\bibnamefont {{Rozza}}},
  \ and\ \bibinfo {author} {\bibfnamefont {M.}~\bibnamefont {{Tacconi}}},\
  }\href {\doibase 10.3847/1538-4357/aa6e4f} {\bibfield  {journal} {\bibinfo
  {journal} {\apj}\ }\textbf {\bibinfo {volume} {840}},\ \bibinfo {eid} {115}
  (\bibinfo {year} {2017}{\natexlab{a}})},\ \Eprint
  {http://arxiv.org/abs/1704.06337} {arXiv:1704.06337 [astro-ph.HE]}
  \BibitemShut {NoStop}%
\bibitem [{\citenamefont {{Genolini}}\ \emph {et~al.}(2017)\citenamefont
  {{Genolini}}, \citenamefont {{Serpico}}, \citenamefont {{Boudaud}},
  \citenamefont {{Caroff}}, \citenamefont {{Poulin}}, \citenamefont {{Derome}},
  \citenamefont {{Lavalle}}, \citenamefont {{Maurin}}, \citenamefont
  {{Poireau}}, \citenamefont {{Rosier}}, \citenamefont {{Salati}},\ and\
  \citenamefont {{Vecchi}}}]{Genolini2017}%
  \BibitemOpen
  \bibfield  {author} {\bibinfo {author} {\bibfnamefont {Y.}~\bibnamefont
  {{Genolini}}}, \bibinfo {author} {\bibfnamefont {P.~D.}\ \bibnamefont
  {{Serpico}}}, \bibinfo {author} {\bibfnamefont {M.}~\bibnamefont
  {{Boudaud}}}, \bibinfo {author} {\bibfnamefont {S.}~\bibnamefont {{Caroff}}},
  \bibinfo {author} {\bibfnamefont {V.}~\bibnamefont {{Poulin}}}, \bibinfo
  {author} {\bibfnamefont {L.}~\bibnamefont {{Derome}}}, \bibinfo {author}
  {\bibfnamefont {J.}~\bibnamefont {{Lavalle}}}, \bibinfo {author}
  {\bibfnamefont {D.}~\bibnamefont {{Maurin}}}, \bibinfo {author}
  {\bibfnamefont {V.}~\bibnamefont {{Poireau}}}, \bibinfo {author}
  {\bibfnamefont {S.}~\bibnamefont {{Rosier}}}, \bibinfo {author}
  {\bibfnamefont {P.}~\bibnamefont {{Salati}}}, \ and\ \bibinfo {author}
  {\bibfnamefont {M.}~\bibnamefont {{Vecchi}}},\ }\href@noop {} {\bibfield
  {journal} {\bibinfo  {journal} {ArXiv e-prints}\ } (\bibinfo {year}
  {2017})},\ \Eprint {http://arxiv.org/abs/1706.09812} {arXiv:1706.09812
  [astro-ph.HE]} \BibitemShut {NoStop}%
\bibitem [{\citenamefont {{Blasi}}\ \emph {et~al.}(2012)\citenamefont
  {{Blasi}}, \citenamefont {{Amato}},\ and\ \citenamefont
  {{Serpico}}}]{Blasi2012}%
  \BibitemOpen
  \bibfield  {author} {\bibinfo {author} {\bibfnamefont {P.}~\bibnamefont
  {{Blasi}}}, \bibinfo {author} {\bibfnamefont {E.}~\bibnamefont {{Amato}}}, \
  and\ \bibinfo {author} {\bibfnamefont {P.~D.}\ \bibnamefont {{Serpico}}},\
  }\href {\doibase 10.1103/PhysRevLett.109.061101} {\bibfield  {journal}
  {\bibinfo  {journal} {Physical Review Letters}\ }\textbf {\bibinfo {volume}
  {109}},\ \bibinfo {eid} {061101} (\bibinfo {year} {2012})},\ \Eprint
  {http://arxiv.org/abs/1207.3706} {arXiv:1207.3706 [astro-ph.HE]} \BibitemShut
  {NoStop}%
\bibitem [{\citenamefont {{Tomassetti}}(2012)}]{Tomassetti2012}%
  \BibitemOpen
  \bibfield  {author} {\bibinfo {author} {\bibfnamefont {N.}~\bibnamefont
  {{Tomassetti}}},\ }\href {\doibase 10.1088/2041-8205/752/1/L13} {\bibfield
  {journal} {\bibinfo  {journal} {\apjl}\ }\textbf {\bibinfo {volume} {752}},\
  \bibinfo {eid} {L13} (\bibinfo {year} {2012})},\ \Eprint
  {http://arxiv.org/abs/1204.4492} {arXiv:1204.4492 [astro-ph.HE]} \BibitemShut
  {NoStop}%
\bibitem [{\citenamefont
  {{Tomassetti}}(2015{\natexlab{b}})}]{Tomassetti2015apjl01}%
  \BibitemOpen
  \bibfield  {author} {\bibinfo {author} {\bibfnamefont {N.}~\bibnamefont
  {{Tomassetti}}},\ }\href {\doibase 10.1088/2041-8205/815/1/L1} {\bibfield
  {journal} {\bibinfo  {journal} {\apjl}\ }\textbf {\bibinfo {volume} {815}},\
  \bibinfo {eid} {L1} (\bibinfo {year} {2015}{\natexlab{b}})},\ \Eprint
  {http://arxiv.org/abs/1511.04460} {arXiv:1511.04460 [astro-ph.HE]}
  \BibitemShut {NoStop}%
\bibitem [{\citenamefont
  {{Tomassetti}}(2015{\natexlab{c}})}]{Tomassetti2015prd}%
  \BibitemOpen
  \bibfield  {author} {\bibinfo {author} {\bibfnamefont {N.}~\bibnamefont
  {{Tomassetti}}},\ }\href {\doibase 10.1103/PhysRevD.92.081301} {\bibfield
  {journal} {\bibinfo  {journal} {\prd}\ }\textbf {\bibinfo {volume} {92}},\
  \bibinfo {eid} {081301} (\bibinfo {year} {2015}{\natexlab{c}})},\ \Eprint
  {http://arxiv.org/abs/1509.05775} {arXiv:1509.05775 [astro-ph.HE]}
  \BibitemShut {NoStop}%
\bibitem [{\citenamefont {{Guo}}\ \emph {et~al.}(2016)\citenamefont {{Guo}},
  \citenamefont {{Tian}},\ and\ \citenamefont {{Jin}}}]{Guo2016}%
  \BibitemOpen
  \bibfield  {author} {\bibinfo {author} {\bibfnamefont {Y.-Q.}\ \bibnamefont
  {{Guo}}}, \bibinfo {author} {\bibfnamefont {Z.}~\bibnamefont {{Tian}}}, \
  and\ \bibinfo {author} {\bibfnamefont {C.}~\bibnamefont {{Jin}}},\ }\href
  {\doibase 10.3847/0004-637X/819/1/54} {\bibfield  {journal} {\bibinfo
  {journal} {\apj}\ }\textbf {\bibinfo {volume} {819}},\ \bibinfo {eid} {54}
  (\bibinfo {year} {2016})}\BibitemShut {NoStop}%
\bibitem [{\citenamefont {{Feng}}\ \emph {et~al.}(2016)\citenamefont {{Feng}},
  \citenamefont {{Tomassetti}},\ and\ \citenamefont {{Oliva}}}]{Feng2016}%
  \BibitemOpen
  \bibfield  {author} {\bibinfo {author} {\bibfnamefont {J.}~\bibnamefont
  {{Feng}}}, \bibinfo {author} {\bibfnamefont {N.}~\bibnamefont
  {{Tomassetti}}}, \ and\ \bibinfo {author} {\bibfnamefont {A.}~\bibnamefont
  {{Oliva}}},\ }\href {\doibase 10.1103/PhysRevD.94.123007} {\bibfield
  {journal} {\bibinfo  {journal} {\prd}\ }\textbf {\bibinfo {volume} {94}},\
  \bibinfo {eid} {123007} (\bibinfo {year} {2016})},\ \Eprint
  {http://arxiv.org/abs/1610.06182} {arXiv:1610.06182 [astro-ph.HE]}
  \BibitemShut {NoStop}%
\bibitem [{\citenamefont {{Vladimirov}}\ \emph {et~al.}(2012)\citenamefont
  {{Vladimirov}}, \citenamefont {{J{\'o}hannesson}}, \citenamefont
  {{Moskalenko}},\ and\ \citenamefont {{Porter}}}]{Vladimirov2012}%
  \BibitemOpen
  \bibfield  {author} {\bibinfo {author} {\bibfnamefont {A.~E.}\ \bibnamefont
  {{Vladimirov}}}, \bibinfo {author} {\bibfnamefont {G.}~\bibnamefont
  {{J{\'o}hannesson}}}, \bibinfo {author} {\bibfnamefont {I.~V.}\ \bibnamefont
  {{Moskalenko}}}, \ and\ \bibinfo {author} {\bibfnamefont {T.~A.}\
  \bibnamefont {{Porter}}},\ }\href {\doibase 10.1088/0004-637X/752/1/68}
  {\bibfield  {journal} {\bibinfo  {journal} {\apj}\ }\textbf {\bibinfo
  {volume} {752}},\ \bibinfo {eid} {68} (\bibinfo {year} {2012})},\ \Eprint
  {http://arxiv.org/abs/1108.1023} {arXiv:1108.1023 [astro-ph.HE]} \BibitemShut
  {NoStop}%
\bibitem [{\citenamefont {{Bernard}}\ \emph {et~al.}(2013)\citenamefont
  {{Bernard}}, \citenamefont {{Delahaye}}, \citenamefont {{Keum}},
  \citenamefont {{Liu}}, \citenamefont {{Salati}},\ and\ \citenamefont
  {{Taillet}}}]{Bernard2013}%
  \BibitemOpen
  \bibfield  {author} {\bibinfo {author} {\bibfnamefont {G.}~\bibnamefont
  {{Bernard}}}, \bibinfo {author} {\bibfnamefont {T.}~\bibnamefont
  {{Delahaye}}}, \bibinfo {author} {\bibfnamefont {Y.-Y.}\ \bibnamefont
  {{Keum}}}, \bibinfo {author} {\bibfnamefont {W.}~\bibnamefont {{Liu}}},
  \bibinfo {author} {\bibfnamefont {P.}~\bibnamefont {{Salati}}}, \ and\
  \bibinfo {author} {\bibfnamefont {R.}~\bibnamefont {{Taillet}}},\ }\href
  {\doibase 10.1051/0004-6361/201321202} {\bibfield  {journal} {\bibinfo
  {journal} {\aap}\ }\textbf {\bibinfo {volume} {555}},\ \bibinfo {eid} {A48}
  (\bibinfo {year} {2013})},\ \Eprint {http://arxiv.org/abs/1207.4670}
  {arXiv:1207.4670 [astro-ph.HE]} \BibitemShut {NoStop}%
\bibitem [{\citenamefont {{Thoudam}}\ and\ \citenamefont
  {{H{\"o}randel}}(2013)}]{Thoudam2013}%
  \BibitemOpen
  \bibfield  {author} {\bibinfo {author} {\bibfnamefont {S.}~\bibnamefont
  {{Thoudam}}}\ and\ \bibinfo {author} {\bibfnamefont {J.~R.}\ \bibnamefont
  {{H{\"o}randel}}},\ }\href {\doibase 10.1093/mnras/stt1464} {\bibfield
  {journal} {\bibinfo  {journal} {\mnras}\ }\textbf {\bibinfo {volume} {435}},\
  \bibinfo {pages} {2532} (\bibinfo {year} {2013})},\ \Eprint
  {http://arxiv.org/abs/1304.1400} {arXiv:1304.1400 [astro-ph.HE]} \BibitemShut
  {NoStop}%
\bibitem [{\citenamefont {{Tomassetti}}\ and\ \citenamefont
  {{Donato}}(2015)}]{Tomassetti2015apjl02}%
  \BibitemOpen
  \bibfield  {author} {\bibinfo {author} {\bibfnamefont {N.}~\bibnamefont
  {{Tomassetti}}}\ and\ \bibinfo {author} {\bibfnamefont {F.}~\bibnamefont
  {{Donato}}},\ }\href {\doibase 10.1088/2041-8205/803/2/L15} {\bibfield
  {journal} {\bibinfo  {journal} {\apjl}\ }\textbf {\bibinfo {volume} {803}},\
  \bibinfo {eid} {L15} (\bibinfo {year} {2015})},\ \Eprint
  {http://arxiv.org/abs/1502.06150} {arXiv:1502.06150 [astro-ph.HE]}
  \BibitemShut {NoStop}%
\bibitem [{\citenamefont {Cuoco}\ \emph {et~al.}(2017)\citenamefont {Cuoco},
  \citenamefont {Kr\"amer},\ and\ \citenamefont {Korsmeier}}]{Cuoco2017}%
  \BibitemOpen
  \bibfield  {author} {\bibinfo {author} {\bibfnamefont {A.}~\bibnamefont
  {Cuoco}}, \bibinfo {author} {\bibfnamefont {M.}~\bibnamefont {Kr\"amer}}, \
  and\ \bibinfo {author} {\bibfnamefont {M.}~\bibnamefont {Korsmeier}},\ }\href
  {\doibase 10.1103/PhysRevLett.118.191102} {\bibfield  {journal} {\bibinfo
  {journal} {Phys. Rev. Lett.}\ }\textbf {\bibinfo {volume} {118}},\ \bibinfo
  {pages} {191102} (\bibinfo {year} {2017})}\BibitemShut {NoStop}%
\bibitem [{\citenamefont {{Cummings}}\ \emph {et~al.}(2016)\citenamefont
  {{Cummings}}, \citenamefont {{Stone}}, \citenamefont {{Heikkila}},
  \citenamefont {{Lal}}, \citenamefont {{Webber}}, \citenamefont
  {{J{\'o}hannesson}}, \citenamefont {{Moskalenko}}, \citenamefont
  {{Orlando}},\ and\ \citenamefont {{Porter}}}]{Cummings2016}%
  \BibitemOpen
  \bibfield  {author} {\bibinfo {author} {\bibfnamefont {A.~C.}\ \bibnamefont
  {{Cummings}}}, \bibinfo {author} {\bibfnamefont {E.~C.}\ \bibnamefont
  {{Stone}}}, \bibinfo {author} {\bibfnamefont {B.~C.}\ \bibnamefont
  {{Heikkila}}}, \bibinfo {author} {\bibfnamefont {N.}~\bibnamefont {{Lal}}},
  \bibinfo {author} {\bibfnamefont {W.~R.}\ \bibnamefont {{Webber}}}, \bibinfo
  {author} {\bibfnamefont {G.}~\bibnamefont {{J{\'o}hannesson}}}, \bibinfo
  {author} {\bibfnamefont {I.~V.}\ \bibnamefont {{Moskalenko}}}, \bibinfo
  {author} {\bibfnamefont {E.}~\bibnamefont {{Orlando}}}, \ and\ \bibinfo
  {author} {\bibfnamefont {T.~A.}\ \bibnamefont {{Porter}}},\ }\href {\doibase
  10.3847/0004-637X/831/1/18} {\bibfield  {journal} {\bibinfo  {journal}
  {\apj}\ }\textbf {\bibinfo {volume} {831}},\ \bibinfo {eid} {18} (\bibinfo
  {year} {2016})}\BibitemShut {NoStop}%
\bibitem [{\citenamefont {{Erlykin}}\ and\ \citenamefont
  {{Wolfendale}}(2015)}]{Erlykin2015}%
  \BibitemOpen
  \bibfield  {author} {\bibinfo {author} {\bibfnamefont {A.~D.}\ \bibnamefont
  {{Erlykin}}}\ and\ \bibinfo {author} {\bibfnamefont {A.~W.}\ \bibnamefont
  {{Wolfendale}}},\ }\href {\doibase 10.1088/0954-3899/42/7/075201} {\bibfield
  {journal} {\bibinfo  {journal} {Journal of Physics G Nuclear Physics}\
  }\textbf {\bibinfo {volume} {42}},\ \bibinfo {eid} {075201} (\bibinfo {year}
  {2015})}\BibitemShut {NoStop}%
\bibitem [{\citenamefont {{Malkov}}\ \emph {et~al.}(2012)\citenamefont
  {{Malkov}}, \citenamefont {{Diamond}},\ and\ \citenamefont
  {{Sagdeev}}}]{Malkov2012}%
  \BibitemOpen
  \bibfield  {author} {\bibinfo {author} {\bibfnamefont {M.~A.}\ \bibnamefont
  {{Malkov}}}, \bibinfo {author} {\bibfnamefont {P.~H.}\ \bibnamefont
  {{Diamond}}}, \ and\ \bibinfo {author} {\bibfnamefont {R.~Z.}\ \bibnamefont
  {{Sagdeev}}},\ }\href {\doibase 10.1103/PhysRevLett.108.081104} {\bibfield
  {journal} {\bibinfo  {journal} {Physical Review Letters}\ }\textbf {\bibinfo
  {volume} {108}},\ \bibinfo {eid} {081104} (\bibinfo {year} {2012})},\ \Eprint
  {http://arxiv.org/abs/1110.5335} {arXiv:1110.5335 [astro-ph.GA]} \BibitemShut
  {NoStop}%
\bibitem [{\citenamefont {{Fisk}}\ and\ \citenamefont
  {{Gloeckler}}(2012)}]{Fisk2012}%
  \BibitemOpen
  \bibfield  {author} {\bibinfo {author} {\bibfnamefont {L.~A.}\ \bibnamefont
  {{Fisk}}}\ and\ \bibinfo {author} {\bibfnamefont {G.}~\bibnamefont
  {{Gloeckler}}},\ }\href {\doibase 10.1088/0004-637X/744/2/127} {\bibfield
  {journal} {\bibinfo  {journal} {\apj}\ }\textbf {\bibinfo {volume} {744}},\
  \bibinfo {eid} {127} (\bibinfo {year} {2012})}\BibitemShut {NoStop}%
\bibitem [{\citenamefont {{Ohira}}\ and\ \citenamefont
  {{Ioka}}(2011)}]{Ohira2011}%
  \BibitemOpen
  \bibfield  {author} {\bibinfo {author} {\bibfnamefont {Y.}~\bibnamefont
  {{Ohira}}}\ and\ \bibinfo {author} {\bibfnamefont {K.}~\bibnamefont
  {{Ioka}}},\ }\href {\doibase 10.1088/2041-8205/729/1/L13} {\bibfield
  {journal} {\bibinfo  {journal} {\apjl}\ }\textbf {\bibinfo {volume} {729}},\
  \bibinfo {eid} {L13} (\bibinfo {year} {2011})}\BibitemShut {NoStop}%
\bibitem [{\citenamefont {{Aloisio}}\ \emph {et~al.}(2015)\citenamefont
  {{Aloisio}}, \citenamefont {{Blasi}},\ and\ \citenamefont
  {{Serpico}}}]{Aloisio2015}%
  \BibitemOpen
  \bibfield  {author} {\bibinfo {author} {\bibfnamefont {R.}~\bibnamefont
  {{Aloisio}}}, \bibinfo {author} {\bibfnamefont {P.}~\bibnamefont {{Blasi}}},
  \ and\ \bibinfo {author} {\bibfnamefont {P.~D.}\ \bibnamefont {{Serpico}}},\
  }\href {\doibase 10.1051/0004-6361/201526877} {\bibfield  {journal} {\bibinfo
   {journal} {\aap}\ }\textbf {\bibinfo {volume} {583}},\ \bibinfo {eid} {A95}
  (\bibinfo {year} {2015})},\ \Eprint {http://arxiv.org/abs/1507.00594}
  {arXiv:1507.00594 [astro-ph.HE]} \BibitemShut {NoStop}%
\bibitem [{\citenamefont {{Boschini}}\ \emph
  {et~al.}(2017{\natexlab{b}})\citenamefont {{Boschini}}, \citenamefont {{Della
  Torre}}, \citenamefont {{Gervasi}}, \citenamefont {{La Vacca}},\ and\
  \citenamefont {{Rancoita}}}]{DellaTorre2017}%
  \BibitemOpen
  \bibfield  {author} {\bibinfo {author} {\bibfnamefont {M.}~\bibnamefont
  {{Boschini}}}, \bibinfo {author} {\bibfnamefont {S.}~\bibnamefont {{Della
  Torre}}}, \bibinfo {author} {\bibfnamefont {M.}~\bibnamefont {{Gervasi}}},
  \bibinfo {author} {\bibfnamefont {G.}~\bibnamefont {{La Vacca}}}, \ and\
  \bibinfo {author} {\bibfnamefont {P.}~\bibnamefont {{Rancoita}}},\ }\href
  {\doibase 10.1016/j.asr.2017.04.017} {\bibfield  {journal} {\bibinfo
  {journal} {Accepted on Adv. Space Res.}\ ,\ \bibinfo {pages} {0}} (\bibinfo
  {year} {2017}{\natexlab{b}})}\BibitemShut {NoStop}%
\bibitem [{\citenamefont {{Boschini}}\ \emph
  {et~al.}(2017{\natexlab{c}})\citenamefont {{Boschini}}, \citenamefont {{Della
  Torre}}, \citenamefont {{Gervasi}}, \citenamefont {{Grandi}}, \citenamefont
  {Johannesson}, \citenamefont {Kachelriess}, \citenamefont {{La Vacca}},
  \citenamefont {Masi}, \citenamefont {Moskalenko}, \citenamefont {Orlando},
  \citenamefont {Ostapchenko}, \citenamefont {Pensotti}, \citenamefont
  {Porter}, \citenamefont {Quadrani},\ and\ \citenamefont
  {{Rancoita}}}]{Boskini2017}%
  \BibitemOpen
  \bibfield  {author} {\bibinfo {author} {\bibfnamefont {M.}~\bibnamefont
  {{Boschini}}}, \bibinfo {author} {\bibfnamefont {S.}~\bibnamefont {{Della
  Torre}}}, \bibinfo {author} {\bibfnamefont {M.}~\bibnamefont {{Gervasi}}},
  \bibinfo {author} {\bibfnamefont {D.}~\bibnamefont {{Grandi}}}, \bibinfo
  {author} {\bibfnamefont {G.}~\bibnamefont {Johannesson}}, \bibinfo {author}
  {\bibfnamefont {M.}~\bibnamefont {Kachelriess}}, \bibinfo {author}
  {\bibfnamefont {G.}~\bibnamefont {{La Vacca}}}, \bibinfo {author}
  {\bibfnamefont {N.}~\bibnamefont {Masi}}, \bibinfo {author} {\bibfnamefont
  {I.}~\bibnamefont {Moskalenko}}, \bibinfo {author} {\bibfnamefont
  {E.}~\bibnamefont {Orlando}}, \bibinfo {author} {\bibfnamefont {S.~S.}\
  \bibnamefont {Ostapchenko}}, \bibinfo {author} {\bibfnamefont
  {S.}~\bibnamefont {Pensotti}}, \bibinfo {author} {\bibfnamefont {T.~A.}\
  \bibnamefont {Porter}}, \bibinfo {author} {\bibfnamefont {L.}~\bibnamefont
  {Quadrani}}, \ and\ \bibinfo {author} {\bibfnamefont {P.}~\bibnamefont
  {{Rancoita}}},\ }\href {\doibase 10.3847/1538-4357/aa6e4f} {\bibfield
  {journal} {\bibinfo  {journal} {Astrophys. J.}\ }\textbf {\bibinfo {volume}
  {840}},\ \bibinfo {pages} {115} (\bibinfo {year}
  {2017}{\natexlab{c}})}\BibitemShut {NoStop}%
\bibitem [{\citenamefont {Foreman-Mackey}\ \emph {et~al.}(2016)\citenamefont
  {Foreman-Mackey}, \citenamefont {Vousden}, \citenamefont {Price-Whelan},
  \citenamefont {Pitkin}, \citenamefont {Zabalza}, \citenamefont {Ryan},
  \citenamefont {Emily}, \citenamefont {Smith}, \citenamefont {Ashton},
  \citenamefont {Cruz}, \citenamefont {Kerzendorf}, \citenamefont {Caswell},
  \citenamefont {Hoyer}, \citenamefont {Barbary}, \citenamefont {Czekala},
  \citenamefont {Hogg},\ and\ \citenamefont {Brewer}}]{corner}%
  \BibitemOpen
  \bibfield  {author} {\bibinfo {author} {\bibfnamefont {D.}~\bibnamefont
  {Foreman-Mackey}}, \bibinfo {author} {\bibfnamefont {W.}~\bibnamefont
  {Vousden}}, \bibinfo {author} {\bibfnamefont {A.}~\bibnamefont
  {Price-Whelan}}, \bibinfo {author} {\bibfnamefont {M.}~\bibnamefont
  {Pitkin}}, \bibinfo {author} {\bibfnamefont {V.}~\bibnamefont {Zabalza}},
  \bibinfo {author} {\bibfnamefont {G.}~\bibnamefont {Ryan}}, \bibinfo {author}
  {\bibnamefont {Emily}}, \bibinfo {author} {\bibfnamefont {M.}~\bibnamefont
  {Smith}}, \bibinfo {author} {\bibfnamefont {G.}~\bibnamefont {Ashton}},
  \bibinfo {author} {\bibfnamefont {K.}~\bibnamefont {Cruz}}, \bibinfo {author}
  {\bibfnamefont {W.}~\bibnamefont {Kerzendorf}}, \bibinfo {author}
  {\bibfnamefont {T.~A.}\ \bibnamefont {Caswell}}, \bibinfo {author}
  {\bibfnamefont {S.}~\bibnamefont {Hoyer}}, \bibinfo {author} {\bibfnamefont
  {K.}~\bibnamefont {Barbary}}, \bibinfo {author} {\bibfnamefont
  {I.}~\bibnamefont {Czekala}}, \bibinfo {author} {\bibfnamefont {D.~W.}\
  \bibnamefont {Hogg}}, \ and\ \bibinfo {author} {\bibfnamefont {B.~J.}\
  \bibnamefont {Brewer}},\ }\href {\doibase 10.5281/zenodo.45906} {\enquote
  {\bibinfo {title} {corner.py: corner.py v1.0.2},}\ } (\bibinfo {year}
  {2016})\BibitemShut {NoStop}%
\bibitem [{\citenamefont {{Maurin}}\ \emph {et~al.}(2014)\citenamefont
  {{Maurin}}, \citenamefont {{Melot}},\ and\ \citenamefont
  {{Taillet}}}]{Maurin2014}%
  \BibitemOpen
  \bibfield  {author} {\bibinfo {author} {\bibfnamefont {D.}~\bibnamefont
  {{Maurin}}}, \bibinfo {author} {\bibfnamefont {F.}~\bibnamefont {{Melot}}}, \
  and\ \bibinfo {author} {\bibfnamefont {R.}~\bibnamefont {{Taillet}}},\ }\href
  {\doibase 10.1051/0004-6361/201321344} {\bibfield  {journal} {\bibinfo
  {journal} {\aap}\ }\textbf {\bibinfo {volume} {569}},\ \bibinfo {eid} {A32}
  (\bibinfo {year} {2014})},\ \Eprint {http://arxiv.org/abs/1302.5525}
  {arXiv:1302.5525 [astro-ph.HE]} \BibitemShut {NoStop}%
\end{thebibliography}


%

\end{document}